%% file: SUS-13-019_temp.tex
\pdfoutput=1

\documentclass[11pt,twoside,a4paper,cmspaper,final,collab]{cms-tdr}
\def\svnVersion{289460}\def\cmsCernNo{2013-037}\def\cmsCernDate{\today}\def\cmsMessage{Published in the Journal of High Energy Physics as \href{http://dx.doi.org/10.1007/JHEP05(2015)078}{\doi{10.1007/JHEP05(2015)078.}}}

\begin{document}\cmsNoteHeader{SUS-13-019}

\hyphenation{had-ron-i-za-tion}
\hyphenation{cal-or-i-me-ter}
\hyphenation{de-vices}
\RCS$Revision: 289439 $
\RCS$HeadURL: svn+ssh://svn.cern.ch/reps/tdr2/papers/SUS-13-019/trunk/SUS-13-019.tex $
\RCS$Id: SUS-13-019.tex 289439 2015-05-19 14:09:31Z haweber $
\newcommand{\MTt}{\ensuremath{M_{\mathrm{T2}}}\xspace}
\newcommand{\MT}{\ensuremath{M_{\mathrm{T}}}\xspace}
\newcommand{\NJ}{\ensuremath{N_\mathrm{j}}\xspace}
\newcommand{\NB}{\ensuremath{N_\PQb}\xspace}
\newcommand{\wjets}{\PW(\ensuremath{\ell\nu})+jets\xspace}
\newcommand{\zinv}{\Z(\ensuremath{\nu\PAGn})+jets\xspace}
\newcommand{\zlljets}{\Z(\ensuremath{\ell^{+}\ell^{-}})+jets\xspace}
\newcommand{\zll}{Z(\ensuremath{\ell^{+}\ell^{-}})\xspace}
\newcommand{\gjets}{\ensuremath{\gamma}+jets\xspace}
\newcommand{\ZG}{\ensuremath{\Z(\nu\PAGn)/\gamma}\xspace}
\newcommand{\ttjets}{\ttbar+jets\xspace}
\newcommand{\mb}{\ensuremath{m_{\PQb}}\xspace}
\newcommand{\mbb}{\ensuremath{M_{\PQb\PQb}}\xspace}
\newcommand{\mlsp}{\ensuremath{m_{\PSGczDo}}\xspace}
\newcommand{\mneutTwo}{\ensuremath{m_{\PSGczDt}}\xspace}
\newcommand{\mcharg}{\ensuremath{m_{\PSGczDopm}}\xspace}
\newcommand{\mgluino}{\ensuremath{m_{\PSg}}\xspace}
\newcommand{\msq}{\ensuremath{m_{\PSQ}}\xspace}
\newcommand{\mstop}{\ensuremath{m_{\PSQt}}\xspace}
\newcommand{\mtop}{\ensuremath{m_{\PQt}}\xspace}
\providecommand{\PSGczDopm}{\ensuremath{\widetilde{\chi}^{\pm}_{1}}\xspace}
\providecommand{\NA}{\text{---}\xspace}
\providecommand{\CLs}{\ensuremath{\textrm{CL}_{\textrm{s}}}\xspace}
\providecommand{\CLsb}{\ensuremath{\textrm{CL}_{\textrm{s+b}}}\xspace}
\providecommand{\CLb}{\ensuremath{\textrm{CL}_{\textrm{b}}}\xspace}
\providecommand{\SOFTSUSY} {{\textsc{softsusy}}\xspace}
\providecommand{\SDECAY} {{\textsc{sdecay}}\xspace}
\providecommand{\Prospino} {{\textsc{Prospino}}\xspace}
\providecommand{\FastJet} {{\textsc{FastJet}}\xspace}

\cmsNoteHeader{SUS-13-019}
\title{Searches for supersymmetry using the \MTt variable in hadronic events produced in pp collisions at 8\TeV}

\date{\today}

\abstract{
Searches for supersymmetry (SUSY) are performed using a sample of hadronic
events produced in 8\TeV pp collisions at the CERN LHC.
The searches are based on the \MTt variable,
which is a measure of the transverse momentum imbalance in an event.
The data were collected with the CMS detector and correspond
to an integrated luminosity of 19.5\fbinv.
Two related searches are performed. The first is an inclusive search based on
signal regions defined by the value of the \MTt variable, the
hadronic energy in the event, the jet multiplicity, and the number of jets
identified as originating from bottom quarks.
The second is a search for a mass peak corresponding to a Higgs boson decaying
to a bottom quark-antiquark pair, where the Higgs boson is produced as a decay
product of a SUSY particle.
For both searches, the principal backgrounds are evaluated with data control
samples.
No significant excess over the expected number of background events
is observed, and exclusion limits on various SUSY  models are derived.
}

\hypersetup{%
pdfauthor={CMS Collaboration},%
pdftitle={Searches for supersymmetry using the MT2 variable in hadronic events produced in pp collisions at 8 TeV},%
pdfsubject={CMS},%
pdfkeywords={CMS, physics, supersymmetry}}

\maketitle

\section{Introduction}
Searches for physics beyond the standard model (SM) based on
final states with jets and large values of transverse momentum
imbalance \ptvecmiss are sensitive to a broad class of new-physics
models. Here, we report the results of such searches
based on the \MTt variable~\cite{Lester:1999tx}. The \MTt variable
characterizes  \ptvecmiss in events with two pair-produced heavy particles,
each of which decays to at least one undetected particle,
leading to \ptvecmiss.  An example is supersymmetry
(SUSY) with R-parity conservation~\cite{Martin:1997ns}, in which pair-produced
SUSY particles each decay to SM particles and to a massive,
neutral, weakly interacting lightest SUSY particle (LSP),
which escapes without detection.
The value of \MTt reflects the masses of the pair-produced
particles, which are much lighter for SM background processes
than expected for SUSY particles such as squarks and gluinos.
The \MTt variable was
previously used for top-quark mass measurements by the
CDF and CMS experiments~\cite{Aaltonen:2009rm,Chatrchyan:2013boa}, and for
SUSY searches by the CMS~\cite{Chatrchyan:2012jx,Khachatryan:2014mma} and
ATLAS~\cite{Aad:2012pxa,Aad:2012uu,Aad:2014nua,Aad:2014qaa,Aad:2014vma,Aad:2014yka,Aad:2014kra} experiments.

This paper describes searches for physics beyond the SM
performed using a data sample of pp collisions collected in 2012
at a centre-of-mass energy of 8\TeV with the CMS detector at the
CERN LHC. The size of the sample, measured by its integrated luminosity, is
19.5\fbinv.

Two different \MTt-based searches are presented. The first search, called the
inclusive-\MTt search, employs
several signal regions defined by
the number of jets (\NJ), the number of tagged bottom-quark jets (\NB),
the value of \MTt,
and the hadronic energy in an event.
This general search aims to cover a large variety of SUSY and other new-physics signatures.
The second search, called the \MTt-Higgs search, is a specialized analysis
targeting events with a Higgs boson produced in
the decay of a heavy SUSY particle.  The SM Higgs boson
decays primarily to a bottom quark-antiquark (\bbbar) pair.
For a large variety of SUSY models, the lightest Higgs boson
(h boson) has SM properties, especially if the
masses of all other SUSY Higgs bosons are much larger.
In the \MTt-Higgs search, we therefore search for an excess
of events at the SM Higgs boson mass of 125\GeV in the
invariant mass distribution of b-tagged jet pairs.

The two searches rely on similar selection criteria for the \MTt variable to enhance the sensitivity to a potential
SUSY signal and to reduce the background from SM multijet events  to a minimal level.
The remaining SM background consists mostly of Z+jets events where the Z boson decays to neutrinos, and
W+jets and \ttbar+jets events where one W boson decays
leptonically.
These backgrounds are mostly estimated by methods using data.

This analysis extends a previous CMS publication~\cite{Chatrchyan:2012jx}, based on pp collisions at 7\TeV,
by exploiting a higher collision energy and a larger data sample.
Alternative inclusive searches in hadronic final states based on the 8\TeV data sample are presented in Refs.~\cite{Chatrchyan:2013lya,Chatrchyan:2013wxa,Chatrchyan:2014lfa,Aad:2013wta,Aad:2014wea}.

This paper is organized as follows.
In Section~\ref{sec:MT2}, the \MTt variable is defined.
A description of the detector and trigger is given in Section~\ref{sec:CMS}.
The data sets and the general event
selection procedures are discussed in Section~\ref{sec:cut-flow}.
Section~\ref{sec:strategy} presents the analysis strategy for the
inclusive-\MTt and \MTt-Higgs searches, and Section~\ref{sec:bkgEstimation}
the background estimation method based on data control samples.
A comparison between the observed numbers
of events and the predicted background yields
is presented for the two searches in Section~\ref{sec:data:geq3jets.results}.
Systematic uncertainties are discussed in Section~\ref{sec:systematics}.
The statistical procedures used to calculate exclusion
limits on SUSY particles are presented in Section~\ref{sec:stats},
with the limits themselves presented in Section~\ref{sec:exclusion}.
Section~\ref{sec:conclusion} contains a summary.

\section{Definition of the \texorpdfstring{\MTt variable and interpretation}{MT2 variable and interpretation}\label{sec:MT2}}

The use of \MTt as a search variable is discussed in our previous publication~\cite{Chatrchyan:2012jx}.
Here, we recapitulate the most salient aspects.
The kinematic mass variable \MTt
was introduced as a means to measure the mass of pair-produced particles
in situations where both particles decay to a final state containing an undetected
particle X of mass $m_\mathrm{X}$.
For each decay chain,
the visible system is defined by the transverse momentum $\ptvec^{\; \mathrm{vis}(i)}$,
transverse energy $E_\mathrm{T}^{\text{vis}(i)}$, and mass $m^{\text{vis}(i)}$ ($i = 1,$ 2) obtained by summing
the four-momenta of all detected particles in the decay chain.
The two visible systems are accompanied by the two undetected particles with unknown transverse momenta $\ptvec^{\; \mathrm{X}(i)}$.
In analogy with the transverse mass used for the W boson mass determination~\cite{Arnison:1983rp},
two transverse masses are defined for the two pair-produced particles:

\begin{equation}
  (\MT^{(i)})^2 = (m^{\text{vis}(i)})^2 + m_{\mathrm{X}}^2
 + 2 \left(
E_\mathrm{T}^{\text{vis}(i)}  E_\mathrm{T}^{\mathrm{X}(i)} - \ptvec^{\; \text{vis}(i)} \cdot \ptvec^{\; \mathrm{X}(i)}
     \right).
\label{eq.MT2.transmass}
\end{equation}

If the correct values of $m_\mathrm{X}$ and $\ptvec^{\;\mathrm{X}(i)}$, $m^{\text{vis}(i)}$, and $\ptvec^{\; \text{vis}(i)}$ are chosen, the transverse
masses $\MT^{(i)}$ do not exceed the mass of the parent particles.
The momenta $\ptvec^{\;\mathrm{X}(i)}$ of the unseen particles, however, are not experimentally accessible individually.
Only their sum, the missing transverse momentum \ptvecmiss, is known.
A generalization of the transverse mass, the \MTt variable, is defined as:
\begin{equation}
\MTt(m_{\mathrm{X}}) = \min_{\ptvec^{\mathrm{X}(1)} + \ptvec^{\mathrm{X}(2)} = \ptvecmiss}
  \left[ \max \left( \MT^{(1)} , \MT^{(2)} \right) \right] ,
\label{eq.MT2.definition}
\end{equation}
where the unknown mass $m_{\mathrm{X}}$ is a free parameter. The minimization is performed over trial momenta of the undetected particles fulfilling the \ptvecmiss constraint.

In this analysis, all visible objects, such as jets, are clustered into two pseudojets.
For this purpose, we use the hemisphere algorithm defined in Section~13.4 of Ref.~\cite{Ball:2007zza}. The algorithm is seeded by the two jets with largest dijet invariant mass. The clustering is performed by minimizing the Lund distance measure~\cite{Sjostrand:1982am,Sjostrand:2006za}.
Standard model multijet events, interpreted as two pseudojets, may give rise to large \MTt
if both pseudojets have large masses.
Setting $m^{\text{vis}(i)} = 0$ in Eq.~(\ref{eq.MT2.transmass})
suppresses the multijet contributions without affecting signal sensitivity,
since the kinematic terms of Eq.~(\ref{eq.MT2.transmass}) are large for
most new-physics scenarios.
In the following, \MTt is computed using $E_\mathrm{T}^{\text{vis}(i)}$, $\ptvec^{\; \text{vis}(i)}$ ($i=1,2$),
and \ptvecmiss, setting both $m^{\text{vis}(i)}$ terms in Eq.~(\ref{eq.MT2.transmass}) to zero.

Although most the background from SM multijet events is thus characterized by small values of \MTt,
a residual background at large \MTt arises from multijet events in which the two pseudojets are not back-to-back because
of jet energy mismeasurements.  Further selection criteria are applied to suppress these events, as discussed in Section~\ref{sec:cut-flow}.

\section{Detector and trigger\label{sec:CMS}}

The central feature of the CMS apparatus is a superconducting solenoid of
6\unit{m} internal diameter, providing a magnetic field of 3.8\unit{T}.
Within the superconducting solenoid volume are a silicon pixel and strip tracker,
a lead-tungstate crystal electromagnetic calorimeter, and a brass{\slash}scintillator
hadron calorimeter, each composed of a barrel and two endcap sections.
Muons are measured in gas-ionization detectors embedded in the steel flux-return
yoke outside the solenoid. Extensive forward calorimetry complements the coverage
provided by the barrel and endcap detectors.
The detector is nearly hermetic, covering $0<\phi<2\pi$ in azimuth, and thus allows
the measurement of momentum balance in the plane transverse to the beam
direction.
The first level of the CMS trigger system, composed of custom hardware
processors, uses information from the calorimeters and muon detectors to
select the most interesting events in a fixed time interval of less than
4\mus. The high level trigger processor farm further decreases the event
rate, from around 100\unit{kHz} to around 300\unit{Hz}, before data storage.
A more detailed description of the CMS detector, together with a definition of the
coordinate system used and the relevant kinematic variables,
can be found in Ref.~\cite{Chatrchyan:2008aa}.

Events are selected using three complementary triggers. A trigger based on the scalar sum of jet
\pt values (\HT) requires $\HT>650\GeV$. A second trigger requires $\MET>150\GeV$, where \MET is the magnitude of~\ptvecmiss.
A third trigger requires $\HT>350\GeV$ and $\MET>100\GeV$.
The trigger efficiency is measured to be larger than 99\% for events that satisfy the event selection criteria outlined in Section~\ref{sec:cut-flow}.

\section{Data sets and event selection}
\label{sec:cut-flow}

The event selection is designed using simulated samples of background and signal processes.
Background events are generated with the \MADGRAPH~5 \cite{Alwall:2011uj}, \PYTHIA 6.4.26 \cite{Sjostrand:2006za},
and \POWHEG 1.0~\cite{Re:2010bp} programs.
Signal event samples based on simplified model scenarios (SMS)~\cite{Alves:2011wf}
are generated using the \MADGRAPH~5 program, with the decay
branching fractions of SUSY particles set either to 0\% or 100\%
depending on the SUSY scenario under consideration.  We also generate
signal events in the context of the constrained minimal
supersymmetric SM (cMSSM{\slash}mSUGRA)~\cite{Kane:1993td}.  The cMSSM{\slash}mSUGRA events
are generated using the \PYTHIA program, with the \SDECAY~\cite{SDECAY}
program used to describe the SUSY particle decay branching fractions and the \SOFTSUSY~\cite{SOFTSUSY} program
to calculate the SUSY particle mass spectrum.
The \PYTHIA program is used to describe the parton shower and hadronization.
While all generated background samples are processed with the detailed simulation of the CMS detector response, based
on \GEANT4 \cite{Agostinelli:2002hh}, for signal samples the detector simulation is performed using
the CMS fast simulation package~\cite{FastSim}. Detailed cross checks are conducted to ensure that the results obtained
with fast simulation are in agreement with the ones obtained with \GEANT-based detector simulation.
For SM backgrounds, the most accurate calculations
of the cross sections available in the literature are used~\cite{Melnikov:2006kv,Kidonakis:2010dk}. These are usually at next-to-leading order (NLO) in \alpS.
For the SUSY signal samples, cross sections are calculated at
NLO~\cite{Kulesza:2008jb, Kulesza:2009kq,Beenakker:2009ha, Beenakker:2011fu, Kramer:2012bx}
using the \Prospino 2.1~\cite{Beenakker:1996ch} program.

The data and simulated events are reconstructed and analyzed in an identical manner.
The event reconstruction is based on the particle-flow (PF) algorithm~\cite{CMS:2009nxa,CMS:2010eua},
which
reconstructs and identifies
charged hadrons, neutral hadrons, photons, muons, and electrons.
Electrons and muons are required to have transverse momentum $\PT > 10\GeV$ and pseudorapidity $\abs{\eta} < 2.4$.
For electrons, the transition region
between barrel and endcaps ($1.442<\abs{\eta}<1.566$) is excluded because the electron reconstruction in this region is not optimal.
An isolation requirement is also employed, requiring that the \pt sum
of photons, charged hadrons, and neutral hadrons, in a cone of $\Delta R = \sqrt{\smash[b]{(\Delta \eta)^2 + (\Delta \phi)^2}} = 0.3$ along the lepton direction, divided by the lepton \pt value, be less than 0.15 for electrons and 0.20 for muons.
The isolation value is corrected for the effects of pileup, that is, multiple pp collisions within the same bunch crossing as the primary interaction.
The electron and muon reconstruction and identification criteria are described in Refs.~\cite{CMS:2010bta}
and~\cite{Chatrchyan:2012xi}, respectively.
All particles, except the isolated electrons and muons, are clustered into PF jets~\cite{Chatrchyan:2011ds}
using the anti-\kt jet-clustering algorithm~\cite{Cacciari:2008gp} with a size parameter of 0.5.
The jet energy is calibrated by applying correction factors as a function of
the \pt and the $\eta$ of the jet~\cite{Chatrchyan:2011ds}.
The effect of pileup on jet energies is treated as follows:
tracks not associated with the primary interaction are removed from the jet;
for the neutral part of the jet, the effect of pileup is reduced using the \FastJet pileup subtraction
procedure \cite{Cacciari:2007fd, Cacciari:2008gn}.
All jets are required to satisfy basic quality criteria
(jet ID~\cite{CMS:2010xta}), which eliminate, for example, spurious events due to calorimeter noise.
Jets are also required to have $\PT > 20\GeV$ and $\abs{\eta} < 2.4$.
Jets are b-tagged using the medium working point of the
combined secondary vertex (CSV) algorithm~\cite{Chatrchyan:2012jua}.
Tau leptons are reconstructed in their decays to one or three charged particles~\cite{Chatrchyan:2012zz}
and are required to have $\PT > 20\GeV$ and $\abs{\eta} < 2.3$.
The $\tau$ leptons are also required to satisfy a loose isolation selection:
the \pt-sum of charged hadrons and photons
that appear within $\Delta R < 0.5$ of the candidate $\tau$-lepton direction
is required to be less than 2\GeV after subtraction of the pileup contribution.
Throughout this paper,
any mention of a $\tau$ lepton refers to its reconstructed hadronic decay.
Photons~\cite{CMS-PAS-EGM-10-005} are required to have $\PT > 20\GeV$, $\abs{\eta} < 2.4$,
and to not appear in the transition region between the barrel and endcap detectors.
Photons are further required to satisfy selection criteria based on the shape of their calorimetric shower,
to deposit little energy in the hadron calorimeter, and to fulfill isolation requirements.

The missing transverse momentum vector \ptvecmiss is defined as the projection onto the plane perpendicular to the beam axis of the negative vector sum of the momenta of all reconstructed  particles in the event. Its magnitude is referred to as \MET.
The hadronic activity in the event, \HT, is defined to be the scalar \pt sum of all accepted jets with $\pt>50\GeV$ and $\abs{\eta}<3.0$.
Events selected with the pure-\HT trigger described in Section~\ref{sec:CMS} are required to satisfy $\HT>750\GeV$.
Events selected with one of the two other triggers are required to satisfy $\HT>450\GeV$ and $\MET>200\GeV$.

Corrections for differences observed between the simulation and data due to the jet energy scale~\cite{Chatrchyan:2011ds}, the b-tagging efficiencies~\cite{Chatrchyan:2012jua}, and the \pt spectrum of the system recoil~\cite{Chatrchyan:2013xna} are applied to simulated events.

Events are required to contain at least two jets that, in addition to the previous general jet requirements,
have $\pt > 100\GeV$.
To reduce the background from events with \wjets and top-quark production,
events are rejected if they contain an isolated electron, muon, or $\tau$ lepton.
Background from multijet events, which mostly arises because of jet energy misreconstruction, is reduced by requiring the
minimum difference $\Delta\phi_\text{min}$ in azimuthal angle between the \ptvecmiss vector and one of the four jets with
highest \pt to exceed 0.3 radians.
To reject events in which \MET arises from unclustered energy or from jets aligned near the beam axis,
a maximum difference of 70\GeV is imposed on the magnitude of the vectorial difference between \ptvecmiss and the negative
vector sum of the \PT of all leptons and jets.
Finally, events with possible contributions from beam halo processes or anomalous noise in the calorimeter or tracking systems are rejected~\cite{Chatrchyan:2011tn}.

\section{Search strategy}
\label{sec:strategy}

The \MTt-based search strategy is outlined in this section.
For both the inclusive-\MTt and the \MTt-Higgs searches, all selected jets
are clustered
into two pseudojets as described in Section~\ref{sec:MT2}.
Several mutually exclusive signal regions are defined to optimize the search
for a wide variety of new-physics models.
The definition of signal regions is based on the event topology and event kinematic variables. The
more general inclusive-\MTt search is described first.

The inclusive-\MTt and \MTt-Higgs searches are not mutually exclusive.  All but 4\% of the events selected by the \MTt-Higgs search are also selected by the inclusive-\MTt search.

\subsection{Inclusive-\texorpdfstring{\MTt search}{MT2 search}}
\label{sec:inclusiveMT2}

\begin{figure}[htb!]
\centering
\includegraphics[width=0.45\textwidth]{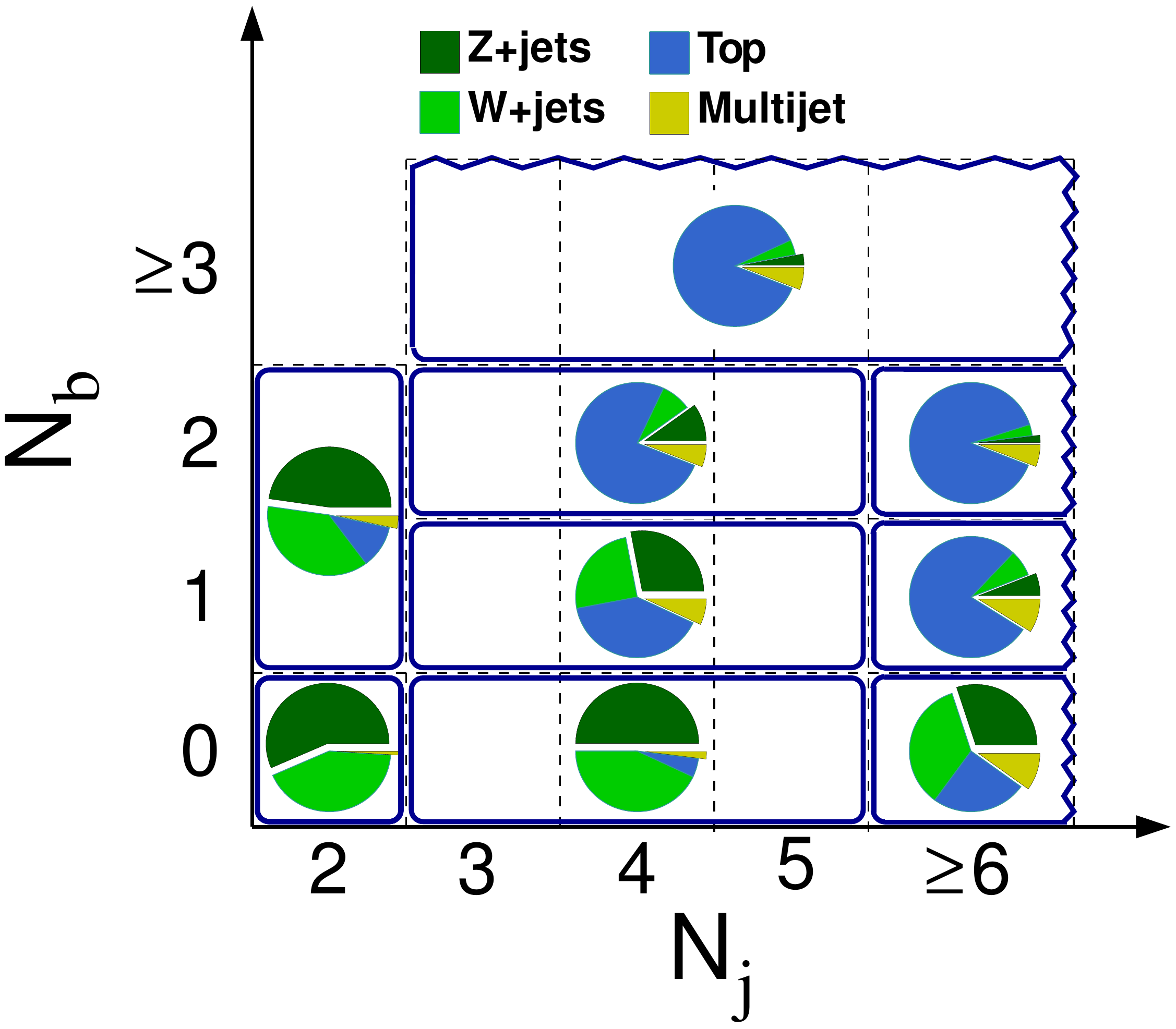}
\includegraphics[width=0.45\textwidth]{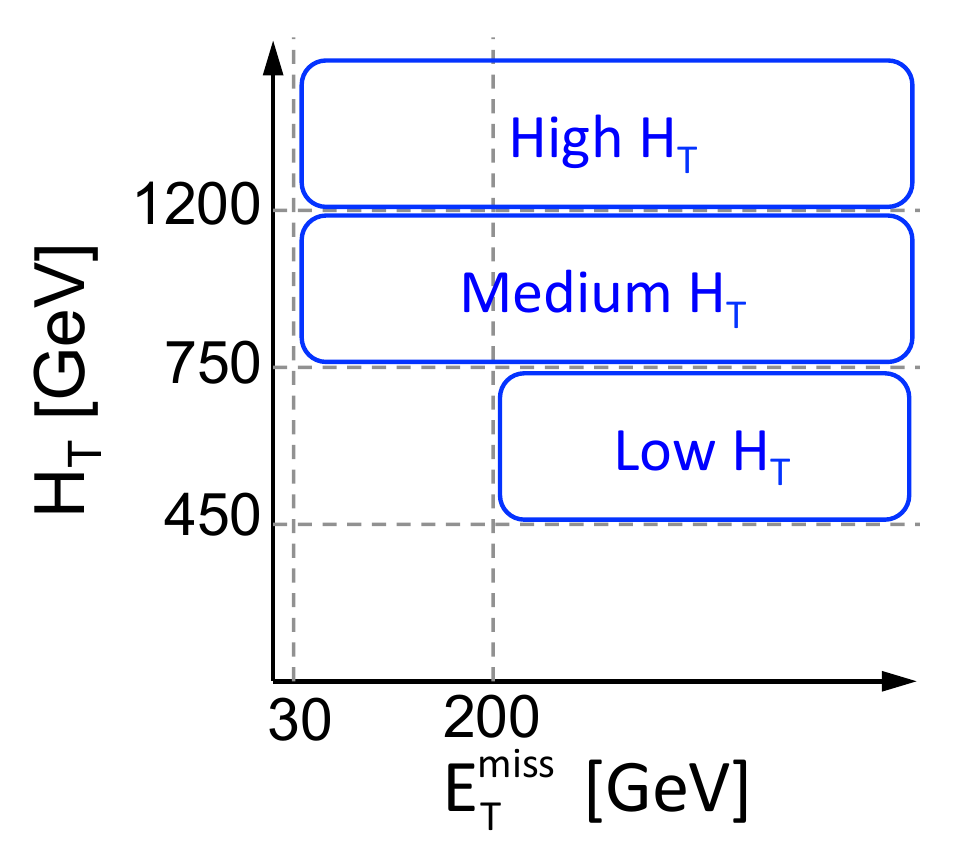}
\caption{Definition of the topological signal regions in terms of the number of jets \NJ and the
number of b-tagged jets \NB (left), and their subsequent division in terms of \HT and \MET (right).
The pie charts illustrate the expected contributions from different SM processes in
the different signal regions;
they are similar in all three \HT regions. }
\label{fig:strategy.toporegions}
\end{figure}

For the inclusive-\MTt search, nine regions, called topological regions, are defined
by \NJ and \NB, the numbers of jets and b-tagged jets in the event with $\pt > 40\GeV$,
as illustrated in Fig.~\ref{fig:strategy.toporegions}~(left).
These regions are chosen after testing the sensitivity of the search to various SUSY SMS models using simulated data.
The regions with $\NB=0$
are the most sensitive to the production of gluinos that do not decay to top and bottom quarks, and to
the production of squarks of the first two generations.
The regions with $\NB>0$ and low (high) values of \NJ are designed
for bottom- and top-squark production with decays to bottom (top) quarks.
Finally, the signal regions with $\NJ\geq3$ and $\NB\geq3$ provide extra sensitivity to final states
with multiple bottom or top quarks, for example from gluino pair-production.
Since the values of \MTt and \HT in a SUSY event depend strongly on the mass
of the initially produced SUSY particles,
a wide range of values in \MTt and \HT is considered.
Each of the nine topological regions is divided into three sub-regions
of \HT, as
shown in Fig.~\ref{fig:strategy.toporegions} (right):
the low-\HT region $450 < \HT \leq 750\GeV$, the medium-\HT region $750 < \HT \leq 1200\GeV$,
and the high-\HT region $\HT > 1200\GeV$.

Each of these regions is
examined in bins of \MTt,
where the number of bins (up to nine) depends
on the specific topological and \HT selection.
By design, the lowest bin in \MTt is chosen such that the multijet background is expected to be less
than $\sim$1--10\% of the total background.
The minimum threshold on \MTt varies between 100 and 200\GeV, depending on the topological region and the \HT requirement.
The edges of the \MTt bins are adjusted to ensure that there are a sufficient number of events in each bin of the
corresponding control samples for the background evaluation (Section~\ref{sec:bkgEstimation}).
The definitions of all signal regions  are specified in Table~\ref{table:signalregiondef}.

\begin{table}[!ht]
\centering
\topcaption{Definition of the signal regions used in the inclusive-\MTt search.}
\label{table:signalregiondef}
\begin{tabular}{l|ccc|ccc|cc}
\hline
 & \multicolumn{3}{c}{Low-\HT region} & \multicolumn{3}{|c|}{Medium-\HT region} & \multicolumn{2}{c}{High-\HT region} \\

\rule[0pt]{0pt}{2.2ex} & \multicolumn{3}{c}{\MTt bin [\GeVns{}]} & \multicolumn{3}{|c|}{\MTt bin [\GeVns{}]} & \multicolumn{2}{c}{\MTt bin [\GeVns{}]} \\
\hline
\multirow{3}{*}{\begin{minipage}{4.25em}$\NJ=2$,\\ $\NB=0$\end{minipage}}
 & 200--240 & 350--420 & 570--650		& 125--150 & 220--270 & 425--580		& 120--150 & 260--350  \\
 & 240--290 & 420--490 & $>$650	& 150--180 & 270--325 & 580--780		& 150--200 & 350--550  \\
 & 290--350 & 490--570 &                       & 180--220 & 325--425 & $>$780	& 200--260 & $>$550  \\
\hline
\multirow{2}{*}{\begin{minipage}{4.25em}$\NJ=2$,\\ $\NB\geq 1$\end{minipage}}
 & 200--250 & 310--380 & 450--550		& 100--135 & 170--260 & $>$450	& 100--180 &  \\
 & 250--310 & 380--450 & $>$550	& 135--170 & 260--450 &               		& $>$180 &  \\
\hline
\multirow{4}{*}{\begin{minipage}{4.25em}$\NJ=3$--5,\\ $\NB=0$\end{minipage}}
 & 200--240 & 420--490 &               		& 160--185 & 300--370 & $>$800	& 160--185 & 350--450 \\
 & 240--290 & 490--570 &               		& 185--215 & 370--480 &               		& 185--220 & 450--650  \\
 & 290--350 & 570--650 &               		& 215--250 & 480--640 &               		& 220--270 & $>$650  \\
 & 350--420 & $>$650   &          	& 250--300 & 640--800 &               		& 270--350 & \\
\hline
\multirow{2}{*}{\begin{minipage}{4.25em}$\NJ=3$--5,\\ $\NB=1$\end{minipage}}
 & 200--250 & 310--380 & 460--550		& 150--175 & 210--270 & 380--600		& 150--180 & 230--350  \\
 & 250--310 & 380--460 & $>$550	& 175--210 & 270--380 & $>$600	& 180--230 & $>$350  \\
\hline
\multirow{2}{*}{\begin{minipage}{4.25em}$\NJ=3$--5,\\ $\NB=2$\end{minipage}}
 & 200--250 & 325--425 &               		& 130--160 & 200--270 & $>$370	& 130--200 &  \\
 & 250--325 & $>$425   &             	& 160--200 & 270--370 &               		& $>$200 &  \\
\hline
\multirow{2}{*}{\begin{minipage}{4.25em}$\NJ\geq 6$,\\ $\NB=0$\end{minipage}}
 & 200--280 & $>$380   &         		& 160--200 & 250--325 & $>$425	& 160--200 & $>$300  \\
 & 280--380 &                &              		& 200--250 & 325--425 &               		& 200--300 &  \\
\hline
\multirow{2}{*}{\begin{minipage}{4.25em}$\NJ\geq 6$,\\$\NB=1$\end{minipage}}
 & 200--250 & $>$325   &         		& 150--190 & 250--350 &               		& 150--200 & $>$300  \\
 & 250--325 &                &               		& 190--250 & $>$350   &               	& 200--300 &  \\
\hline
\multirow{2}{*}{\begin{minipage}{4.25em}$\NJ\geq 6$,\\ $\NB=2$\end{minipage}}
 & 200--250 & $>$300   &               	& 130--170 & 220--300 &               		& 130--200 &  \\
 & 250--300 &                &              		& 170--220 & $>$300   &               	& $>$200 &  \\
\hline
\multirow{2}{*}{\begin{minipage}{4.25em}$\NJ\geq 3$,\\ $\NB\geq 3$\end{minipage}}
 & 200--280 & $>$280   &              		& 125--175 & 175--275 & $>$275  		& $>$125  &  \\
 & & & & & & & & \\
\hline
\end{tabular}
\end{table}

\begin{figure}[!htb]
\centering
\includegraphics[width=0.49\textwidth]{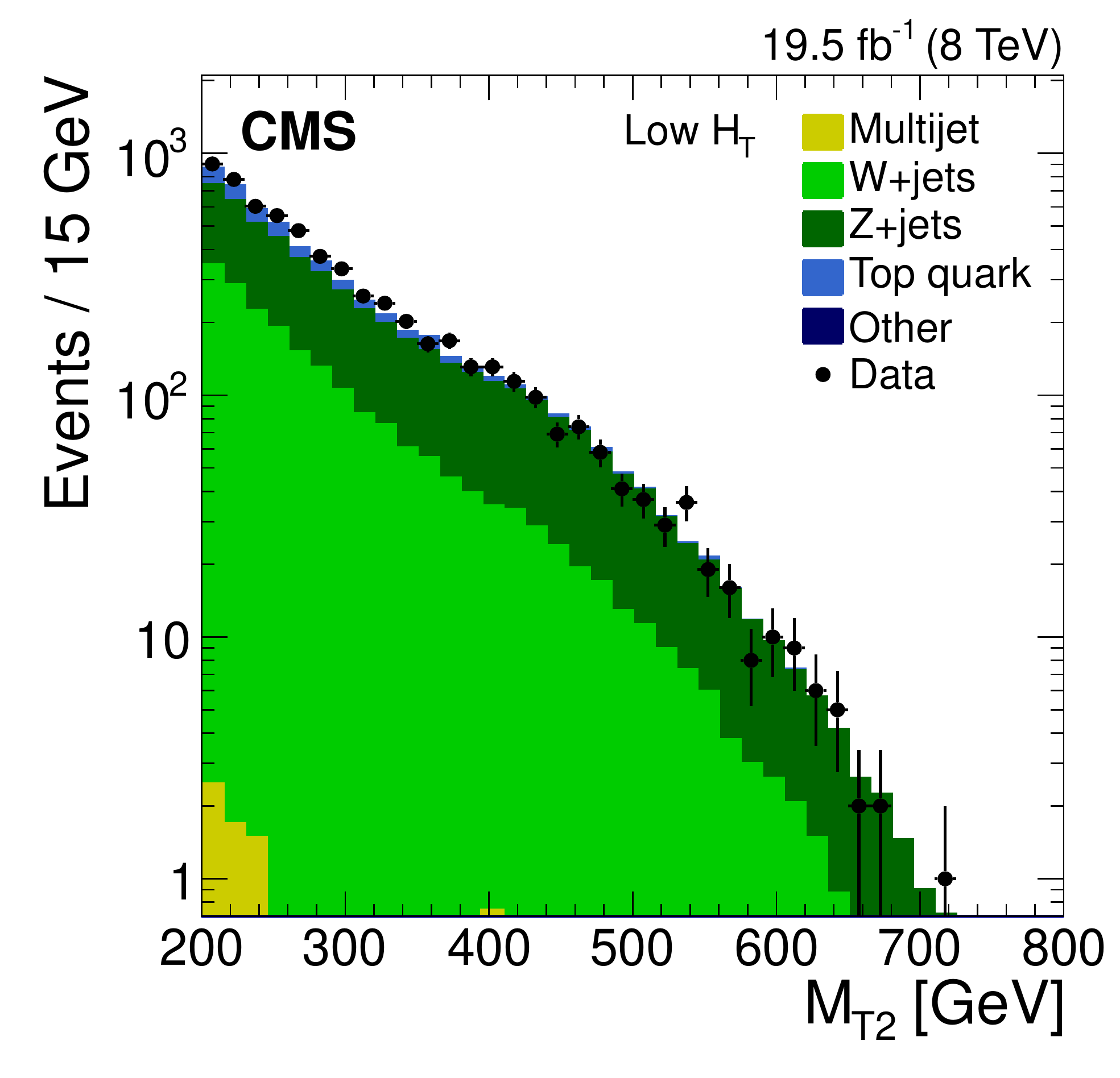} \\
\includegraphics[width=0.49\textwidth]{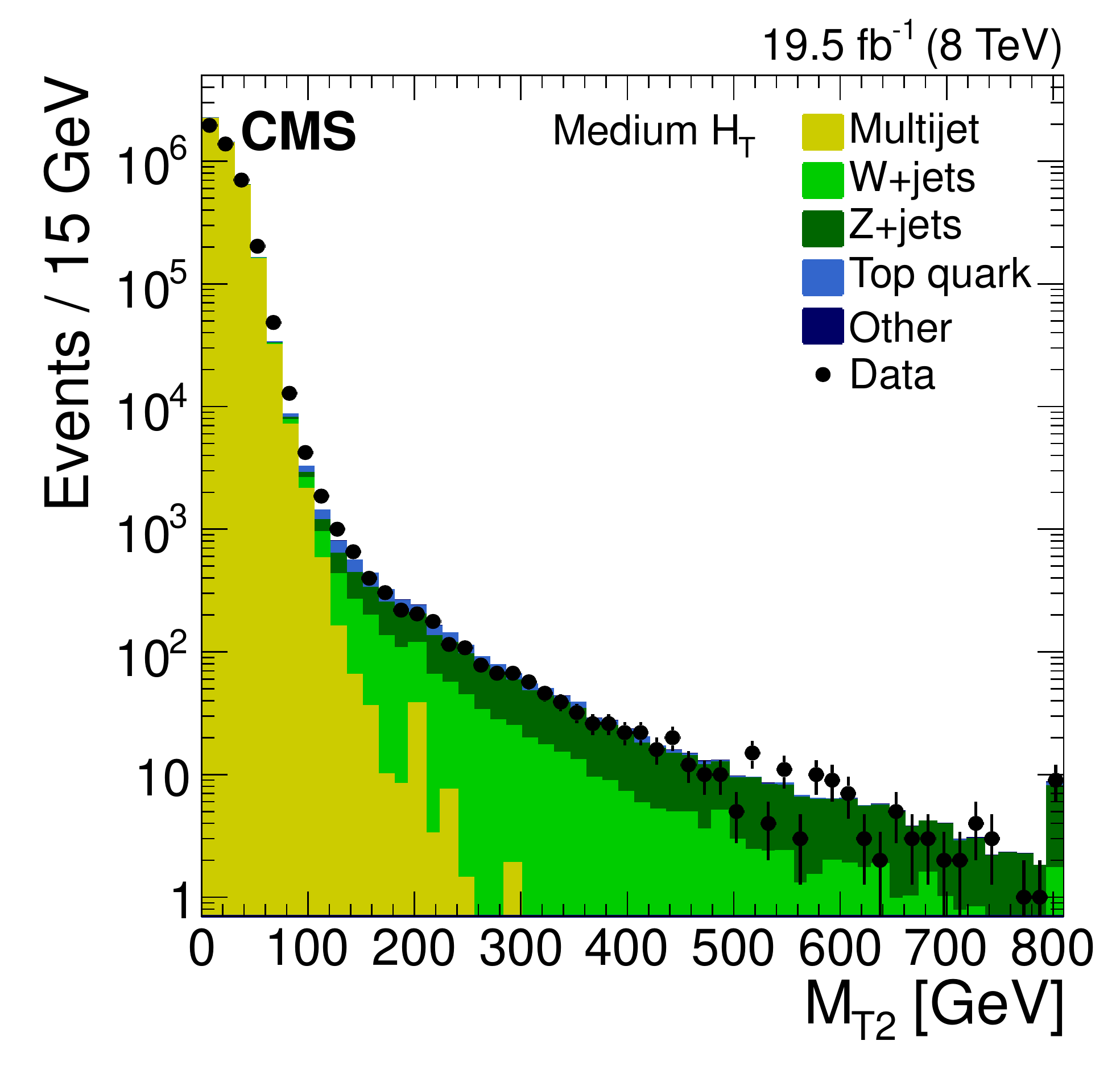}
\includegraphics[width=0.49\textwidth]{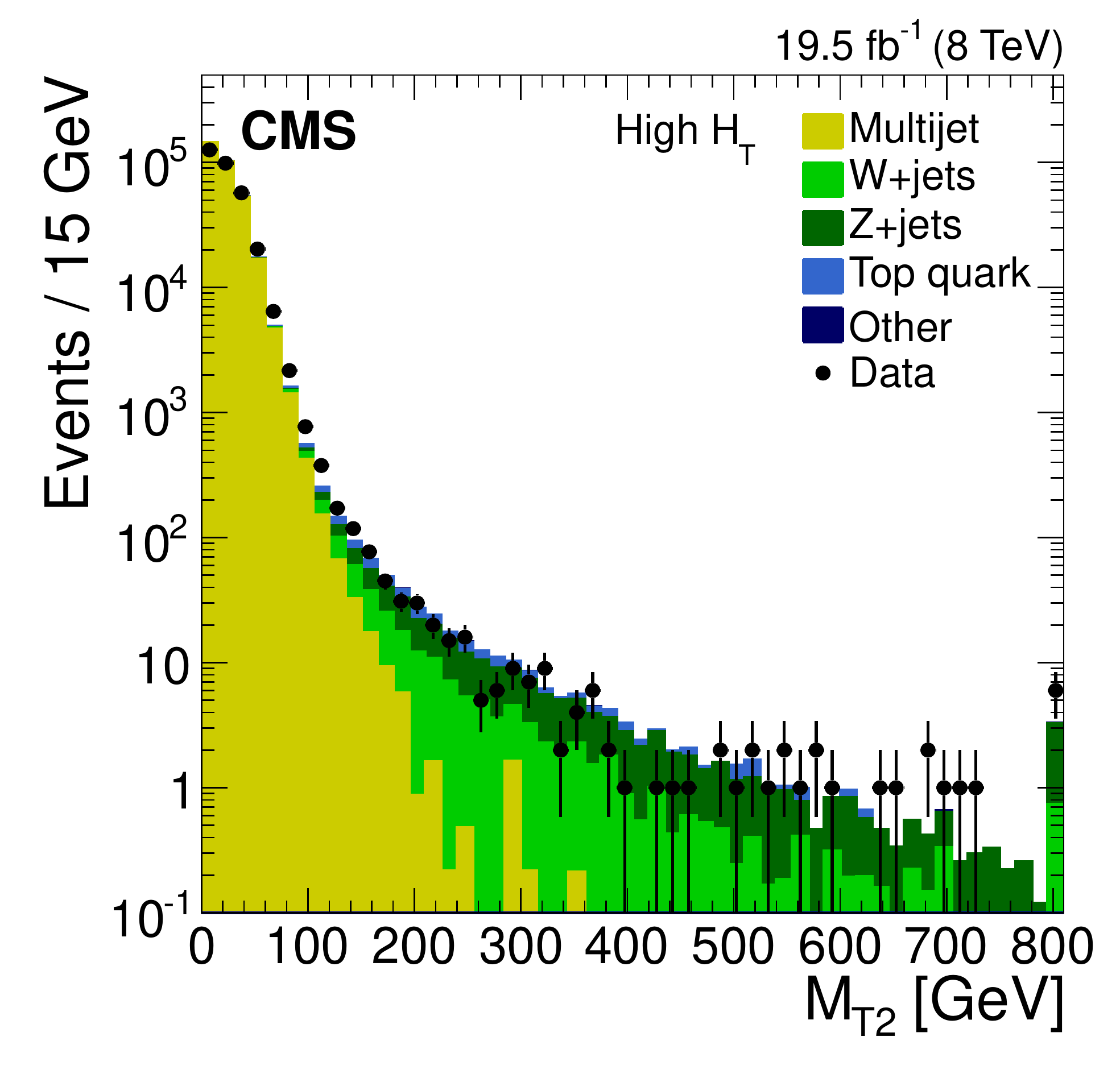}
\caption{Distribution of the \MTt variable after the low-\HT (top), medium-\HT (bottom left), and
high-\HT (bottom right) event selections, respectively. The event yields are integrated in the (\NJ, \NB) plane
over all the topological signal regions, for both simulated and data
samples.
These plots serve as an illustration of the background composition of
the \MTt distributions.
}
\label{fig:MT2_datavsMC_allsrinclusive}
\end{figure}

Figure~\ref{fig:MT2_datavsMC_allsrinclusive} shows the \MTt distributions in simulation and data
for the low-, medium-, and high-\HT selections, inclusively in all signal regions of the \NJ--\NB plane.
For $\MTt < 80\GeV$ the distribution in the medium- and high-\HT regions is completely dominated by multijet events.
For this reason, these bins are used only as control regions.

In the signal regions with $\NJ = 2$ or $\NB = 0$,
the dominant background is from \zinv production. The next-most important background is from \wjets events,
while the background from \ttjets events is small.
In the regions with $\NB = 1$ all three processes (\zinv, \wjets, and \ttjets production) are important.
For all regions requiring multiple b-tagged jets, \ttjets events are the dominant source of background.
The \ttjets contribution to the total background typically increases with the jet multiplicity
and is important for all selections with $\NJ\geq6$, regardless of the \NB selection.
The relative contribution of \ttjets production decreases with increasing \MTt because of the natural cutoff of \MTt above the top-quark mass for these events.

Contributions from other backgrounds, such as \gjets, \zlljets, and diboson production, are found to be negligible.

\subsection{\texorpdfstring{\MTt-Higgs search}{MT2-Higgs search}}
\label{sec:MT2Higgs}

The \MTt-Higgs search is designed to select events with a light \Ph boson produced in a cascade
of supersymmetric particles initiated through the strong pair production of squarks or gluinos.
As the dominant decay mode of the \Ph boson in many SUSY models is $\Ph \to \bbbar$, a signature of a SUSY signal
would be an excess in the invariant mass distribution of the selected b-tagged jet pairs, $\mbb$.
An excess could help identify a preferred new-physics model, as the associated new particles would couple to the Higgs sector. Such an identification is not possible with the inclusive-\MTt search.

Within a cascade of SUSY particles, the \Ph boson is produced together with the LSP in the decays of neutralinos,
such as $\PSGczDt \to \PSGczDo + \Ph$.
As the neutralino $\PSGczDt$ can be a typical decay product of squarks and gluinos, the cross section
for this kind of processes is among the largest in a large part of the SUSY parameter space.
The final state contains at least two b-tagged jets,
multiple hard jets, and a large value of \MTt.

For the \MTt-Higgs search, b-tagged jets are required to have $\pt>20\GeV$.
The event selection requires at least two b-tagged jets, along with $\NJ \ge 4$.
The two b-tagged jets stemming from the \Ph boson decay are generally expected to appear within the same pseudojet, as they originate from the same decay chain.
Using b-tagged jets within the same pseudojet, a b-tagged jet pair is selected if it has $\Delta R(\PQb_1, \PQb_2) <  1.5$.
If multiple pairs are found in one or both pseudojets, the pair with the smallest $\Delta R(\PQb_1, \PQb_2)$ is chosen.
If no pair is found within the same pseudojet, pairs with b-tagged jets in different pseudojets are considered.
If none of the pairs has ${\Delta R(\PQb_1, \PQb_2) < 1.5}$, the event is rejected.
For signal events containing b quarks from the \Ph boson decay, the efficiency to find the correct pair of b-tagged jets is about 70\%.

Using the known \Ph boson mass of 125\GeV~\cite{Aad:2012tfa,Chatrchyan:2012ufa}, 12 signal regions are defined as 15\GeV-wide bins in the $20 < \mbb < 200\GeV$
range. Each of these signal regions is further divided into two sub-regions as follows:
a low-\HT selection requiring $450 < \HT\leq 750\GeV$, $\MET > 200\GeV$,
and $\MTt > 200\GeV$;
and a high-\HT selection requiring $\HT > 750\GeV$
and $\MTt > 125\GeV$.

The overall yields of the main SM backgrounds (\ttjets, \wjets, and \zinv) are estimated
using the same methods as for the inclusive-\MTt analysis. The contribution of
the SM production of the Higgs boson is negligible in the search regions of
this analysis.
The shapes of the $\mbb$ distributions for signal and the various backgrounds are obtained from simulation.
Since in simulation we observe no appreciable correlation between \MTt and $\mbb$
in either the signal or background sample,
the shape of the \mbb distribution is obtained from large simulated samples with relaxed \MTt requirements. An uncertainty due to the
looser \MTt selection is taken into account.
Further uncertainties in the shapes are assessed by varying several modelling parameters of the simulation.

\section{Background estimation}
\label{sec:bkgEstimation}

This section describes the procedures used to estimate the main backgrounds:
multijet events, Z+jets events where the Z boson decays to neutrinos, and W+jets and \ttbar+jets events
where one W boson decays leptonically but the corresponding charged lepton lies outside the acceptance of the analysis, is not reconstructed, or is not isolated.
The same background estimation procedures are used for both the inclusive-\MTt and \MTt-Higgs searches.

\subsection{Determination of the multijet background}
\label{sec:bkgQCD}

The multijet background consists of direct multijet production, but also of events with \ttbar pairs or vector bosons that
decay hadronically.
From Fig.~\ref{fig:MT2_datavsMC_allsrinclusive}, the multijet background is expected to be negligible at large values of \MTt.
This background, arising from difficult-to-model jet energy mismeasurements, is nonetheless subject to considerable uncertainty.
A method based on data control samples is used to predict this background.
The method relies on \MTt and the variable $\Delta \phi_\text{min}$, described in Section~\ref{sec:cut-flow}.
In general terms, the multijet background entering each of the signal regions,
for which a selection requirement is $\Delta\phi_\text{min} > 0.3$,
is estimated from a corresponding control region defined by the same criteria as the signal regions except for $\Delta\phi_\text{min}$, which is required to be less than 0.2.
The control regions are dominated by multijet event production.

The transfer factor between control and signal regions, and our parameterization thereof, are given by
\begin{equation}
 r(\MTt) \equiv \frac{N(\Delta\phi_\text{min} > 0.3)}{N(\Delta\phi_\text{min} < 0.2)}
  = \exp(a - b\, \MTt) + c \qquad \text{for } \MTt > 50\GeV.
\label{eq.bkgdpred.ratio}
\end{equation}

The parameters $a$ and $b$ are obtained from a fit to data in the region $50<\MTt<80\GeV$, where the contributions
of electroweak and top-quark (mainly \ttjets events) production are small.
The constant term $c$ is only measurable in control samples requiring high-\MTt values.
For these events, however, the non-multijet contribution is dominant, and so $c$ cannot be obtained from a fit to data.
Therefore, the parameterization of $r(\MTt)$ is fixed to a constant for $\MTt > 200\GeV$. This constant
is chosen as the value of the exponential fit to $r(\MTt)$ at $\MTt = 200\GeV$.

\begin{figure}[!ht]
\centering
\includegraphics[width=0.45\textwidth]{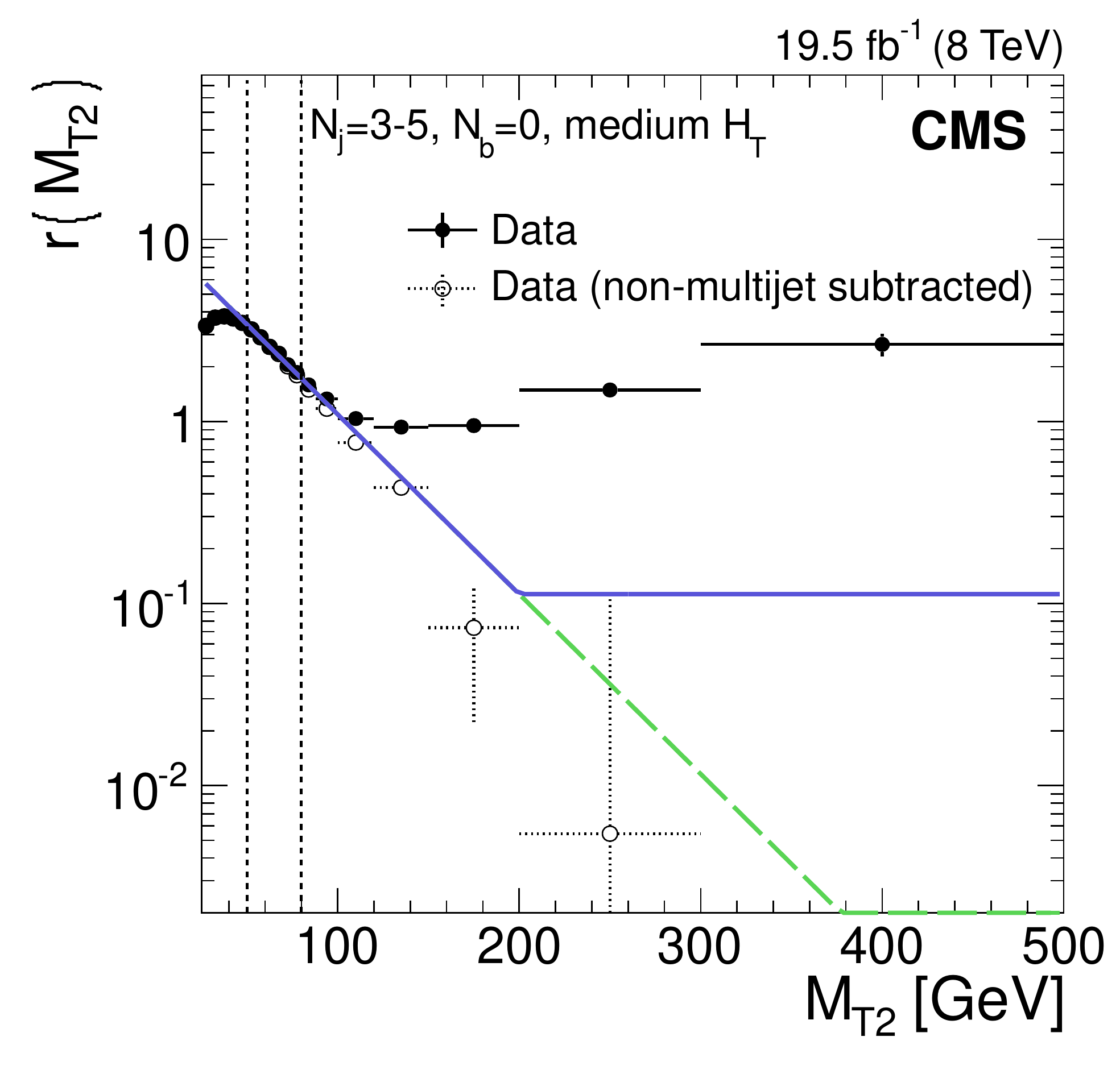}
\caption{The ratio $r(\MTt)$, described in the text, as a function of \MTt for events
satisfying the medium-\HT and the $(\NJ= 3$--5,$\NB= 0)$ requirements of the inclusive-\MTt search.
The solid circle points correspond to simple data yields, while the points with open circles
correspond to data after the subtraction of the
non-multijet backgrounds, as estimated from simulation.
Two different functions, whose exponential components are fitted to the data in the
region $50 < \MTt < 80 \GeV$, are shown. The green dashed line presents an exponential function, while the blue solid line is the parameterization used in the estimation method.
}
\label{fig:ratioMT2data-mHT}
\end{figure}

The parameterization is validated by fitting $r(\MTt)$ to a sample of simulated multijet events, and multiplying this ratio by the number of events found in data with $\Delta\phi_\text{min}<0.2$.
The result is compared to the number of events in data with $\Delta\phi_\text{min}>0.3$, after subtraction of the non-multijet contribution using simulation.
An example is shown in Fig.~\ref{fig:ratioMT2data-mHT}.
The prediction is seen to provide a conservative estimate of the expected multijet background.
The robustness of the method is further validated by varying the range of \MTt in which the exponential term is fitted, and by changing the $\Delta\phi_\text{min}$ requirement used to define the control regions.

In the low-\HT regions, the \MET requirement of the triggers
distorts $r(\MTt)$ for low values of \MTt. Therefore, the data selected by the standard triggers
cannot be used to obtain $r(\MTt)$.
Other triggers, based on \HT only, are used instead.
These triggers accept only a small fraction of the events that satisfy the trigger criteria (``prescaled''), allowing access to the low-\HT region without a \MET requirement.

The dominant sources of uncertainty for this method include the
statistical uncertainty of the fit, the stability of the fit under
variations of the fit conditions, the statistical uncertainty of the
control region with $\Delta\phi_\text{min} < 0.2$ used for the extrapolation,
and a 50\% uncertainty assigned to the choice of the \MTt value used to define
the constant term in the functional form of $r(\MTt)$.
In signal regions with low \MTt, where the exponential term of
Eq.~(\ref{eq.bkgdpred.ratio}) dominates the constant term,
this method provides a relatively accurate estimate of the background,
with uncertainties as small as 10\% that increase to around 50\% for signal regions with less statistical precision.
For signal regions with large \MTt, the constant term
dominates and the uncertainty increases to 50--100\%.
Note that at large \MTt, the estimate of the multijet background provided
by this method, while conservative, is nonetheless negligible compared
to the contributions of the other backgrounds.

\subsection{Determination of the \texorpdfstring{$\mathrm{W}(\ell\nu)+$jets and leptonic top-quark background}{W+jets and leptonic top-quark background}}
\label{sec:data:geq3jets:LostLepton}

The background from \wjets and top-quark production (mainly \ttjets events, but also single top-quark production)
stems from events with a leptonically decaying W boson in which the charged lepton either lies outside the detector acceptance, or lies within the acceptance but fails to satisfy the lepton reconstruction, identification, or isolation criteria.
Since these events arise from a lepton ($\Pe$, $\mu$, or $\tau$ lepton) that is not found, we call them ``lost-lepton'' events.
\begin{figure}[!ht]
\centering
\includegraphics[width=0.32\textwidth]{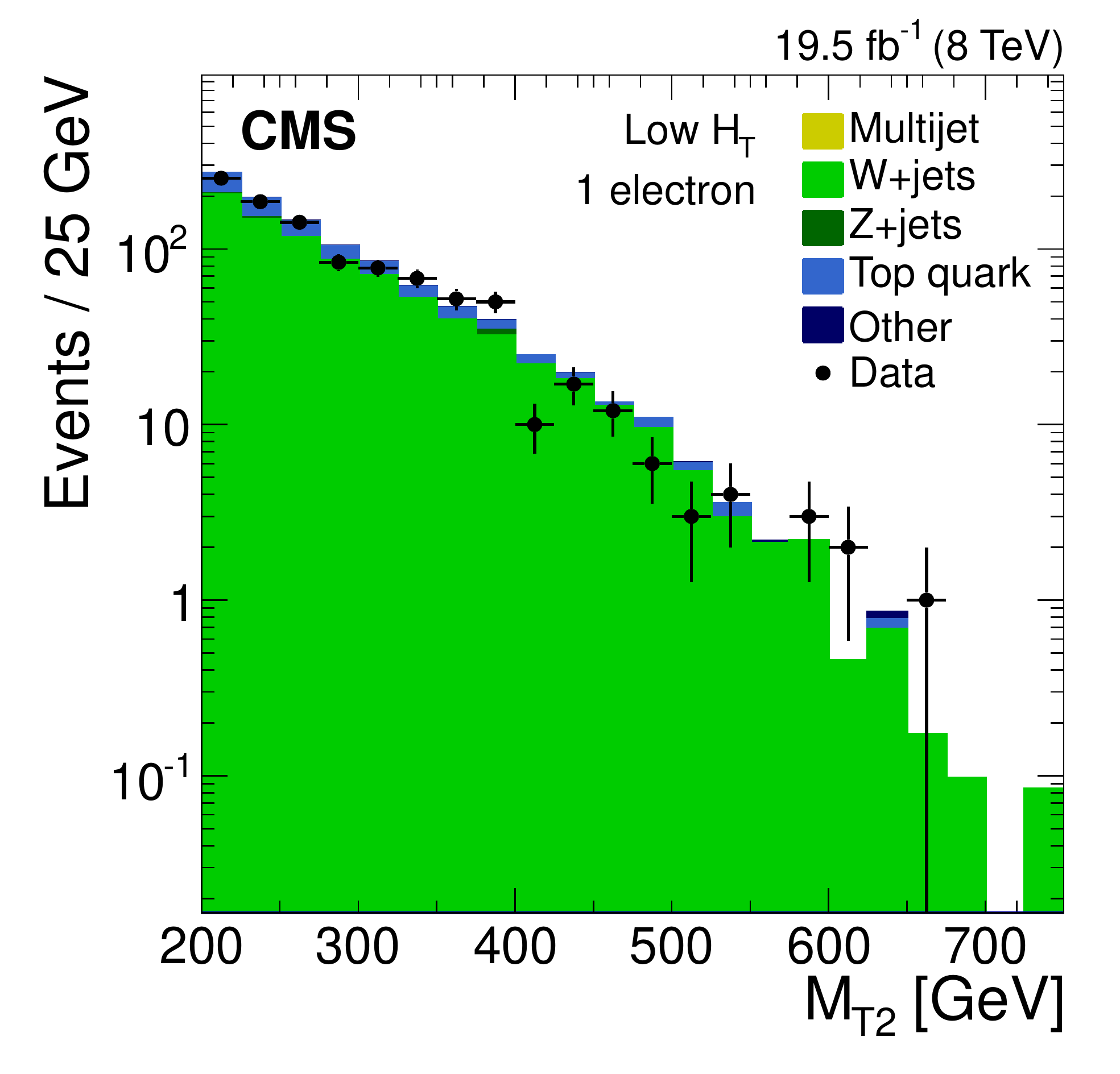}
\includegraphics[width=0.32\textwidth]{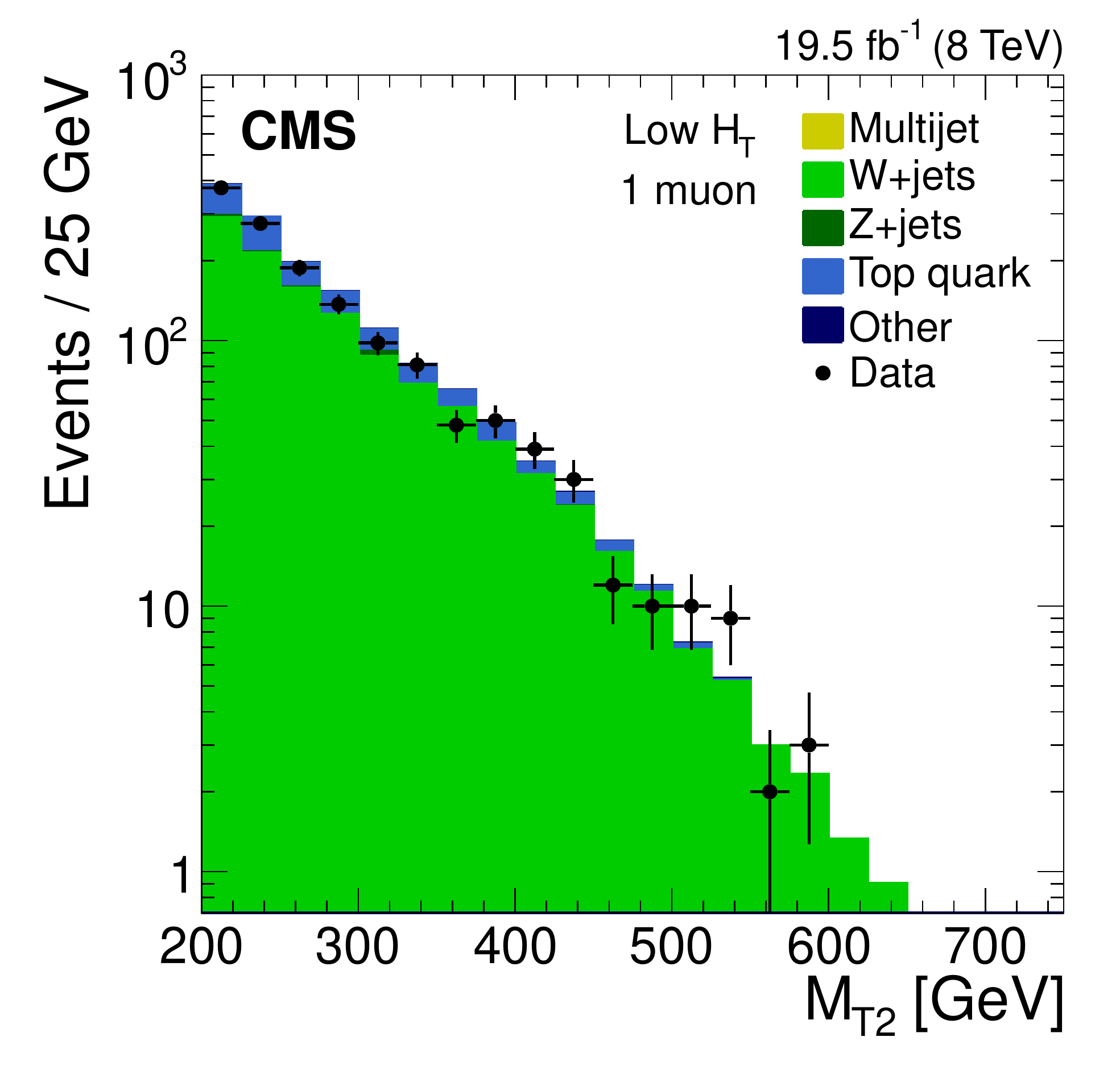}
\includegraphics[width=0.32\textwidth]{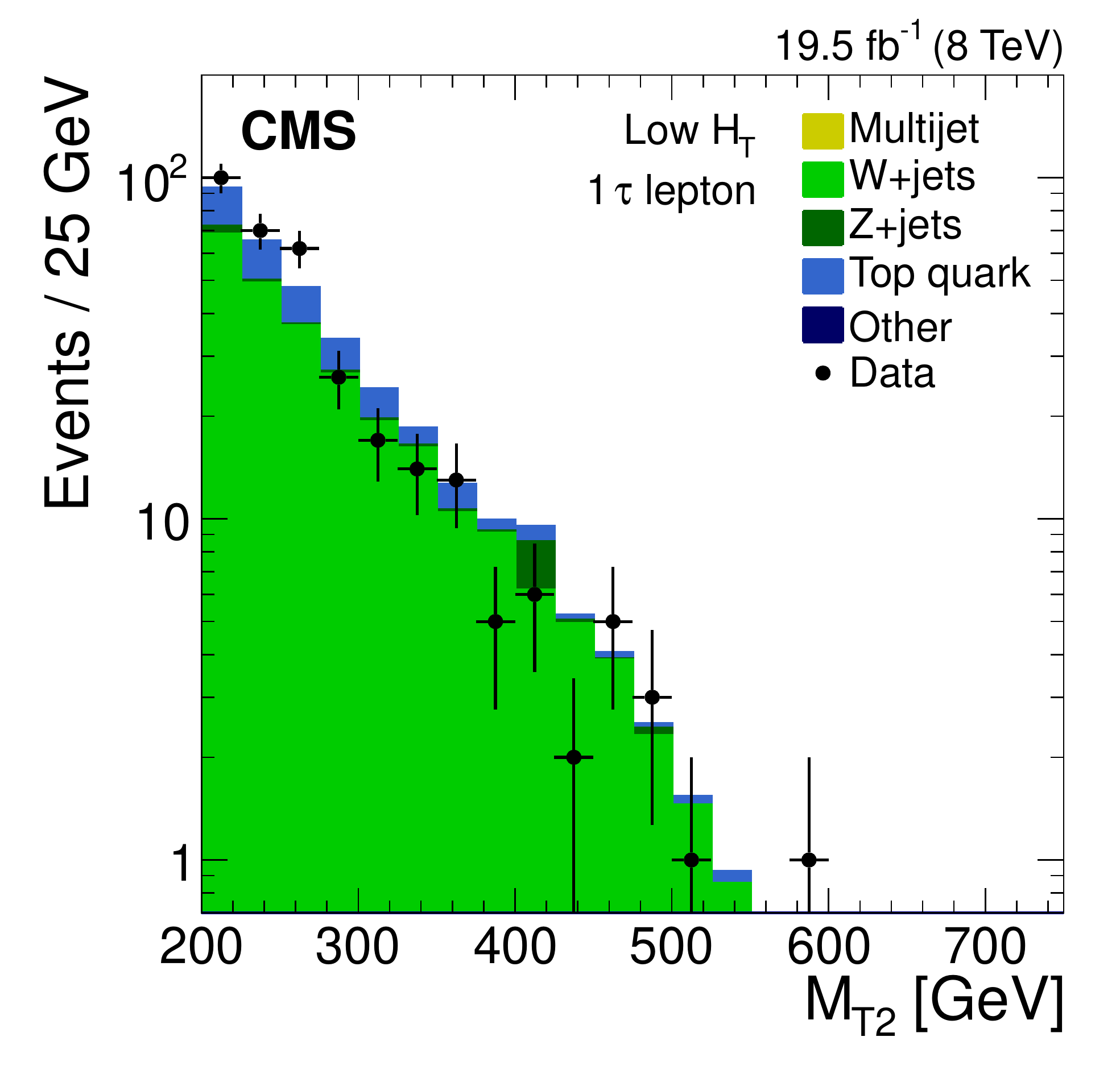}
\includegraphics[width=0.32\textwidth]{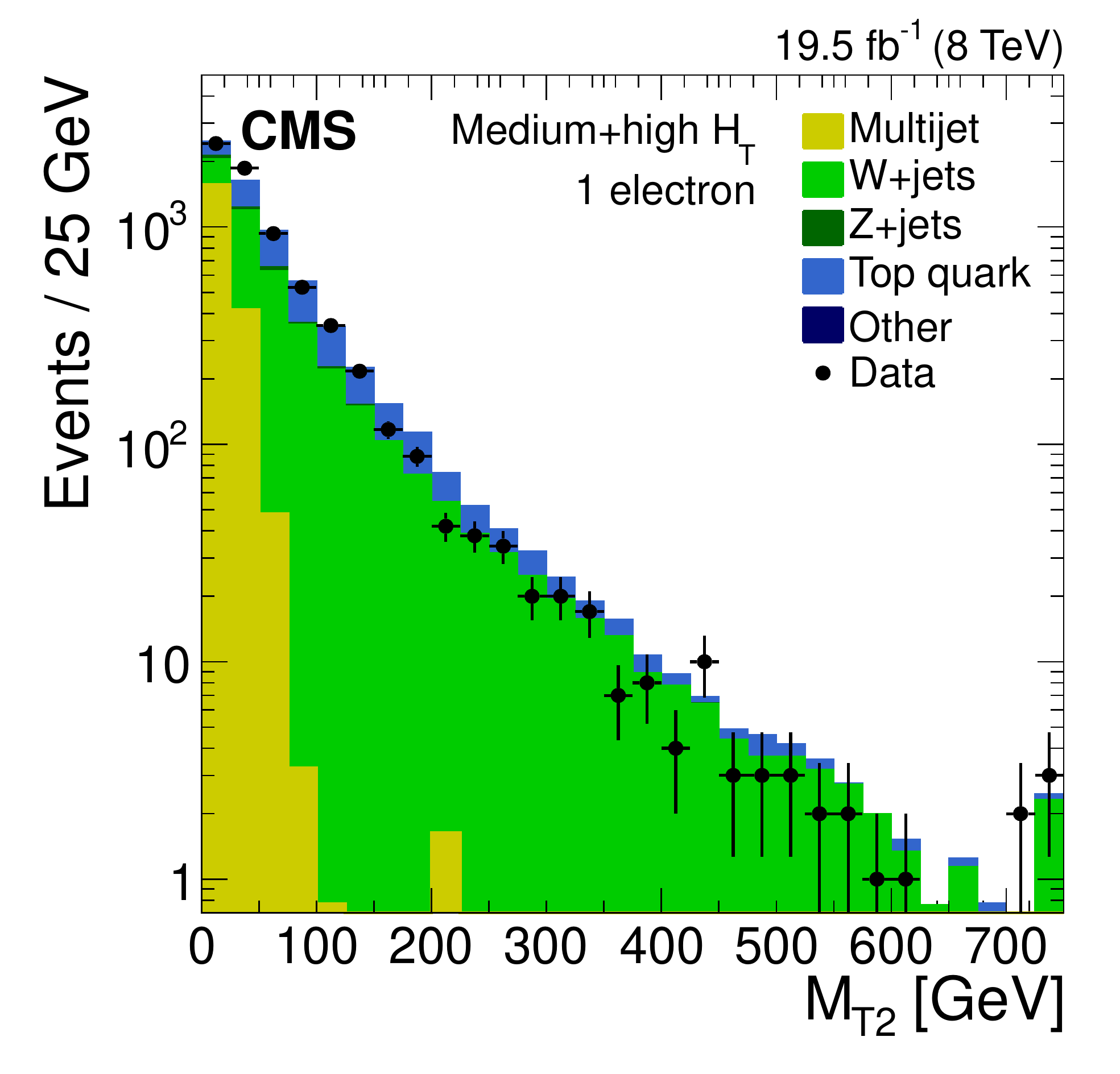}
\includegraphics[width=0.32\textwidth]{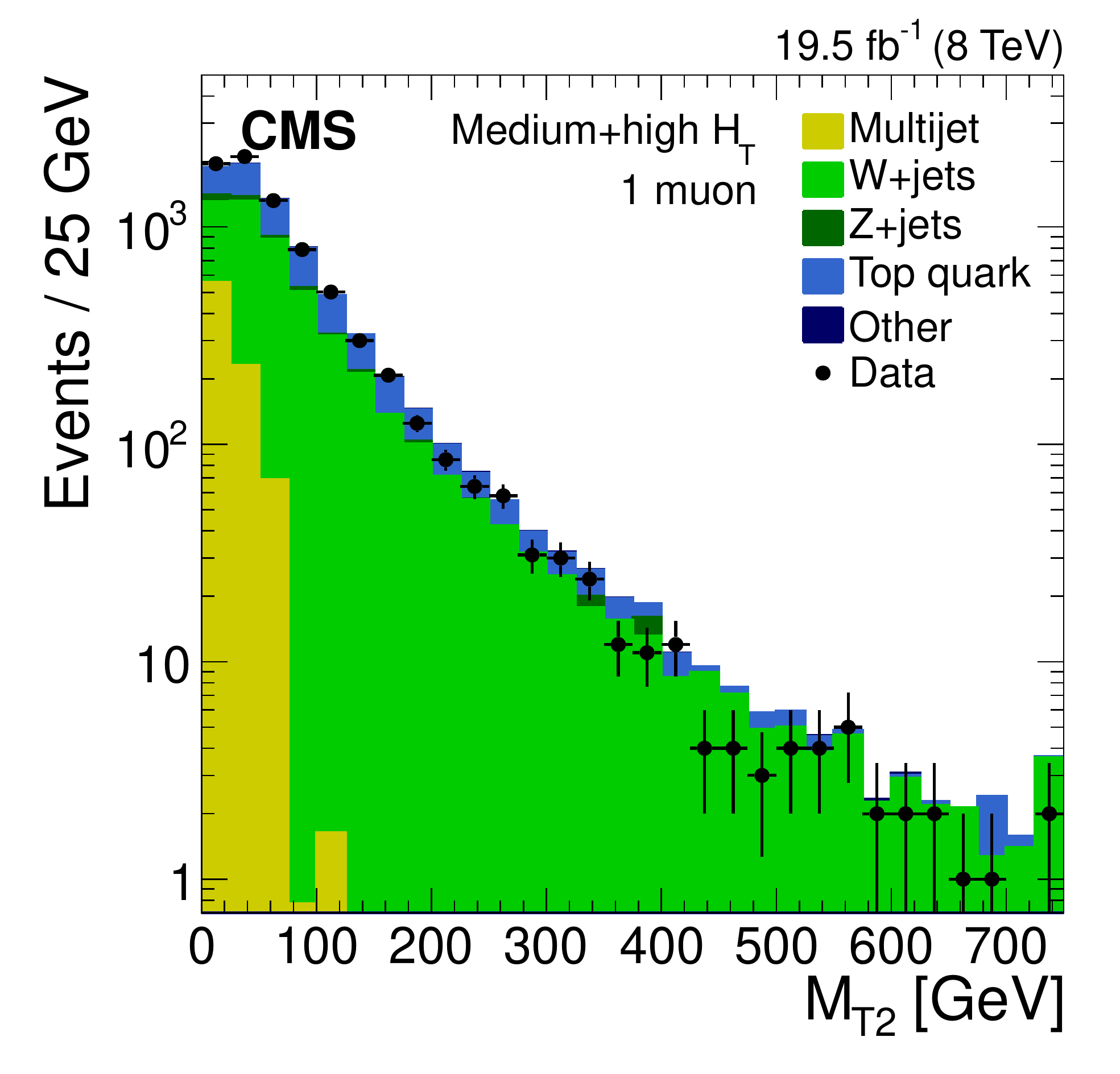}
\includegraphics[width=0.32\textwidth]{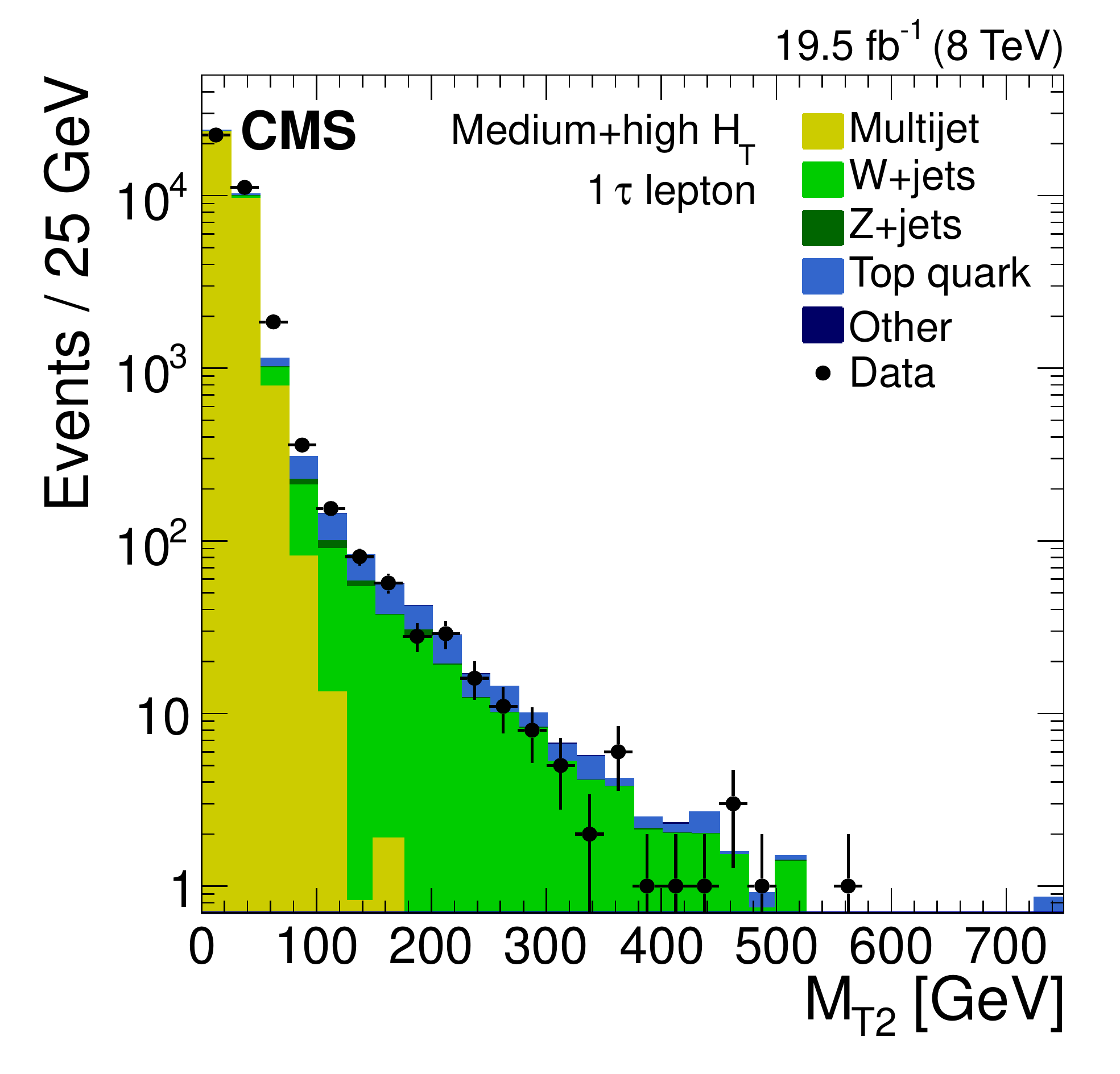}
\caption{Distribution of the \MTt variable for events with one electron
  (left), one muon (middle), or one $\tau$ lepton (right) in data and simulation.
The events satisfy either the low-\HT selection (top) or either of the medium- and high-\HT selections (bottom). They also satisfy
the remaining inclusive-\MTt selection requirements, with the exception of the lepton veto. Finally, the condition
$\MT < 100 \GeV$ is imposed on the charged lepton-\MET system.}
\label{fig:MT2leptdata}
\end{figure}
For both sources of lost leptons, the contribution from $\tau$ leptons is slightly higher than from electrons
or muons since the reconstruction efficiency for $\tau$ leptons is smaller
and the acceptance criteria are more stringent than for the other two types of leptons.
According to simulation, around 40\% of this background can be attributed to
events containing a lost {$\tau$~lepton}. The contribution of electron and muon events is of equal size.

For each signal region, the lost-lepton background is estimated in a corresponding data control sample
for which the full event selection is applied, with the exception of the lepton veto, \ie exactly
one charged lepton ($\Pe$, $\mu$, or $\tau$ lepton) is required instead of zero.
To reduce the potential contribution of signal events to the control samples, the transverse mass of the lepton-\MET system is required to satisfy $\MT < 100 \GeV$.
The \MTt distributions of events satisfying the selection as outlined in Section~\ref{sec:cut-flow},
but after requiring one reconstructed and identified lepton, are shown in Fig.~\ref{fig:MT2leptdata}
for both data and simulation.

After subtracting the number of events expected due to the misidentification of hadrons as leptons and due to leptons from hadron decays, the numbers of events in the one-lepton control samples are scaled by a lost-lepton factor
$R_{\ell\ell} = [1-\varepsilon(\ell)]/[\varepsilon(\ell)\varepsilon(\MT)]$,
where $\varepsilon(\ell)$ is the combined lepton efficiency and acceptance,
and $\varepsilon(\MT)$ is the efficiency of the \MT selection.
This factor $R_{\ell\ell}$ is therefore the transfer factor from the control region to the signal region, obtained in simulation. 

For large values of \MTt, we expect very few events with a single reconstructed charged lepton. Therefore,
the estimation of the lost-lepton background is performed in data for all topological regions
in (\NJ, \NB) and for the different \HT selections, but integrating over all \MTt bins.
The factor $R_{\ell\ell}$ is recalculated for each topological signal region and for the different selections in \HT.
The estimated number of background events is divided among the different \MTt bins using the shape
of the \MTt distribution as predicted by simulation.

The systematic uncertainty in the integrated lost-lepton background estimate includes the uncertainties in the lepton efficiencies,
acceptance, and the
subtraction of the lepton events associated with misidentification and hadron decays.
These uncertainties are obtained by studying the differences between data and simulation
using so-called tag-and-probe~\cite{CMS:2011aa} and tight-to-loose~\cite{Khachatryan:2010ez}
methods.
These uncertainties amount to about 10--20\%. Including
the statistical uncertainties from the data control regions, the
total uncertainty of the lost-lepton background ranges from 10 to 65\%.
The uncertainty in the shape of the \MTt distribution is estimated by varying parameters in the simulation.
The most important of these uncertainties are the recoil modelling \cite{Chatrchyan:2013xna} (20\%), the matching scale, the renormalization and
factorization scales (10--20\%), and the jet energy scale~\cite{Chatrchyan:2011ds} (10\%).
The numbers in parentheses correspond to maximal variations in the \MTt shape, but
the overall normalization is not affected since it is predicted
using the aforementioned method.
The differences in shape between the distributions in data and simulation, shown in Fig.~\ref{fig:MT2leptdata}, lie within these uncertainties.

The effect of signal contributions to the lost-lepton control samples can be
significant and is taken into account before the interpretations
presented in Sections~\ref{sec:stats} and~\ref{sec:exclusion} are performed.
Specifically, the predicted yield in the signal regions is corrected by subtracting the
additional signal contribution caused by the possible presence of the signal in the
lost-lepton control sample.

\subsection{Determination of the \texorpdfstring{$Z(\nu \bar{\nu})$+jets background}{Z+jets background}}
\label{sec:data:geq3jets:Zinv}

The \zinv background is estimated by selecting a control sample of \gjets events and then subtracting the photon momentum in the computation of all the relevant event quantities, such as \MTt,  
in order to replicate the decay of a Z boson into undetected neutrinos. After the subtraction of the photon momentum, the \ptvec and the \MTt variables are recalculated and the event selections
corresponding to the different signal regions are applied. The number of selected events, which
is rescaled as described below, provides the background estimate for the \zinv process.

As discussed in Ref.~\cite{Ask:2011xf}, the Z+jets and \gjets processes differ
because of the different electroweak couplings and the non-zero Z-boson mass $m_{\Z}$.
For vector boson $\pt \gg m_{\Z}$, however,
the ratio of cross sections for prompt-photon to Z-boson production is
determined by the ratio of the couplings of the respective boson
to quarks, and thus approaches a constant value.
In this range of the boson \pt,
the distributions of \HT and other kinematic observables
are very similar for the Z+jets and \gjets processes.
The \gjets process, with its relatively large event yield, 
is thus well suited to provide an estimate of the \zinv background. 

\begin{figure}[!htb]
\centering
\includegraphics[width=0.46\textwidth]{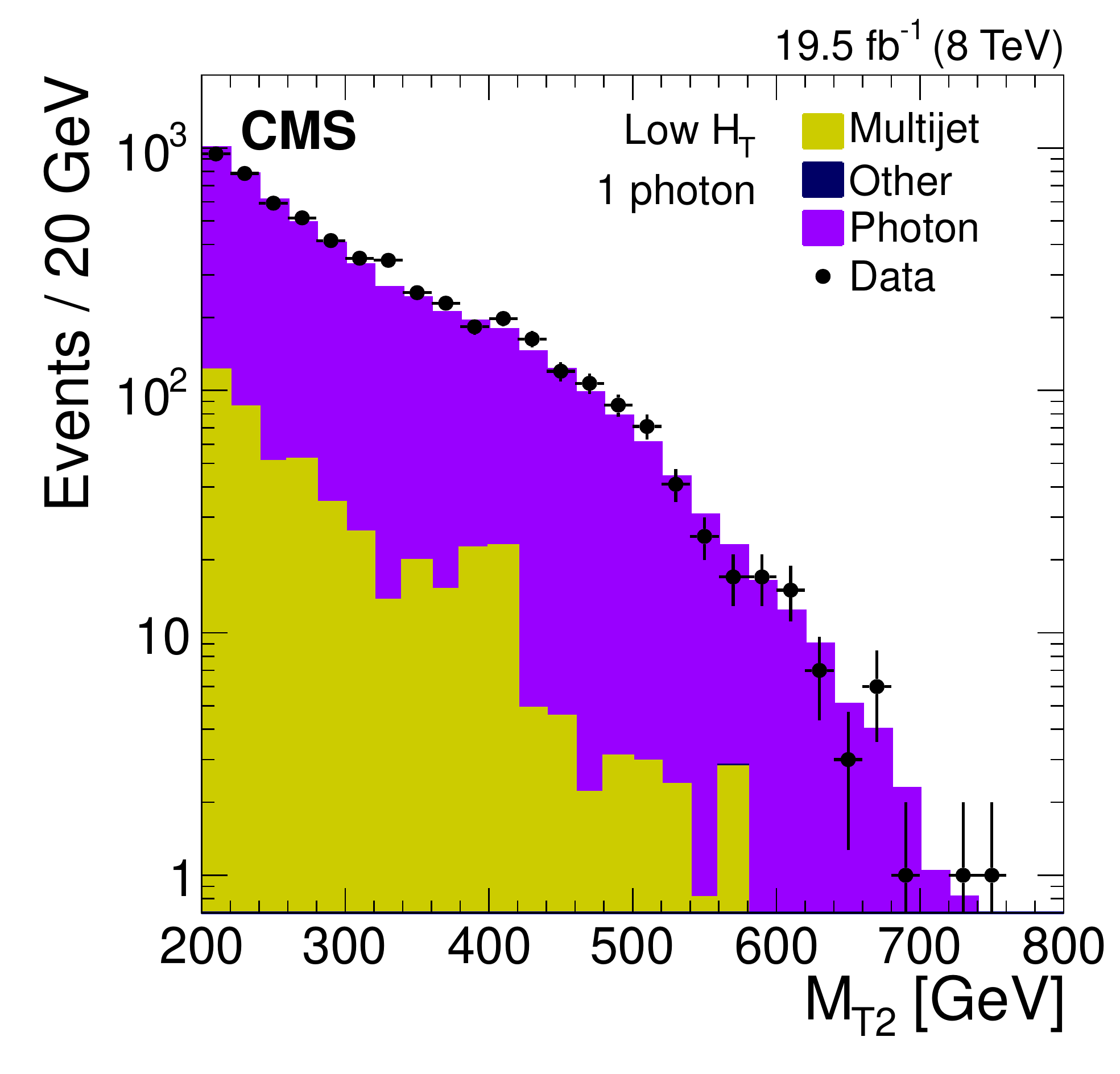}
\includegraphics[width=0.46\textwidth]{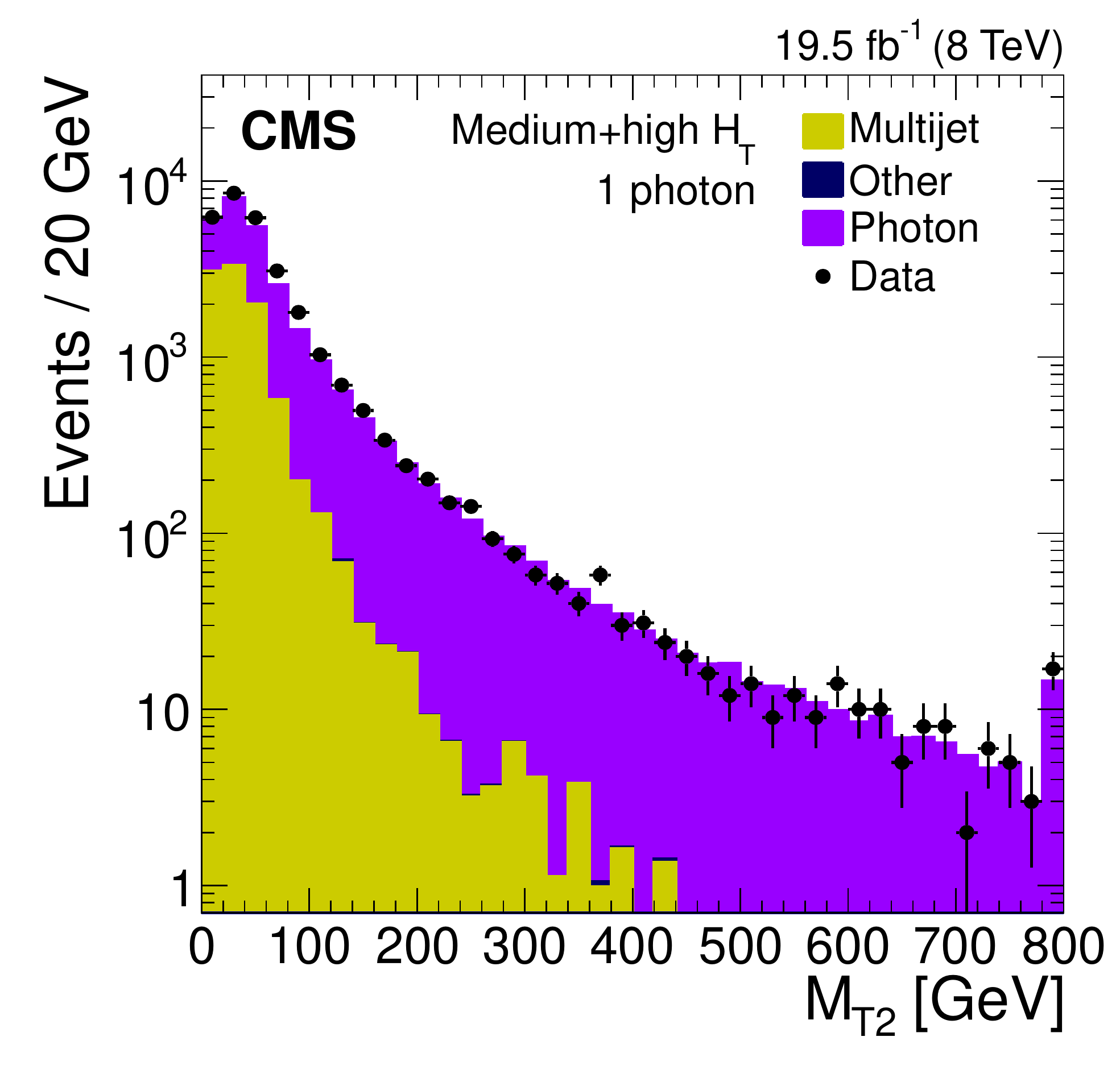}
\caption{Distribution of the \MTt variable for data and simulation after requiring the presence of one
photon, $\NB=0$, and the remainder of the inclusive-\MTt selection criteria. Events satisfying the
low-\HT selection (left), and the medium- and high-\HT selections (right) are shown.
For these results, \MTt is calculated after adding the photon \PT  to the \MET vector.
}
\label{fig:photon_mt2}
\end{figure}

Figure~\ref{fig:photon_mt2} shows a comparison between data and simulation for the \MTt distribution in
\gjets control samples, for which \NB=0 is required.
The photon \ptvec is added to the \ptvecmiss vector and all event variables are recalculated.
To reduce the potential contribution of signal events to these control samples, we require the reconstructed \MET to be less than 100\GeV prior to including the reconstructed photon momentum.
For the low-\HT signal regions, the \gjets events are selected with a single-photon trigger, which requires the photon \pt to exceed 150\GeV.
The single-photon trigger is used because the triggers discussed in Section~\ref{sec:CMS} are unable to select events with low enough \MET.
For the medium- and high-\HT signal regions, the triggers discussed in Section~\ref{sec:CMS} are used.

\begin{figure}[!htb]
\centering
\includegraphics[width=0.49\textwidth]{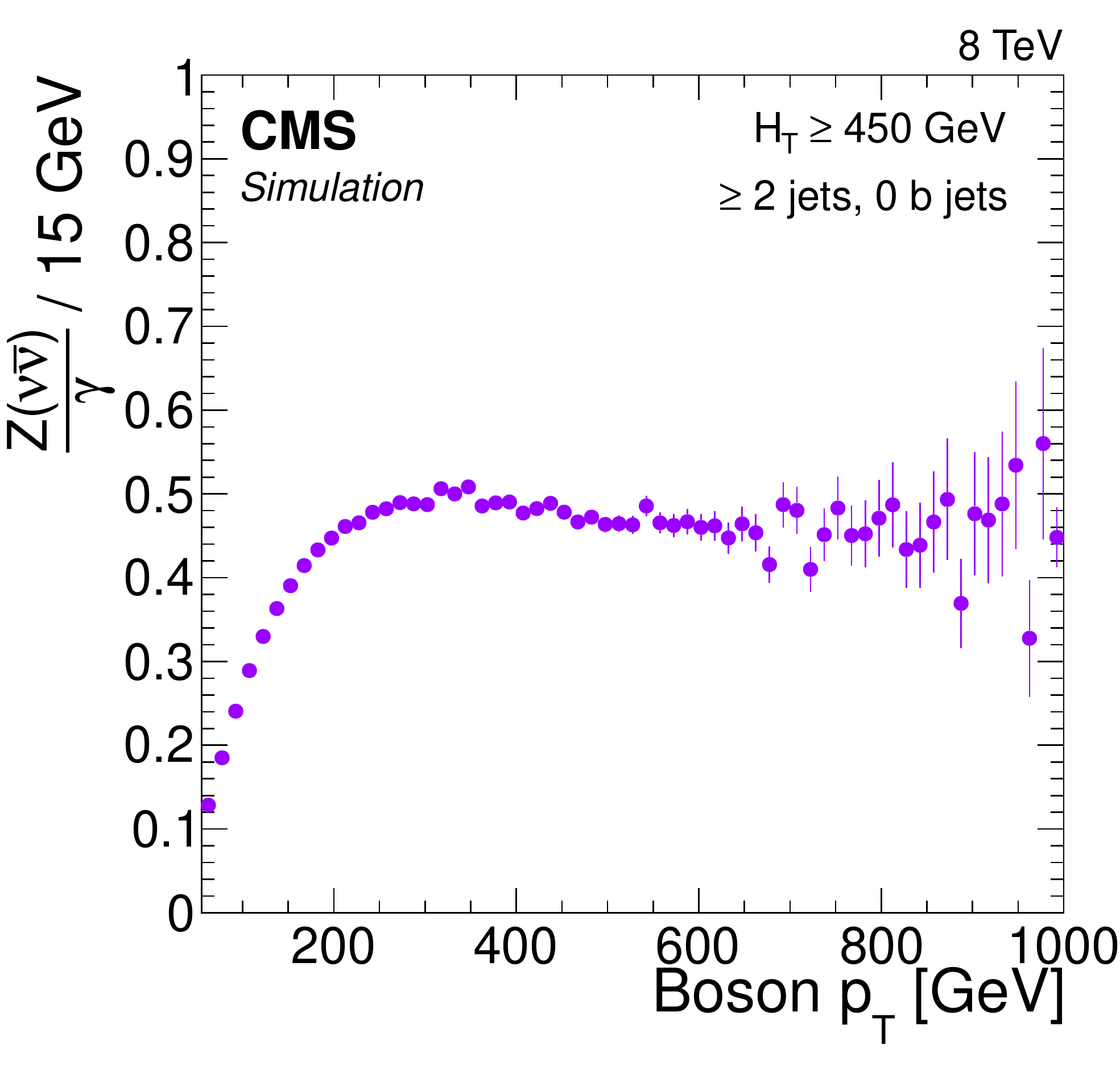}
\caption{Ratio \ZG of events satisfying the event selection of the ($\NJ\ge 2$, $\NB=0$) signal region
as a function of the boson \PT. The events are summed inclusively in all \HT sub-regions
with $\HT\ge450\GeV$.
The ratio is obtained in simulated events after the photon momentum is included in the \MET calculation.
}
\label{fig:PT_PhotonVsZnunuAllcuts}
\end{figure}

The selected photon control samples contain both genuine prompt-photon events and events
with collinear pairs of photons that stem from neutral-meson decays within jets and
are reconstructed as single photons.
The prompt-photon fraction in the control samples is obtained
by means of a maximum likelihood fit of templates from simulation
to a photon shower shape variable in data.
The shower shape variable that we use is $\sigma_{\eta\eta}$,
which is a measure of the lateral extent in $\eta$ of the photon energy cluster in the calorimeter~\cite{CMS-PAS-EGM-10-005}.
The fit is performed separately in the electromagnetic calorimeter barrel
and endcap detectors, for events with $\NB=0$ and with no requirement on \MTt.
This sample of events is dominated by low-\PT photons, for which the shower shape variable provides
high discrimination between prompt photons and photons from neutral-meson decays.
Starting from the overall prompt-photon fraction observed in data, we use simulation to extrapolate the contributions of the two types of photon events in \MTt.
For each signal region with $\NB=0$, the final \zinv background estimate is obtained from the number
of prompt-photon events, rescaled by the \MTt-dependent ratio of \zinv to \gjets events
from simulation. The \ZG ratio increases as a function of the photon \PT
and reaches a constant value above 350\GeV, as shown in Fig.~\ref{fig:PT_PhotonVsZnunuAllcuts}.

The accuracy of the Z-boson \pt distribution in simulation is validated using a control sample of dileptonic Z-boson
events, \ie $\Z \to \Pep\Pem\text{ or }\Pgmp\Pgmm$, selected with dilepton triggers. 
Here, analogously to the photon control sample, the dilepton momentum is subtracted in the computation of all relevant event quantities, such as \MTt, in order to model the $\Z \to \nu\PAGn$ decay. 
From the data-to-simulation comparison of the $\zll/\gamma$ ratio as a function of the search variables, a systematic uncertainty of 20\% is assigned to the \ZG ratio.
For the signal region bins corresponding to $\MTt > 350 \GeV$, this uncertainty increases to 30\%
because of large statistical uncertainty in the ratio for events with large \MTt.
Compared to these uncertainties, the normalization uncertainty associated with the shower shape fit is negligible.

The \ZG ratio may not be well modelled in simulation for $\NB\ge1$ as the coupling of Z bosons and photons differs for b quarks. If the b-quark content in simulation is mismodelled (for example the modelling of gluon splitting $\Pg \to \PQb \PAQb$), the \ZG ratio might be biased in b-quark enriched events. Another biasing effect might be the treatment of the b-quark mass in simulation, which affects the coupling of b quarks to Z bosons and photons. 
Therefore, the previous procedure is only applied in signal regions with $\NB=0$.
For the $\NB=1$ case, the results obtained from the $\NB=0$ control samples are scaled
by $\Z_{\ell\ell}(1\PQb) / \Z_{\ell\ell}(0\PQb)$, the ratio of the numbers of events containing
dileptonic decays of the Z boson and $\NB = 1$ or $\NB = 0$, respectively.
This ratio is obtained using data from the dilepton control sample for different values of \NJ. As the ratio
is found to depend neither on \MTt nor \HT, its value is measured
without any requirement on these two variables, in order to increase the statistical precision of the control samples.

Uncertainties in the $\Z_{\ell\ell}(1\PQb) / \Z_{\ell\ell}(0\PQb)$ ratio are evaluated by varying the kinematic selections to test the stability of the ratio.
The resulting uncertainties are mostly determined by the statistical limitations of the control samples.
The size of the uncertainty is 10--30\% for the regions with $\NJ\leq5$, while it is 50--75\% for regions with $\NJ \geq 6$.

For the signal regions with $\NB\ge 2$, the \zinv background is estimated from simulation and
is assigned an uncertainty of 100\%.
We verified that using an uncertainty twice as large, or twice as small,
has a negligible impact on the final results.
The explanation for this is that for $\NB\geq 2$, the \zinv background
is very small compared to the \ttjets background.

\section{Results}
\label{sec:data:geq3jets.results}

This section reports the number of events observed in the signal regions.
The yields are compared with the estimated number
of background events as predicted by the methods described in Section~\ref{sec:bkgEstimation}.

\subsection{Results for the inclusive-\texorpdfstring{\MTt analysis}{MT2 analysis}}
\label{sec:data:incl:results}

\begin{figure}[!hp]
  \centering
\includegraphics[width=0.32\textwidth]{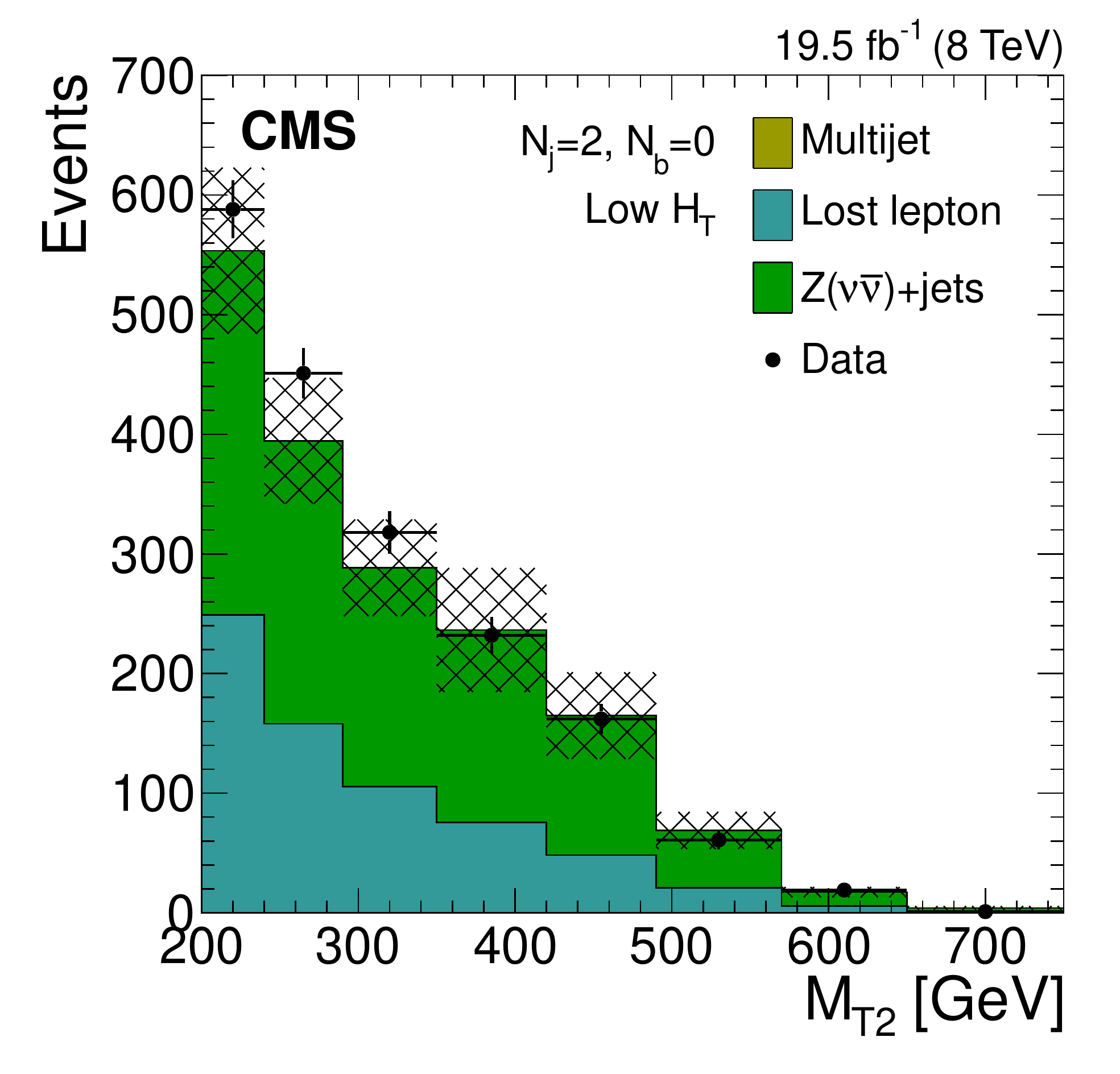}
\includegraphics[width=0.32\textwidth]{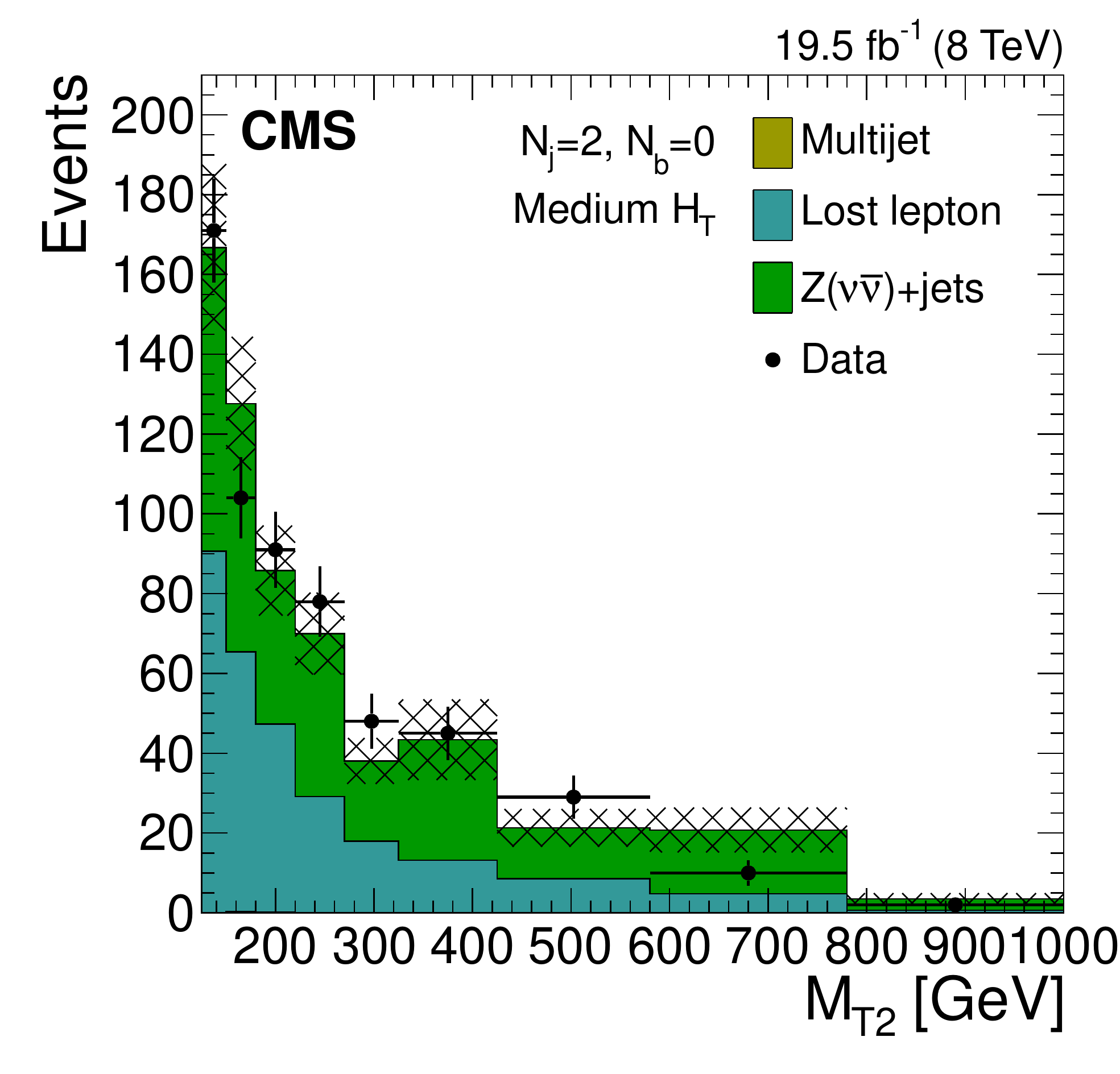}
\includegraphics[width=0.32\textwidth]{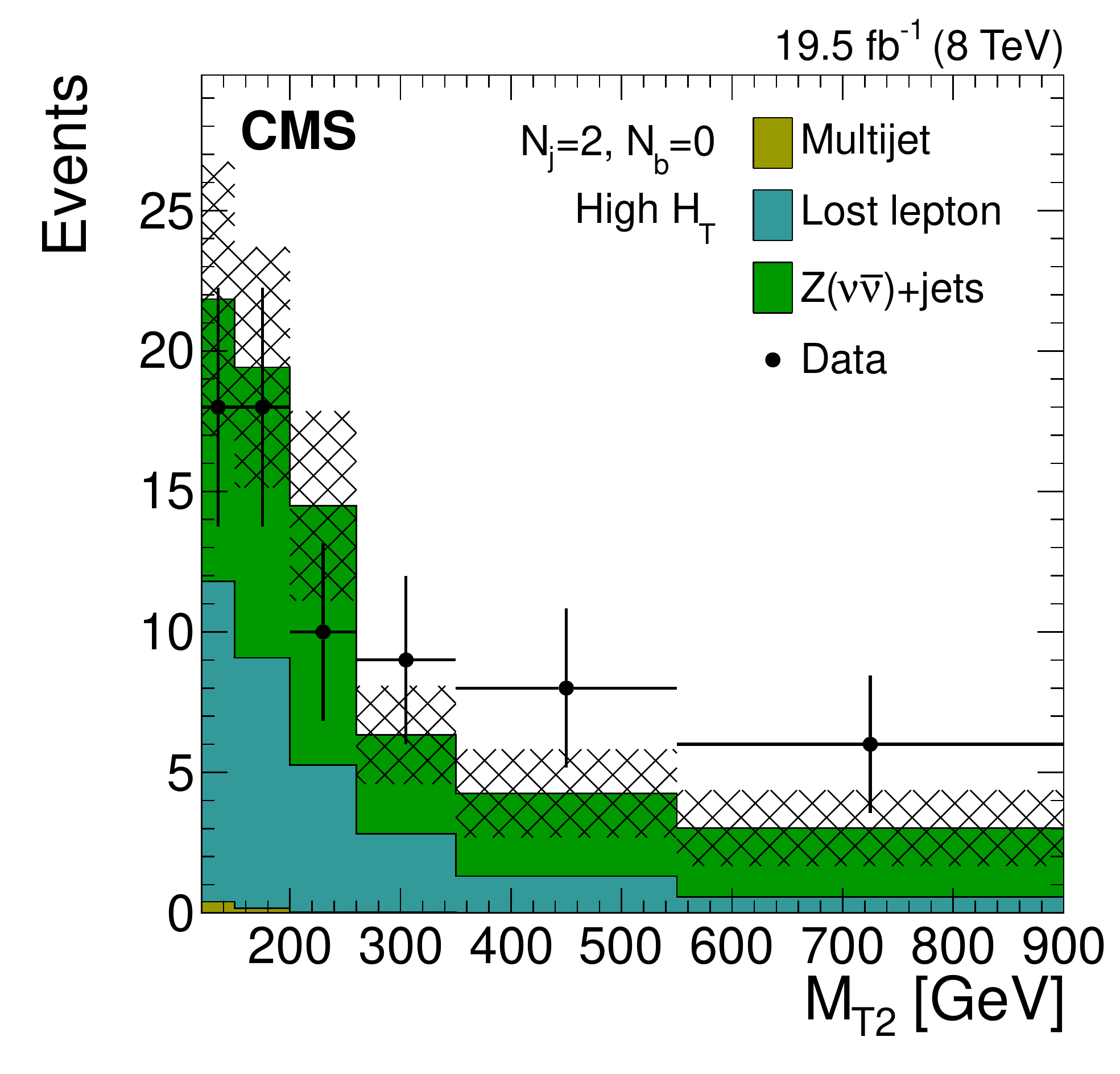}
\includegraphics[width=0.32\textwidth]{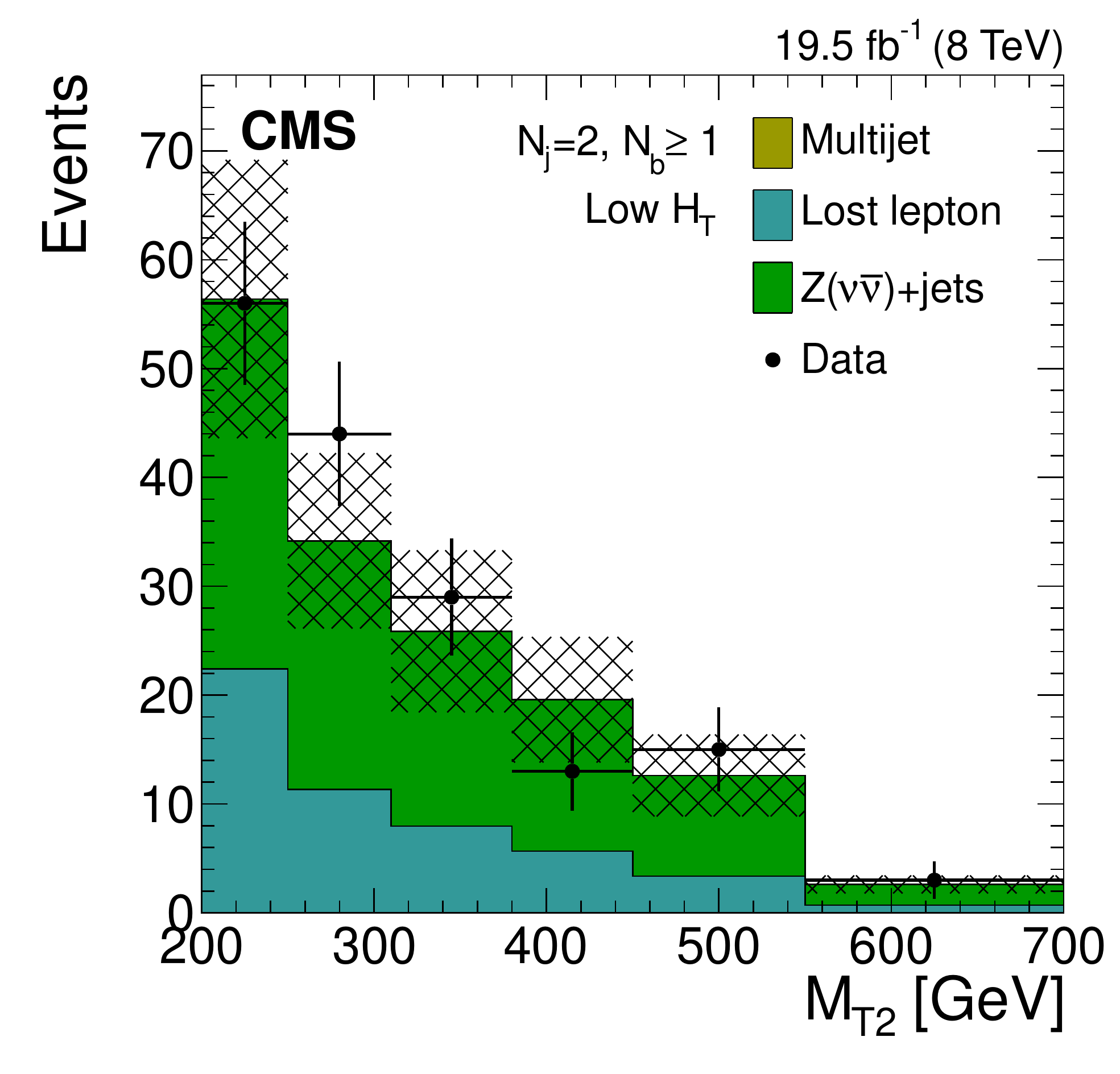}
\includegraphics[width=0.32\textwidth]{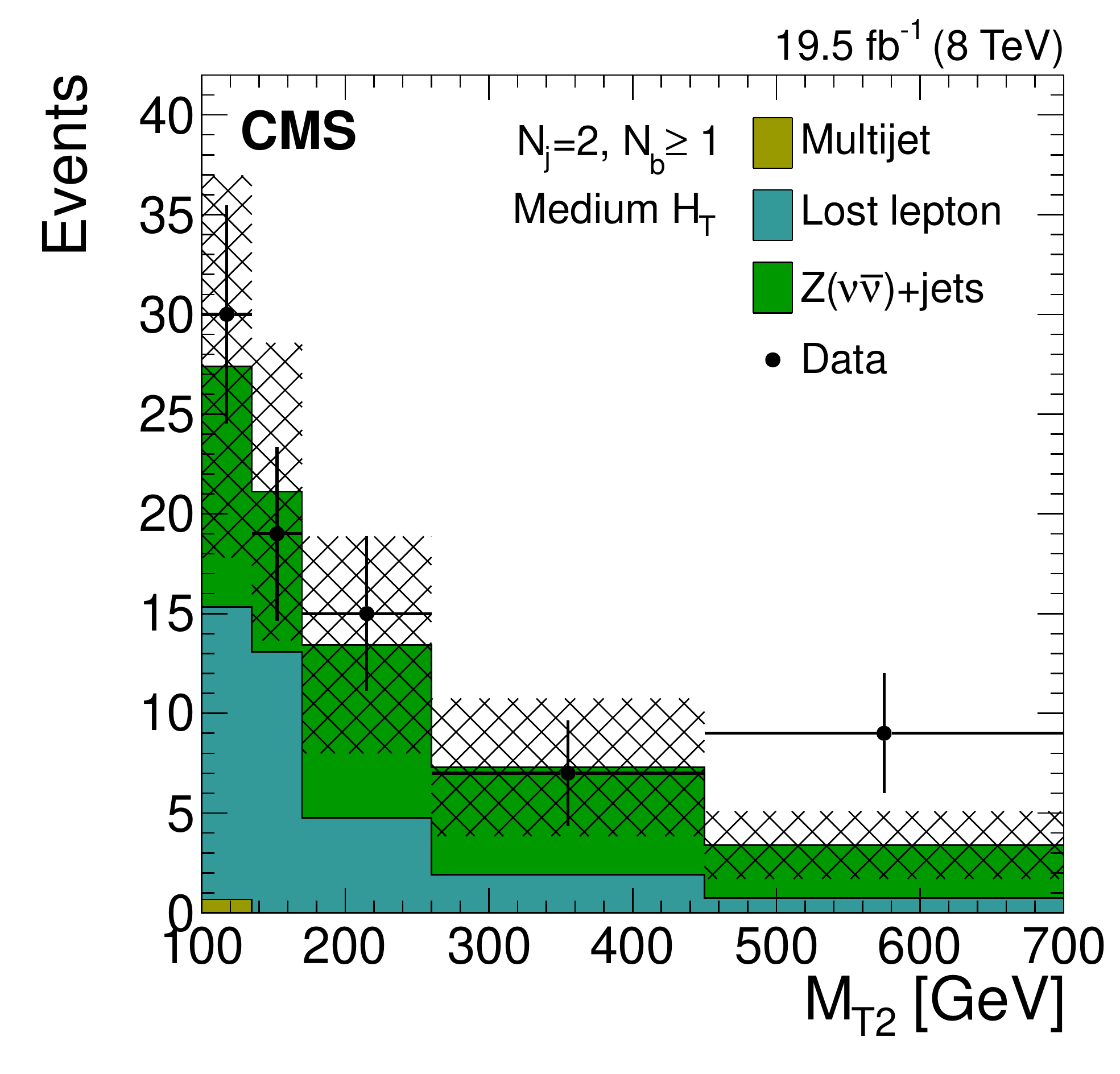}
\includegraphics[width=0.32\textwidth]{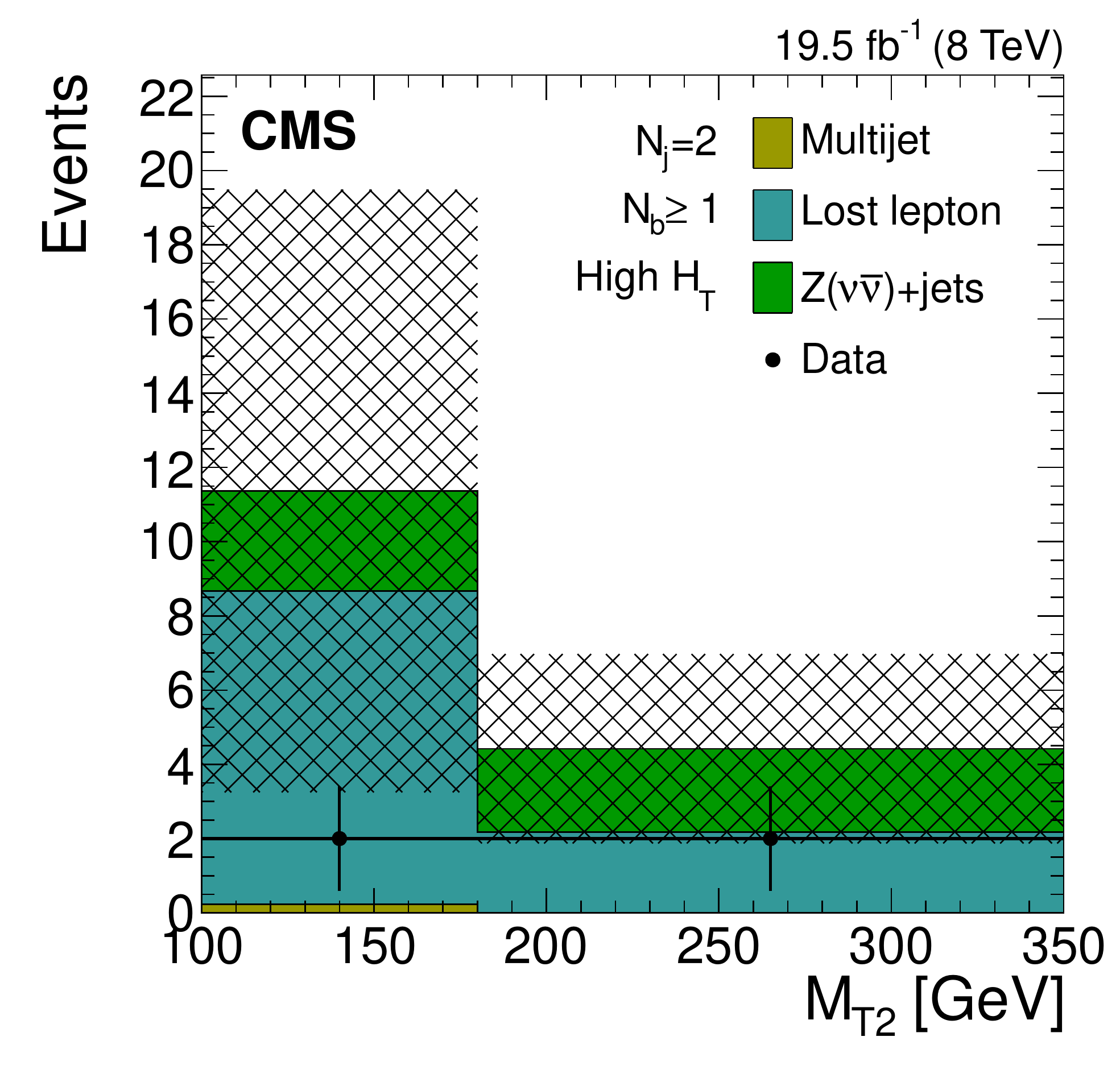}
\includegraphics[width=0.32\textwidth]{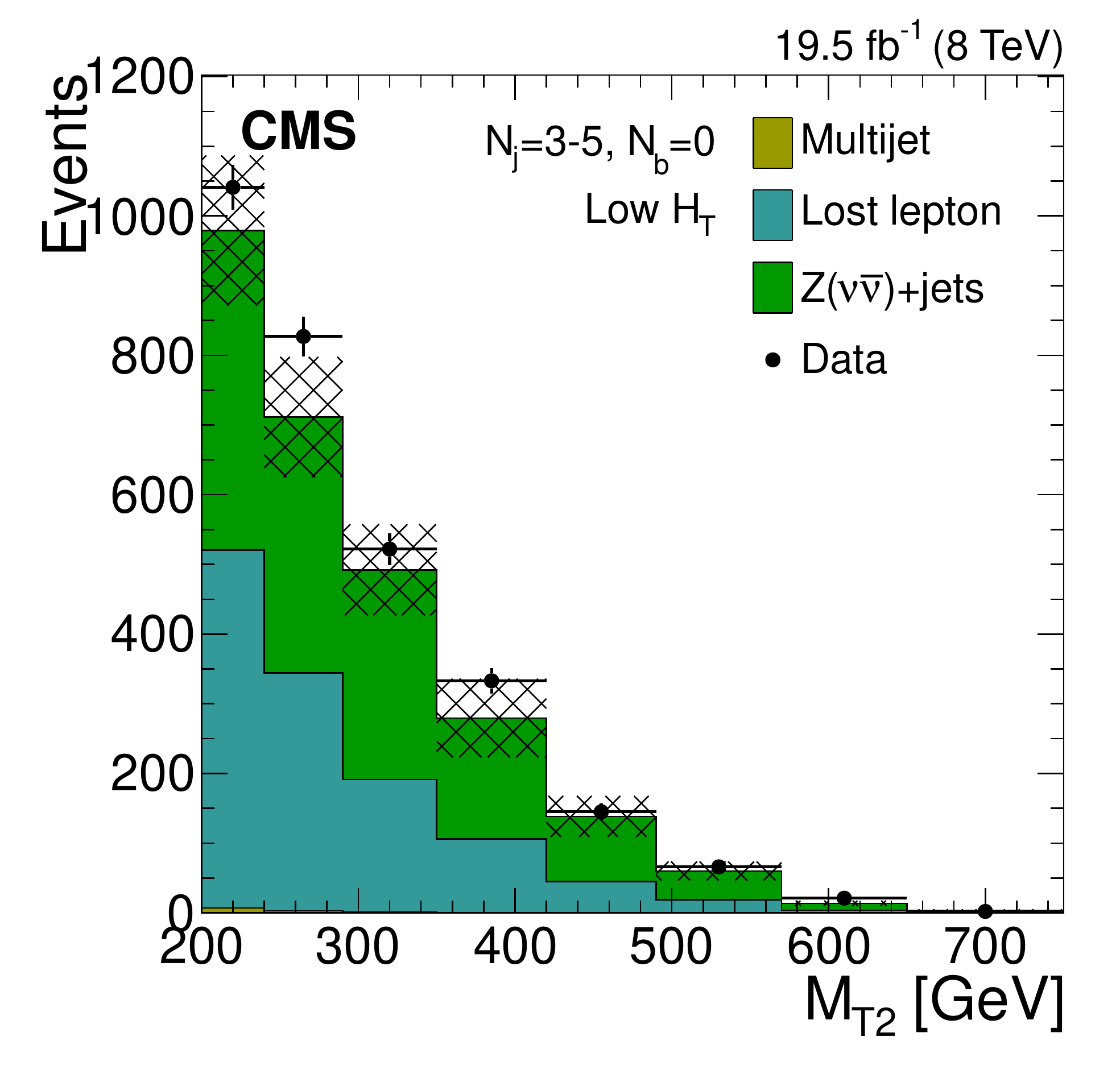}
\includegraphics[width=0.32\textwidth]{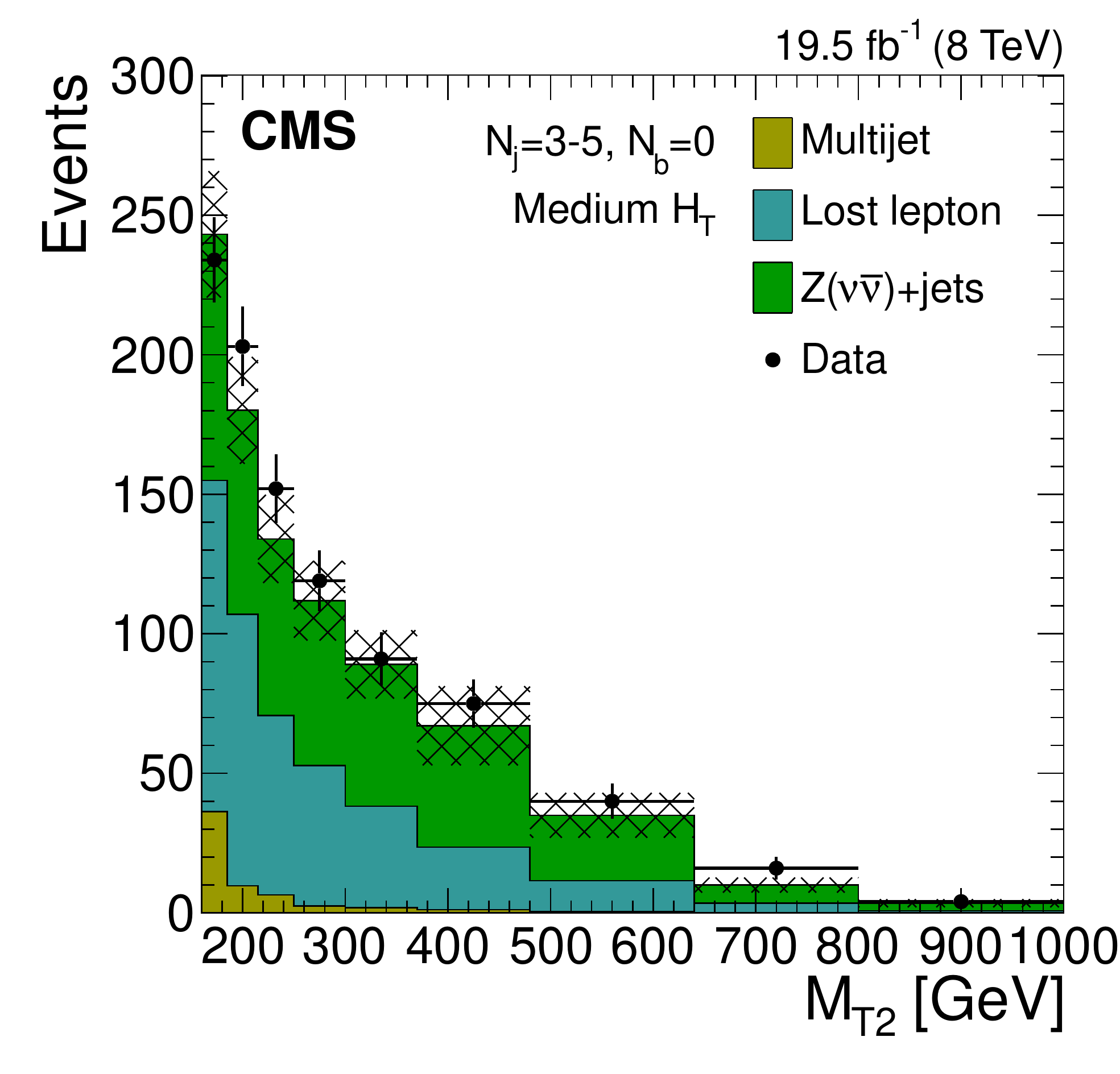}
\includegraphics[width=0.32\textwidth]{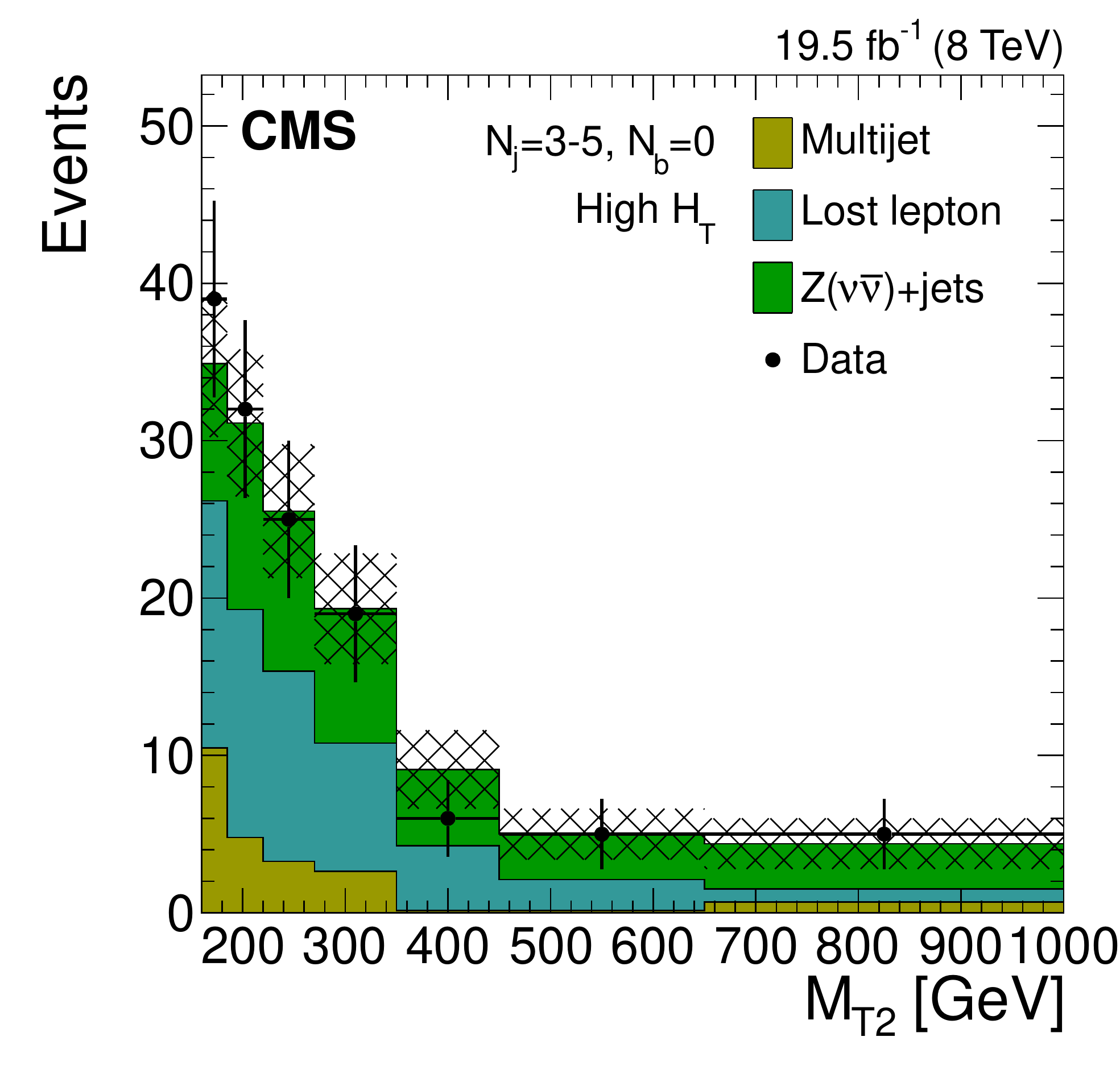}
\caption{
  Distributions of the \MTt variable for the estimated background processes and for data.
  Plots are shown for events satisfying the low-\HT (left), the medium-\HT (middle), and the
  high-\HT (right) selections, and for different topological signal regions (\NJ, \NB) of the
  inclusive-\MTt event selection.
  From top to bottom, these are $(\NJ=2,\NB=0)$, $(\NJ=2,\NB\geq1)$, and $(3\leq \NJ\leq5,\NB=0)$.
  The uncertainties in each plot are drawn as the shaded band and do not include the uncertainty in the shape of the lost-lepton background.
}
\label{fig:datadrivenMT2_1}
\end{figure}

\begin{figure}[!hp]
  \centering
\includegraphics[width=0.32\textwidth]{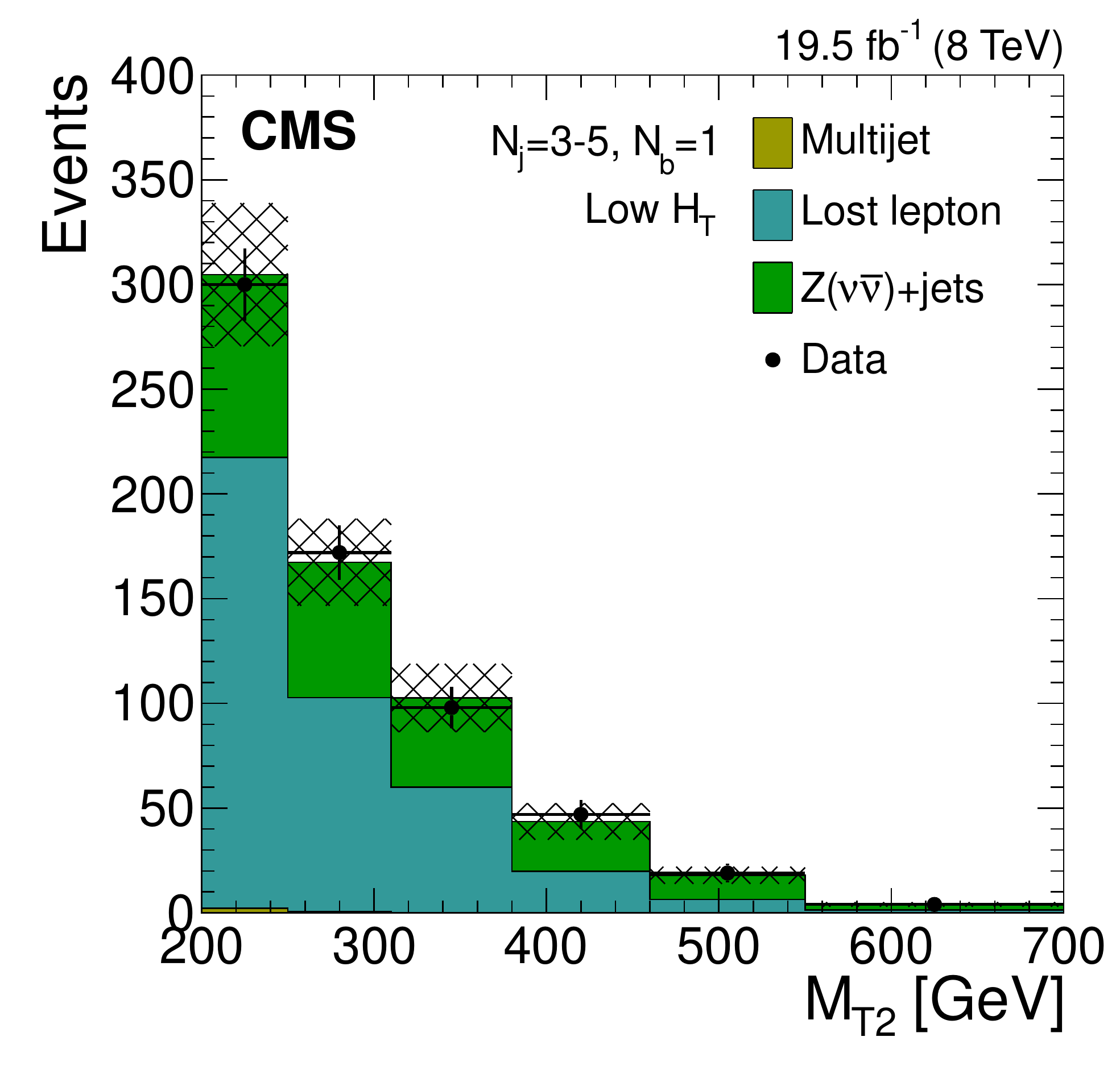}
\includegraphics[width=0.32\textwidth]{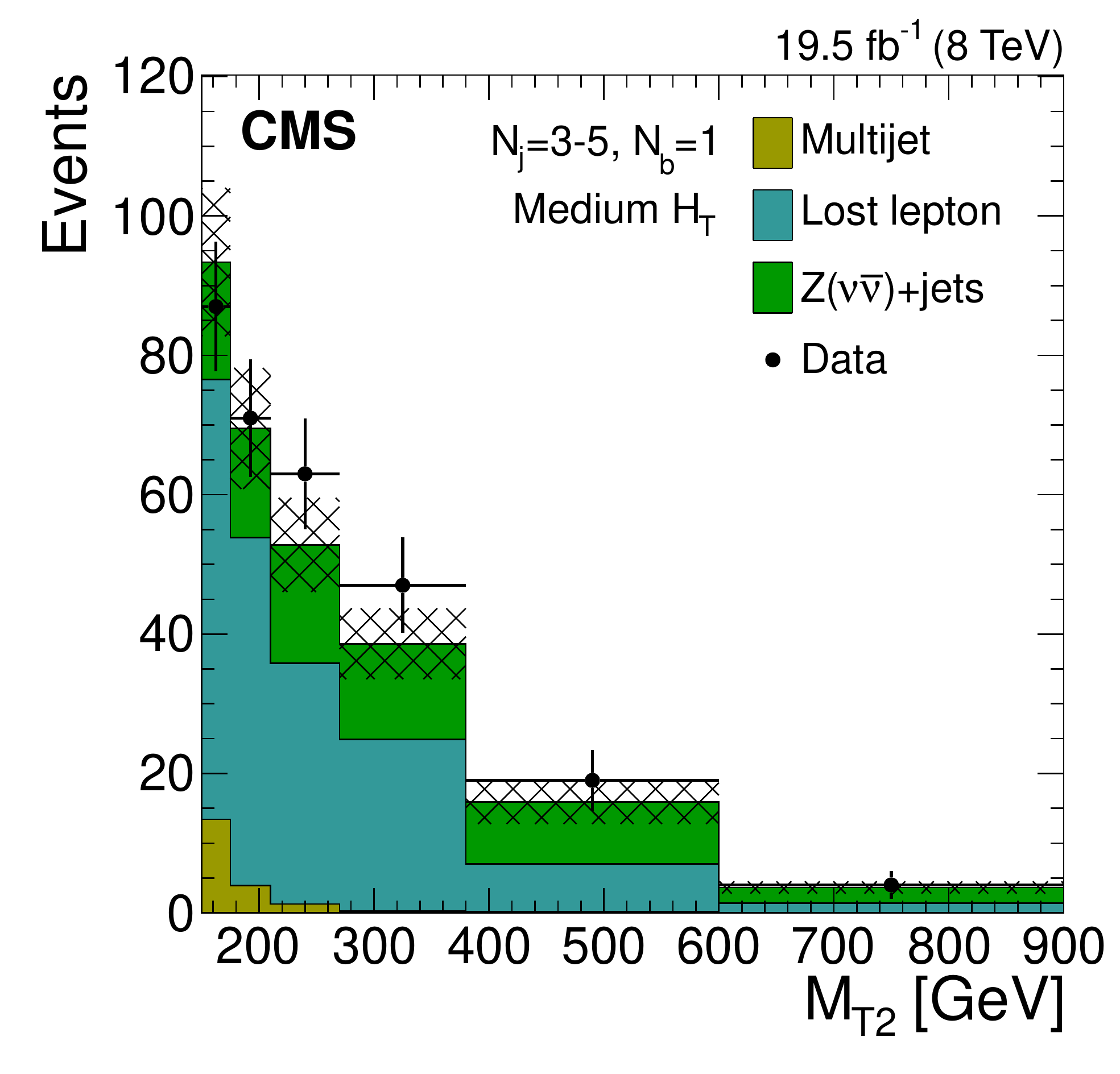}
\includegraphics[width=0.32\textwidth]{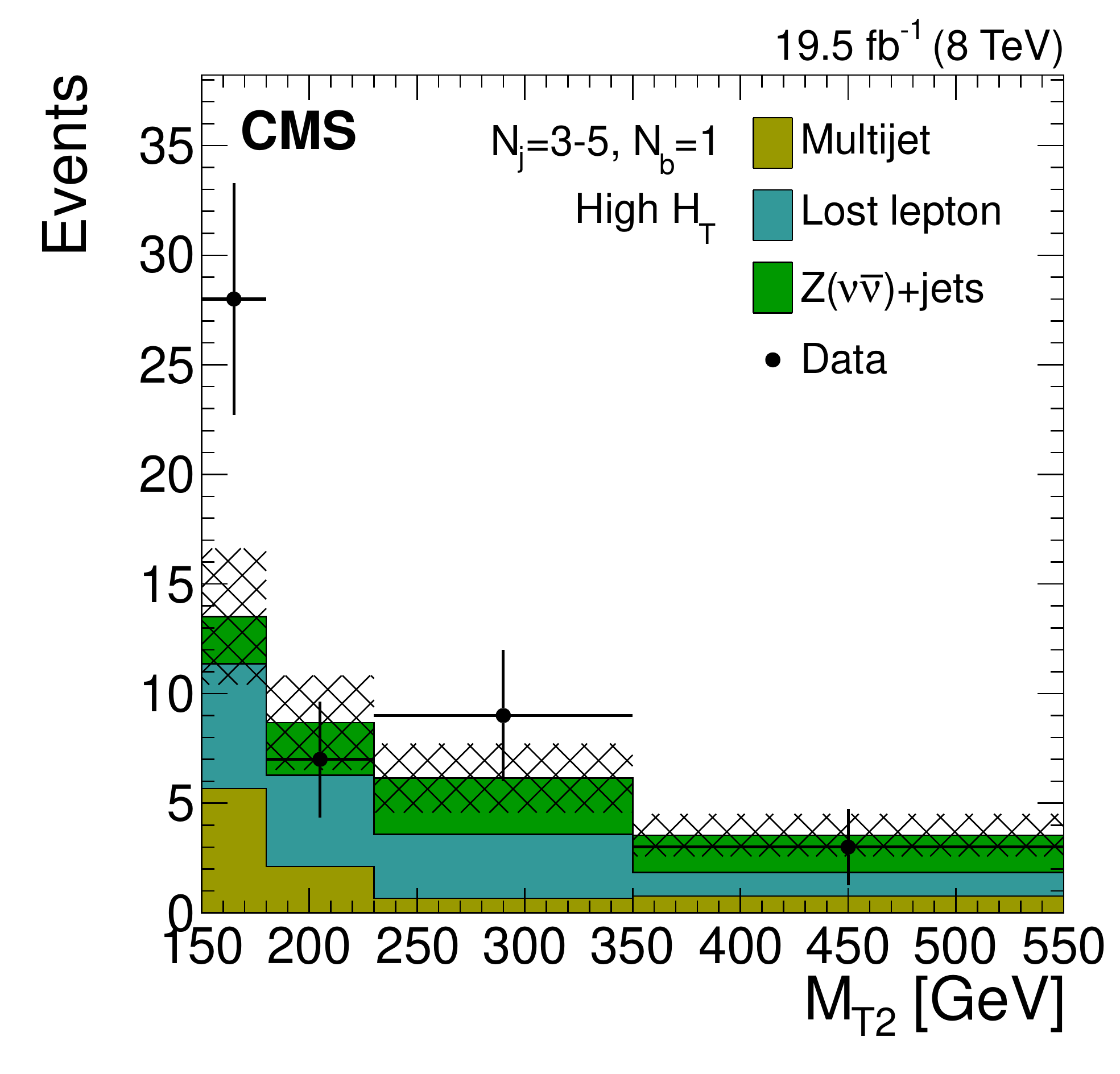}
\includegraphics[width=0.32\textwidth]{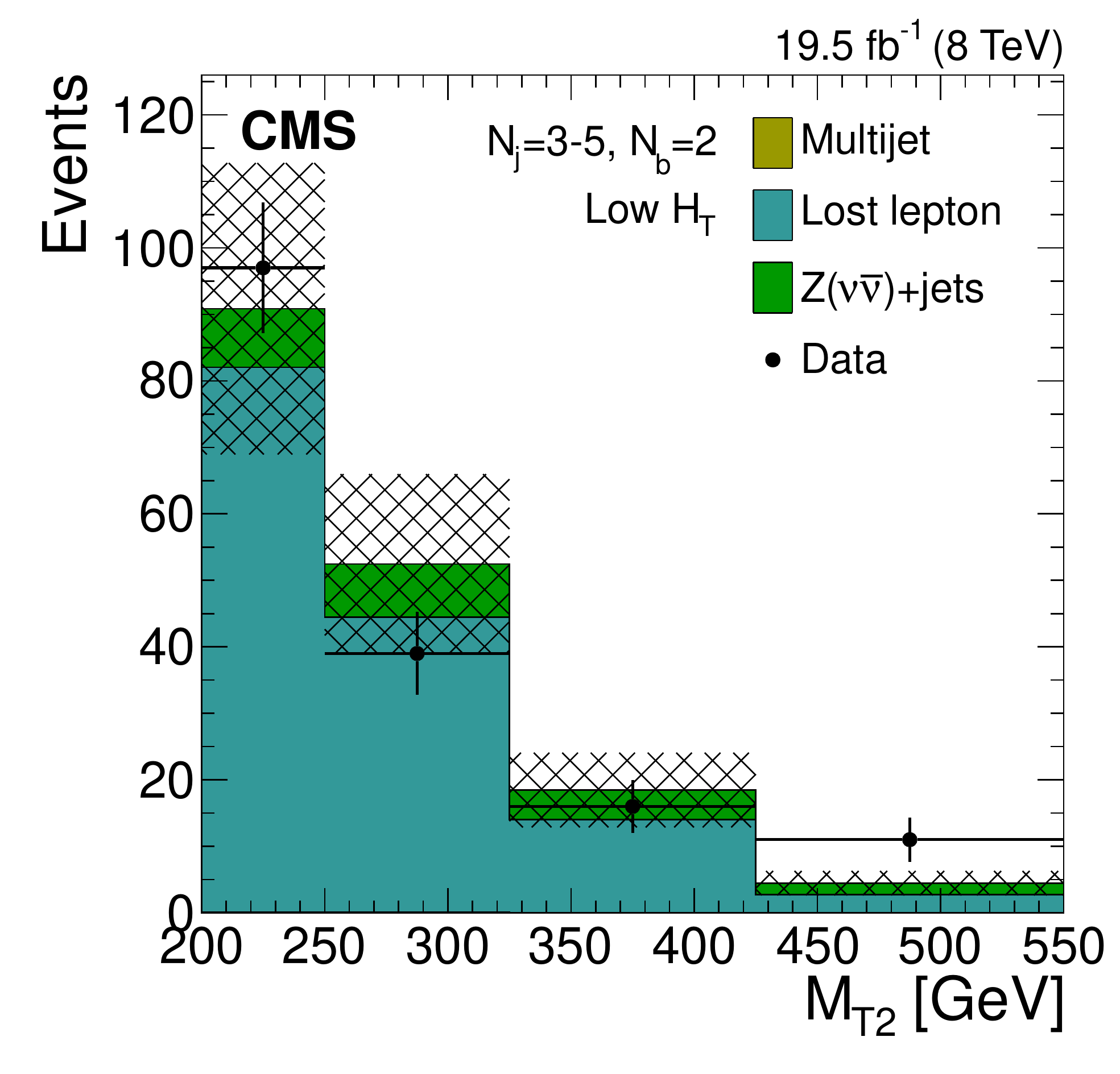}
\includegraphics[width=0.32\textwidth]{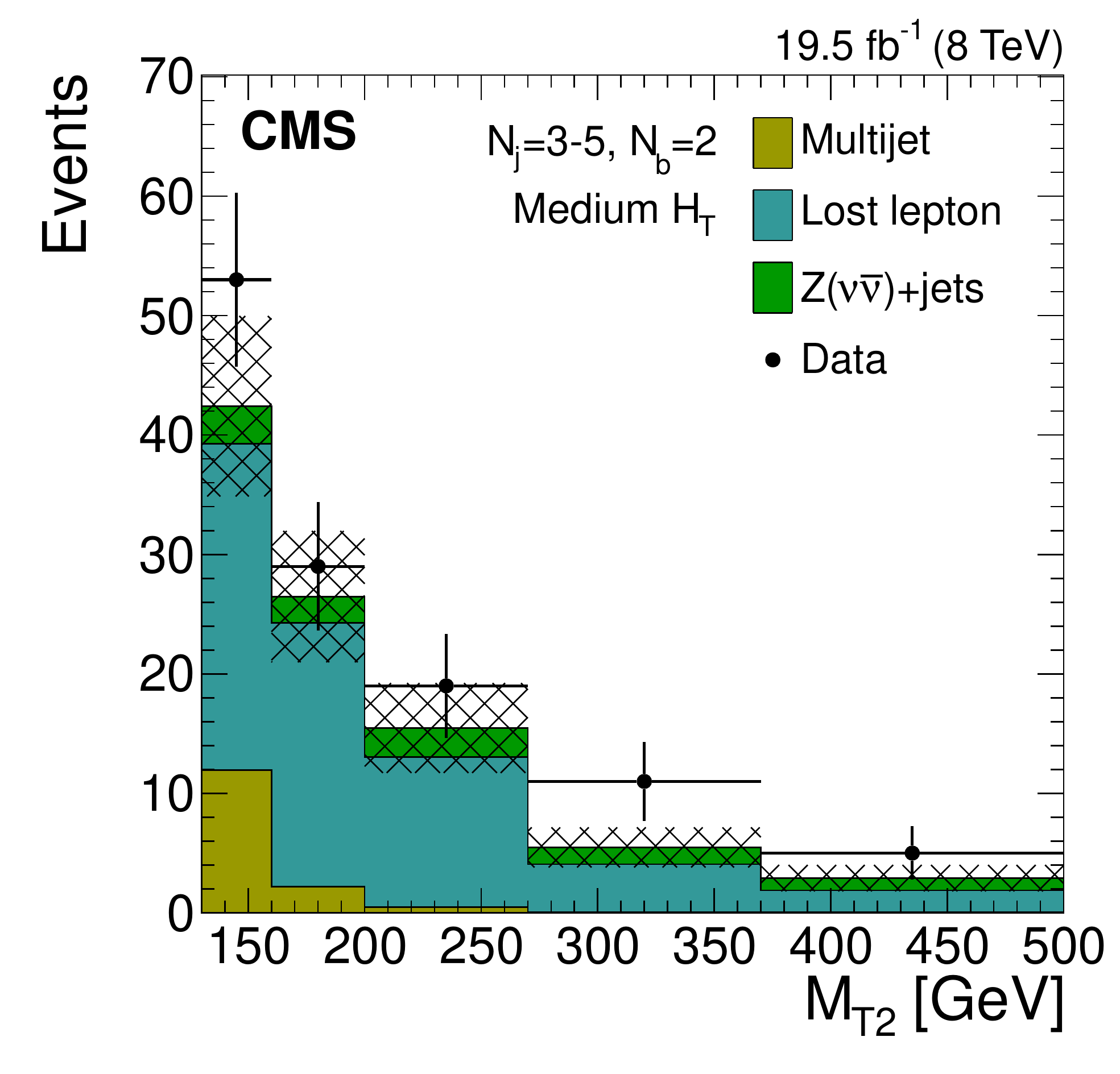}
\includegraphics[width=0.32\textwidth]{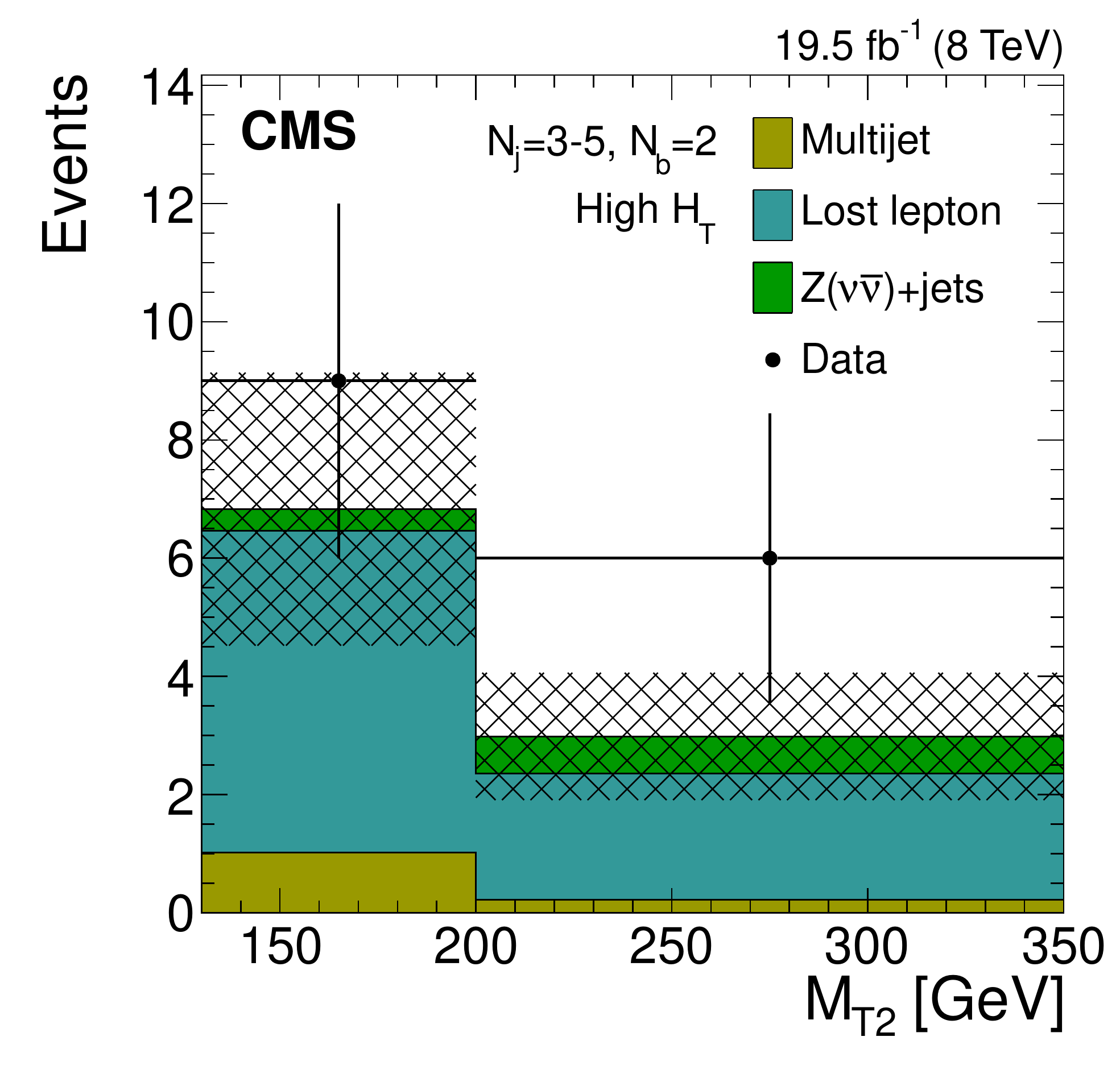}
\includegraphics[width=0.32\textwidth]{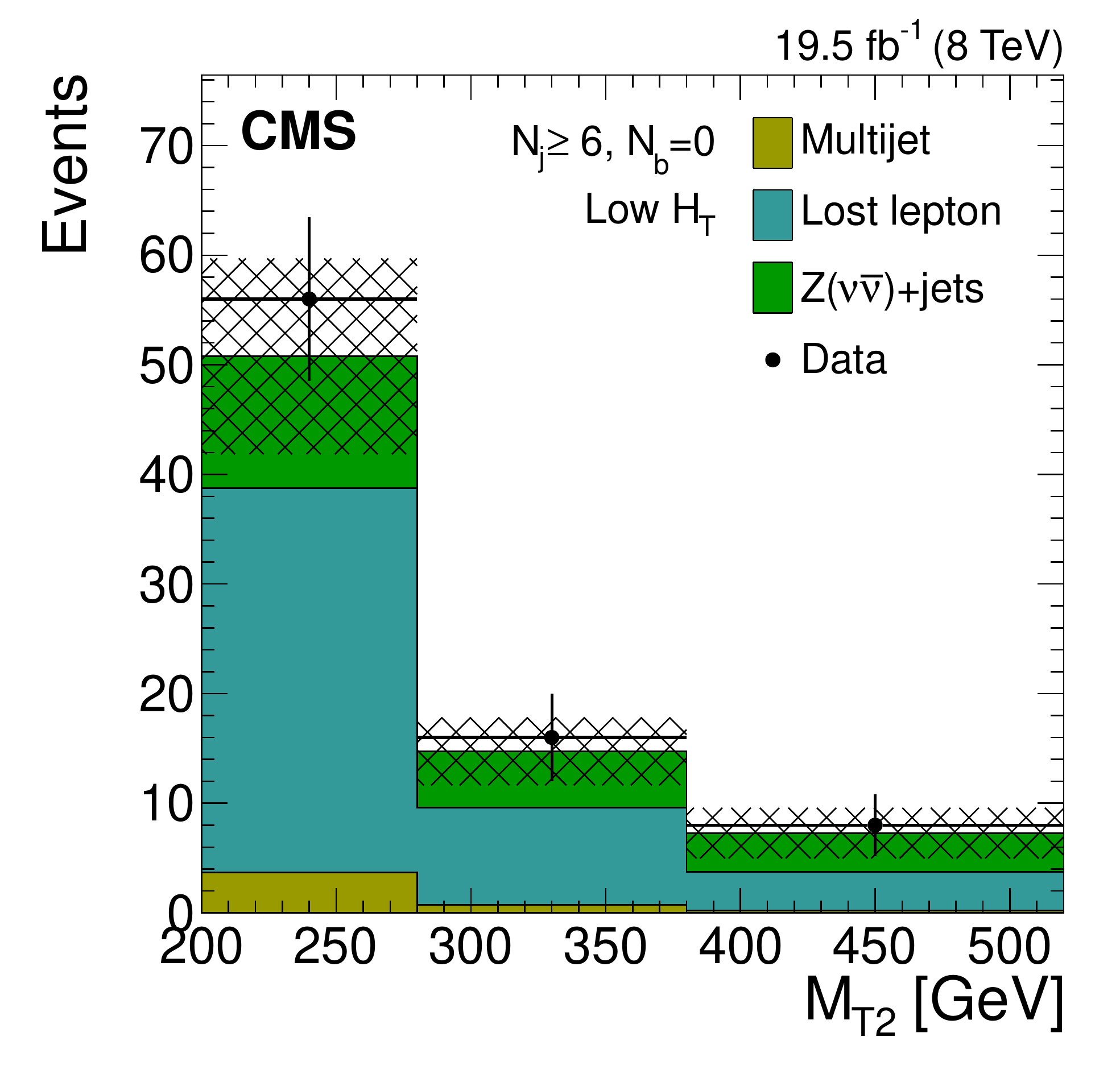}
\includegraphics[width=0.32\textwidth]{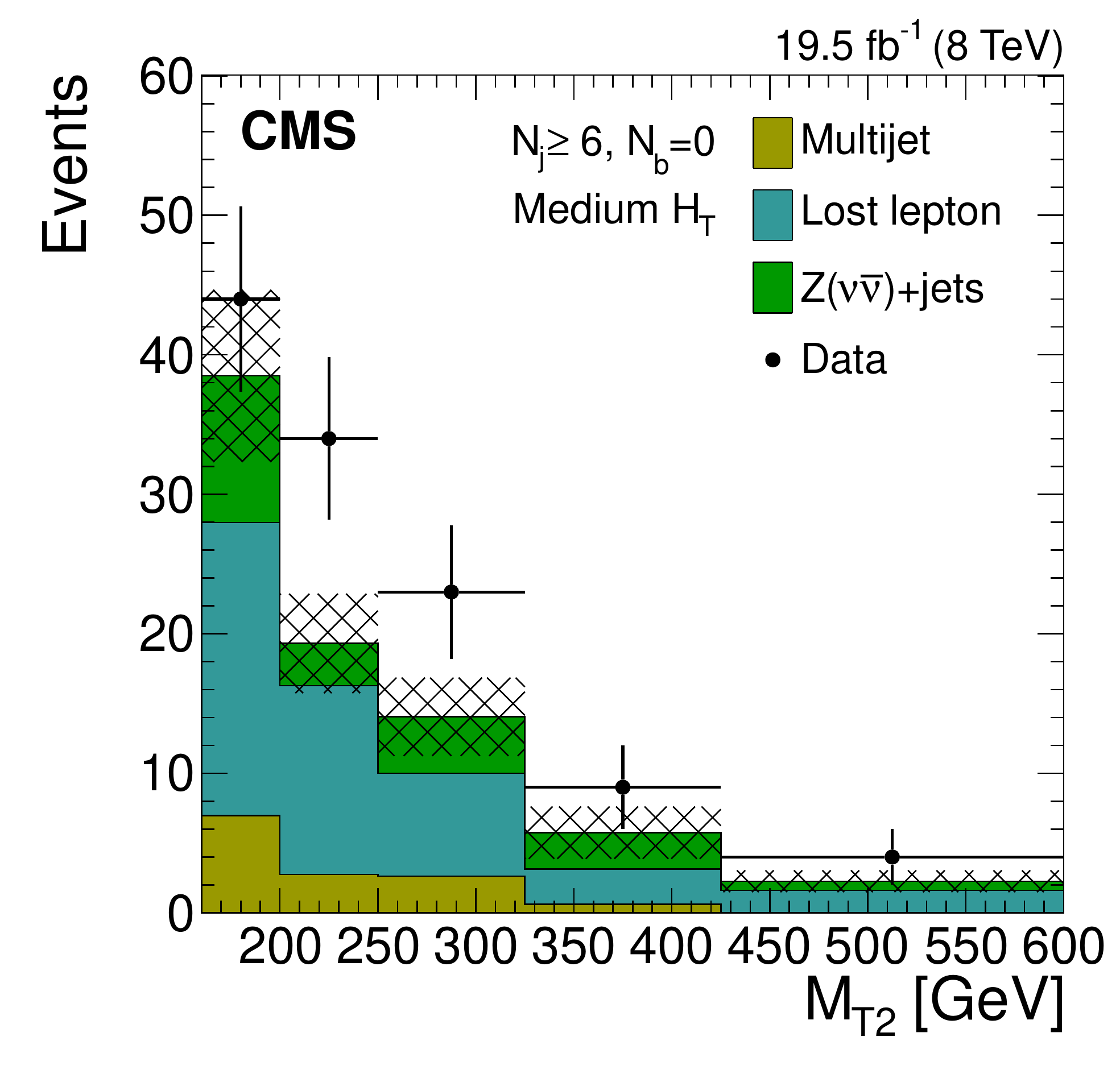}
\includegraphics[width=0.32\textwidth]{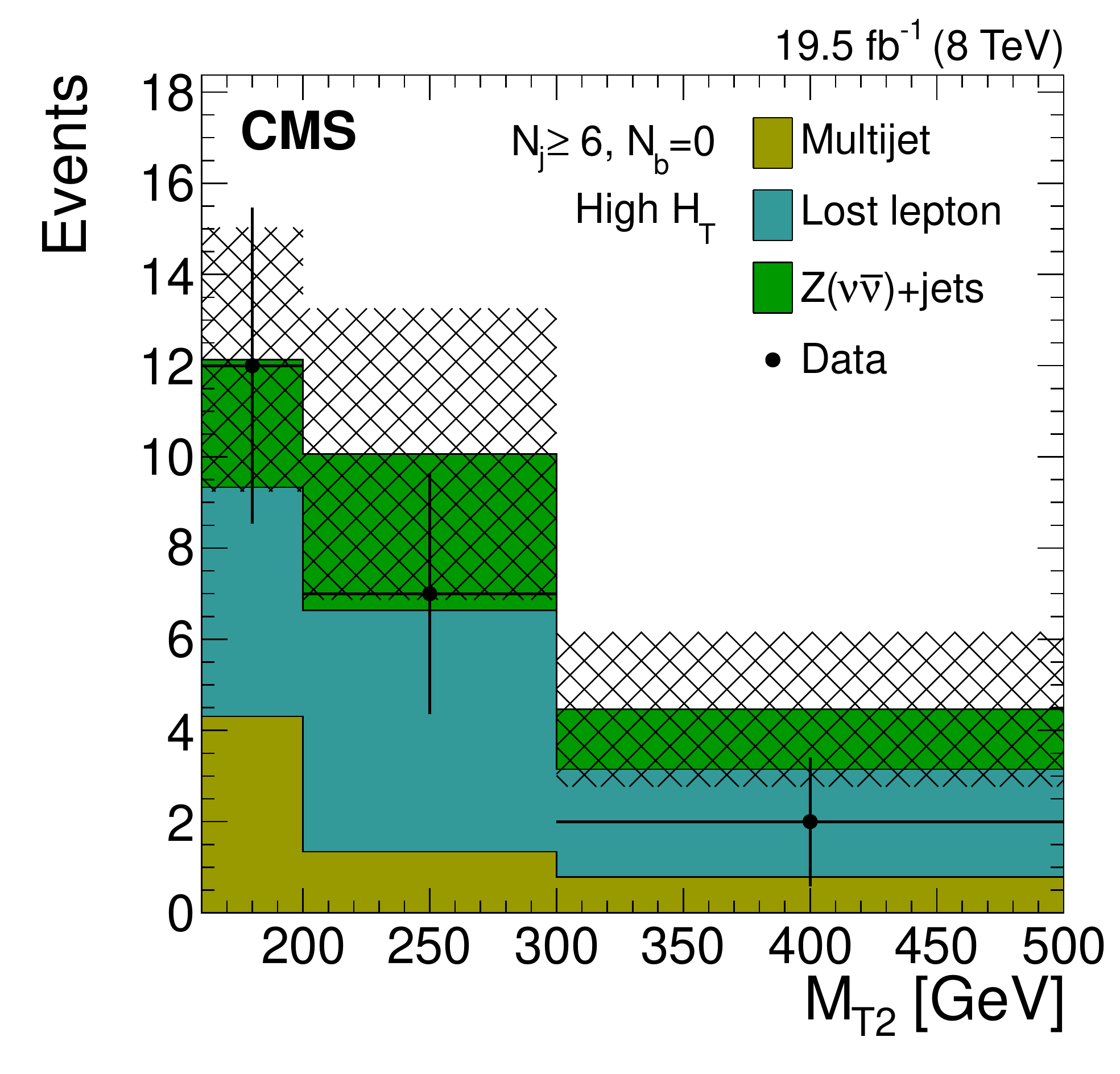}
\caption{
  Distributions of the \MTt variable for the estimated background processes and for data.
  Plots are shown for events satisfying the low-\HT (left), the medium-\HT (middle), and the
  high-\HT (right) selections, and for different topological signal regions (\NJ, \NB) of the
  inclusive-\MTt event selection.
  From top to bottom, these are $(3\leq \NJ \leq5,\NB=1)$, $(3\leq \NJ \leq5,\NB=2)$, $(\NJ\geq6,\NB=0)$.
  The uncertainties in each plot are drawn as the shaded band and do not include the uncertainty in the shape of the lost-lepton background.
}
  \label{fig:datadrivenMT2_2}
\end{figure}

\begin{figure}[!hp]
  \centering
\includegraphics[width=0.32\textwidth]{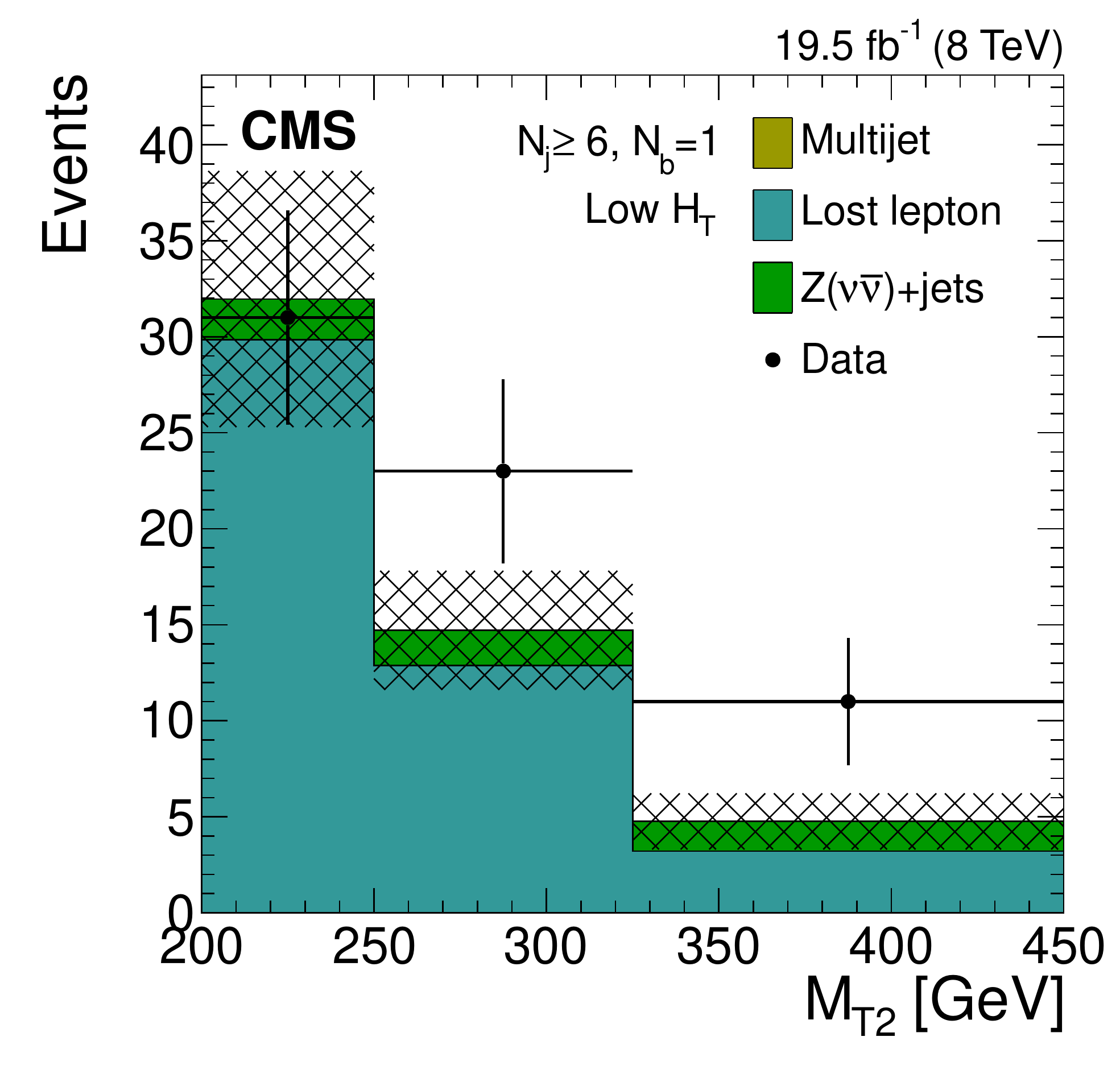}
\includegraphics[width=0.32\textwidth]{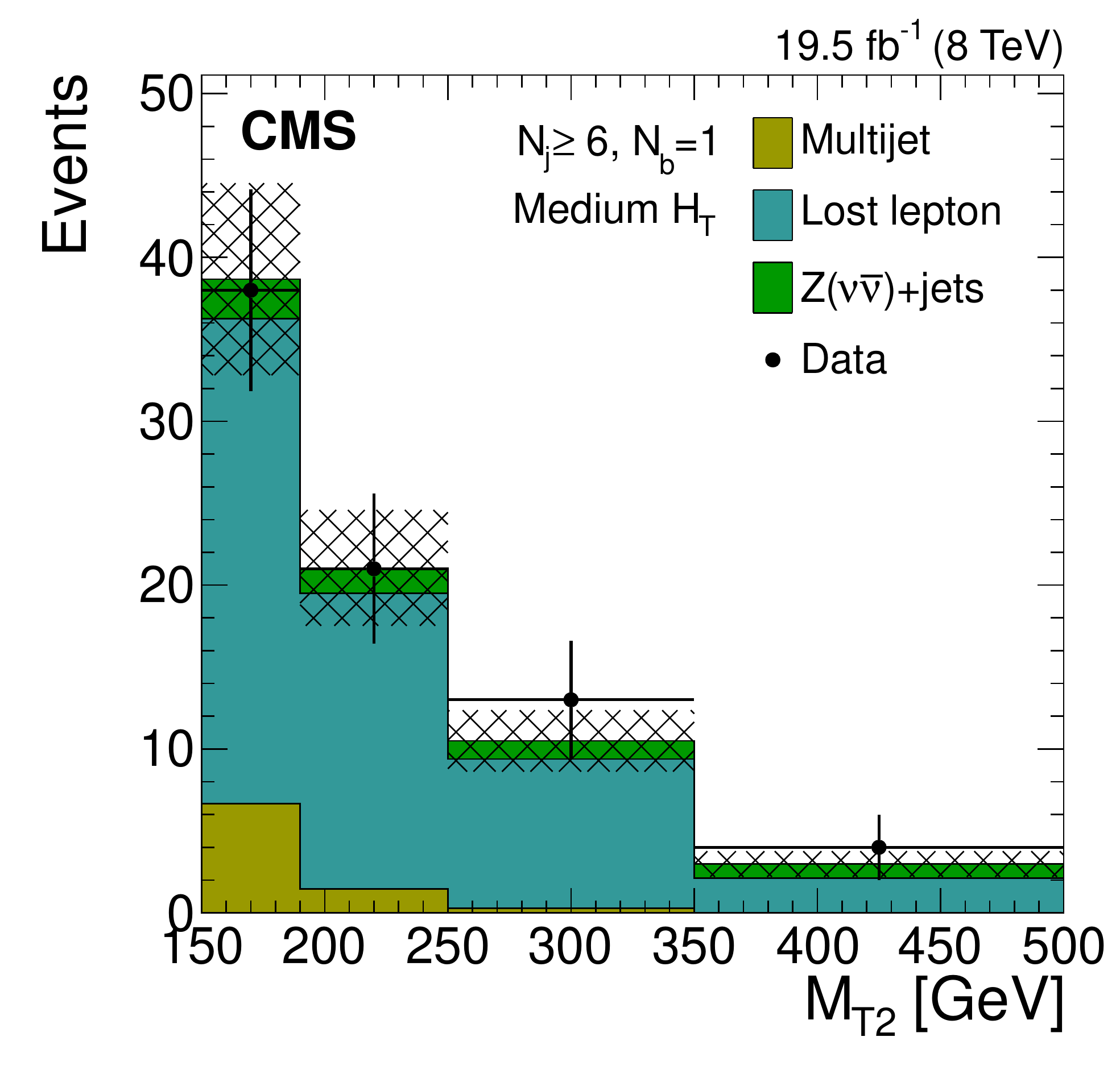}
\includegraphics[width=0.32\textwidth]{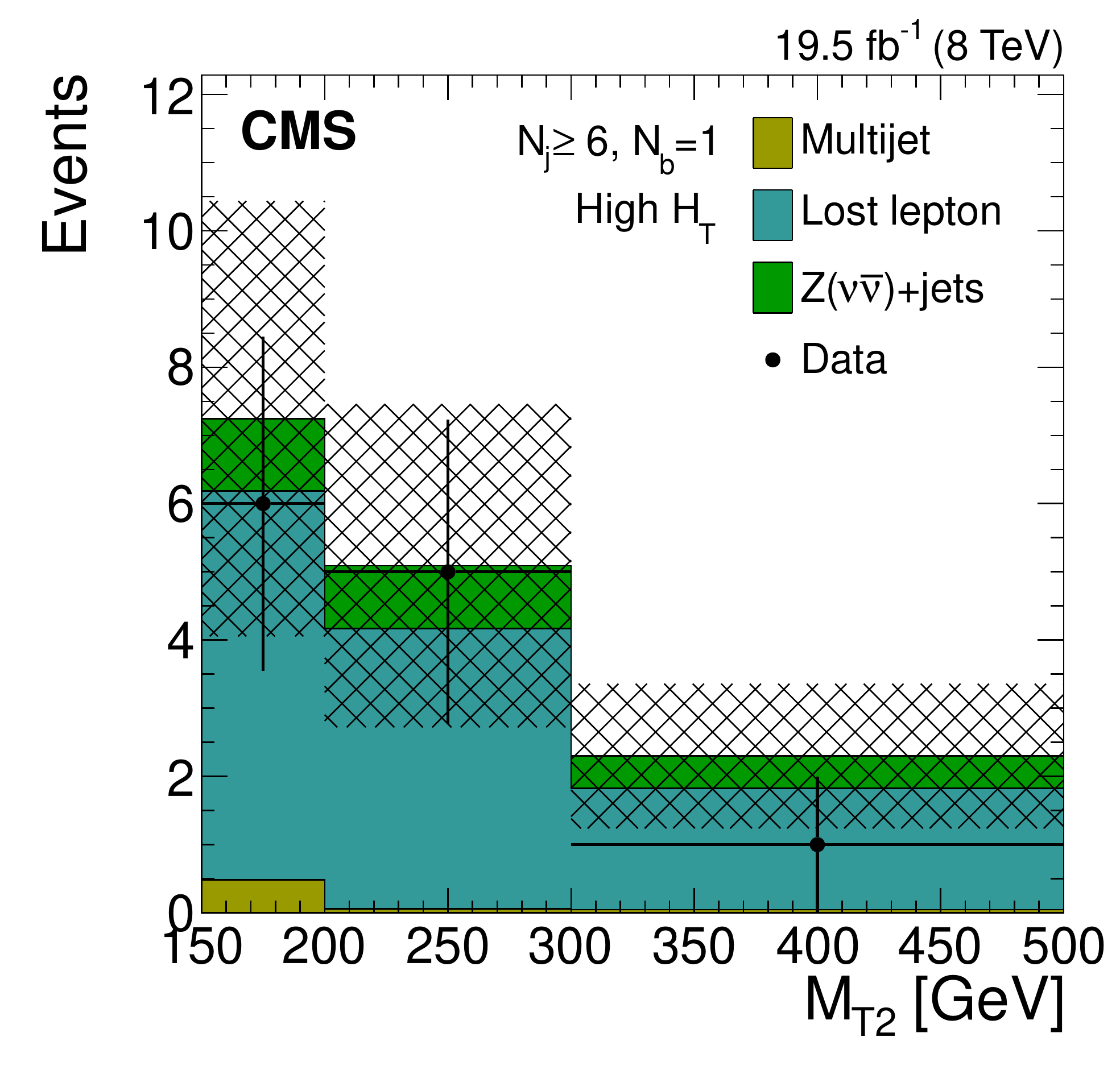}
\includegraphics[width=0.32\textwidth]{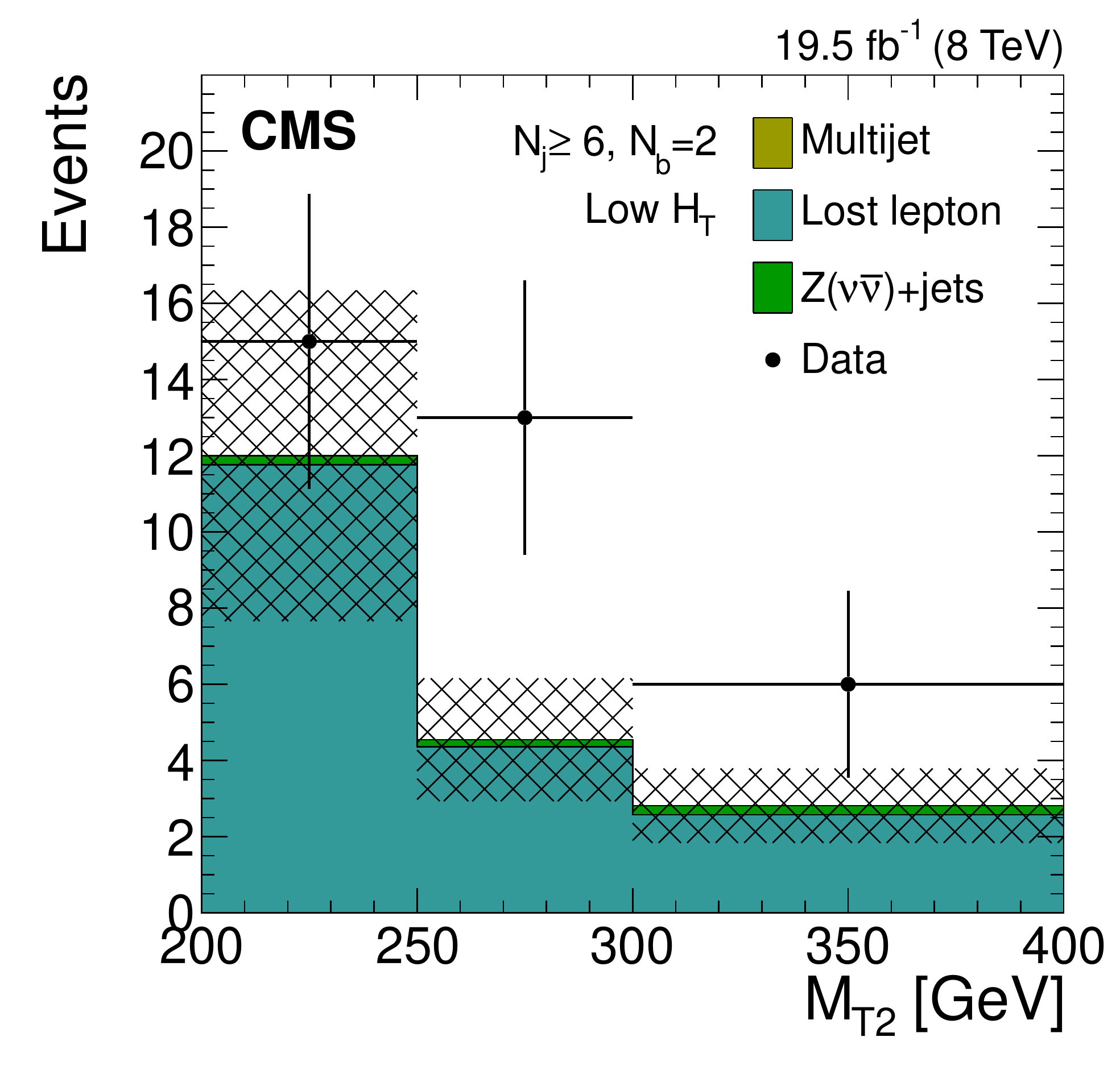}
\includegraphics[width=0.32\textwidth]{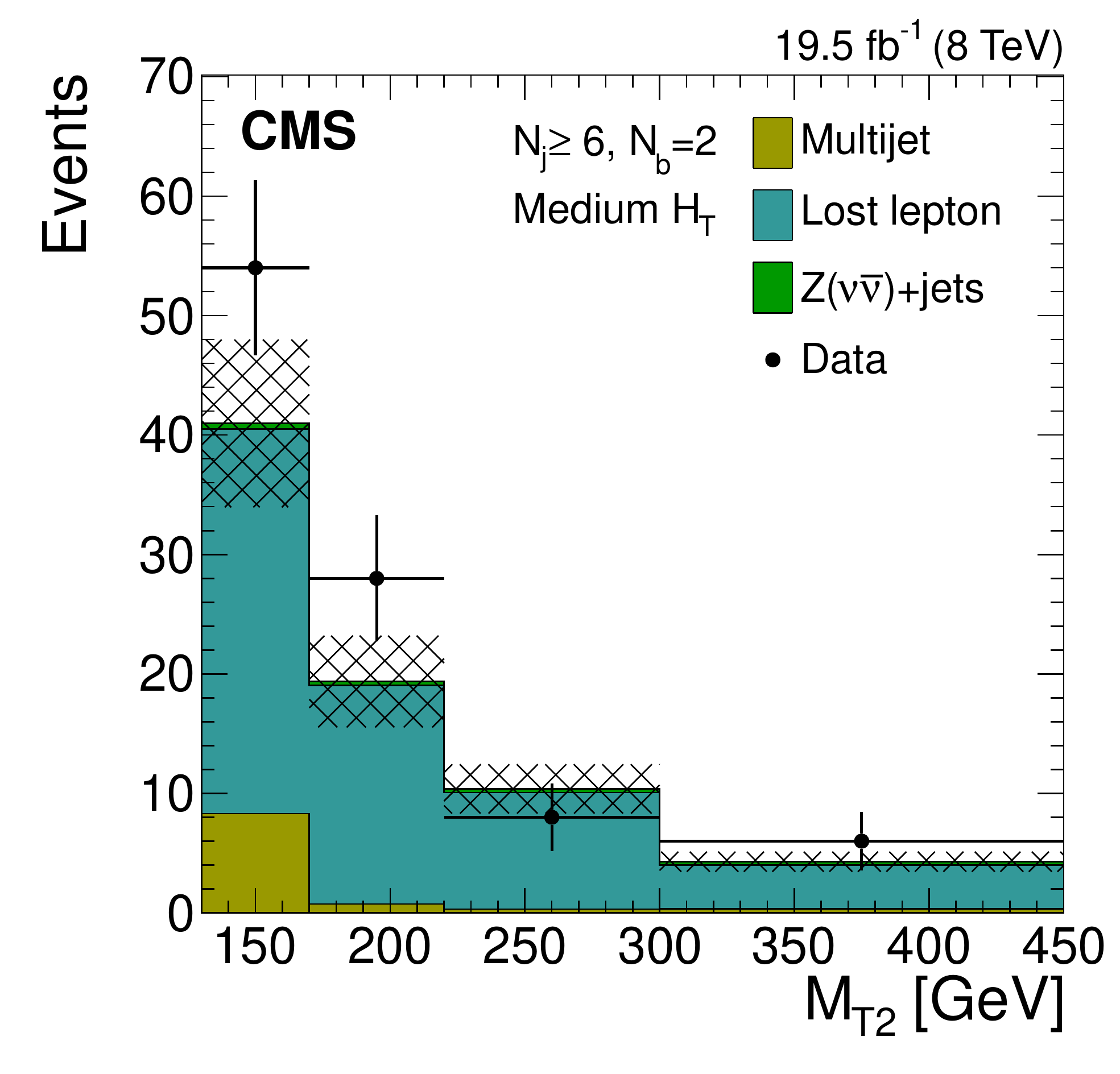}
\includegraphics[width=0.32\textwidth]{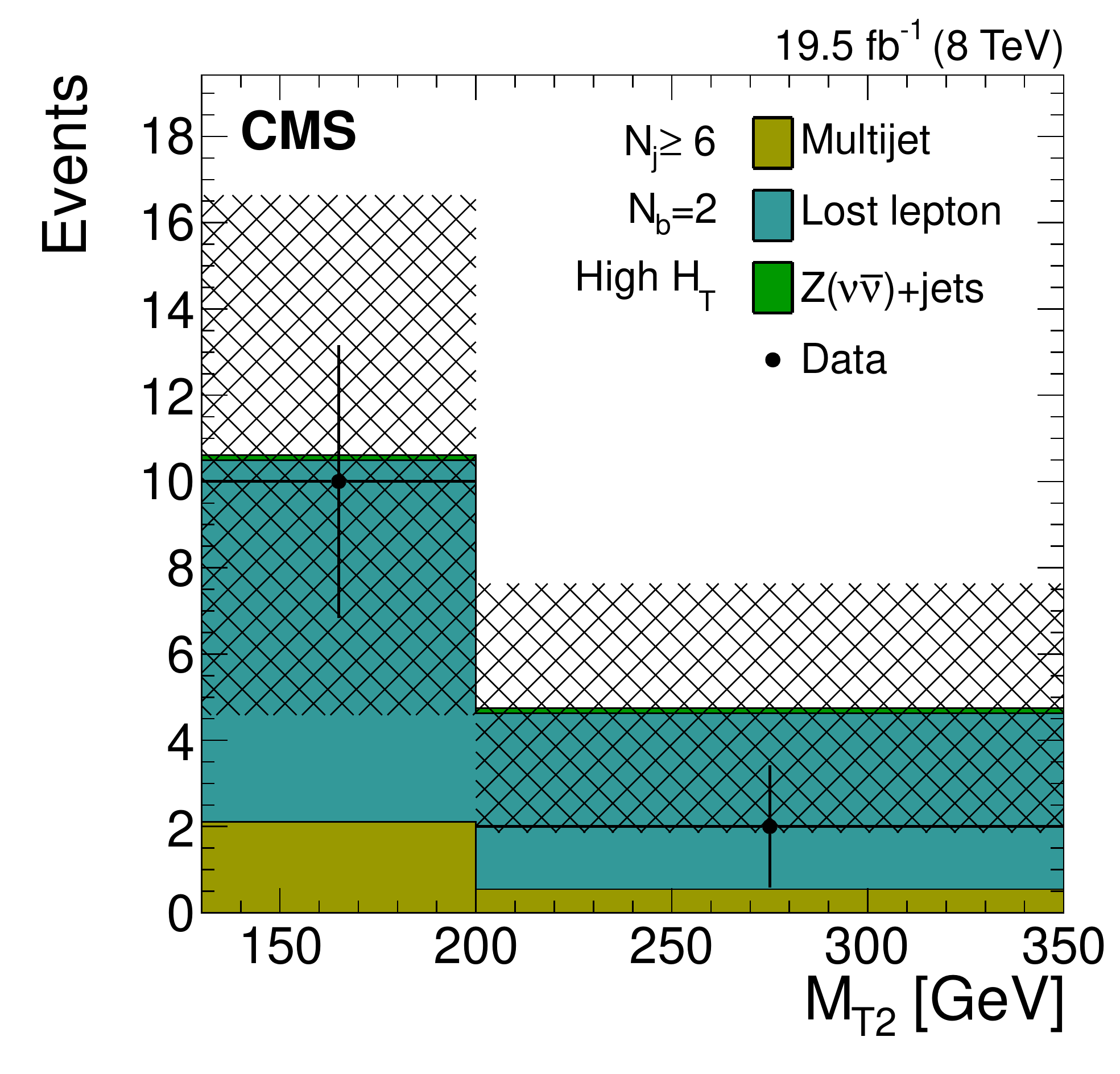}
\includegraphics[width=0.32\textwidth]{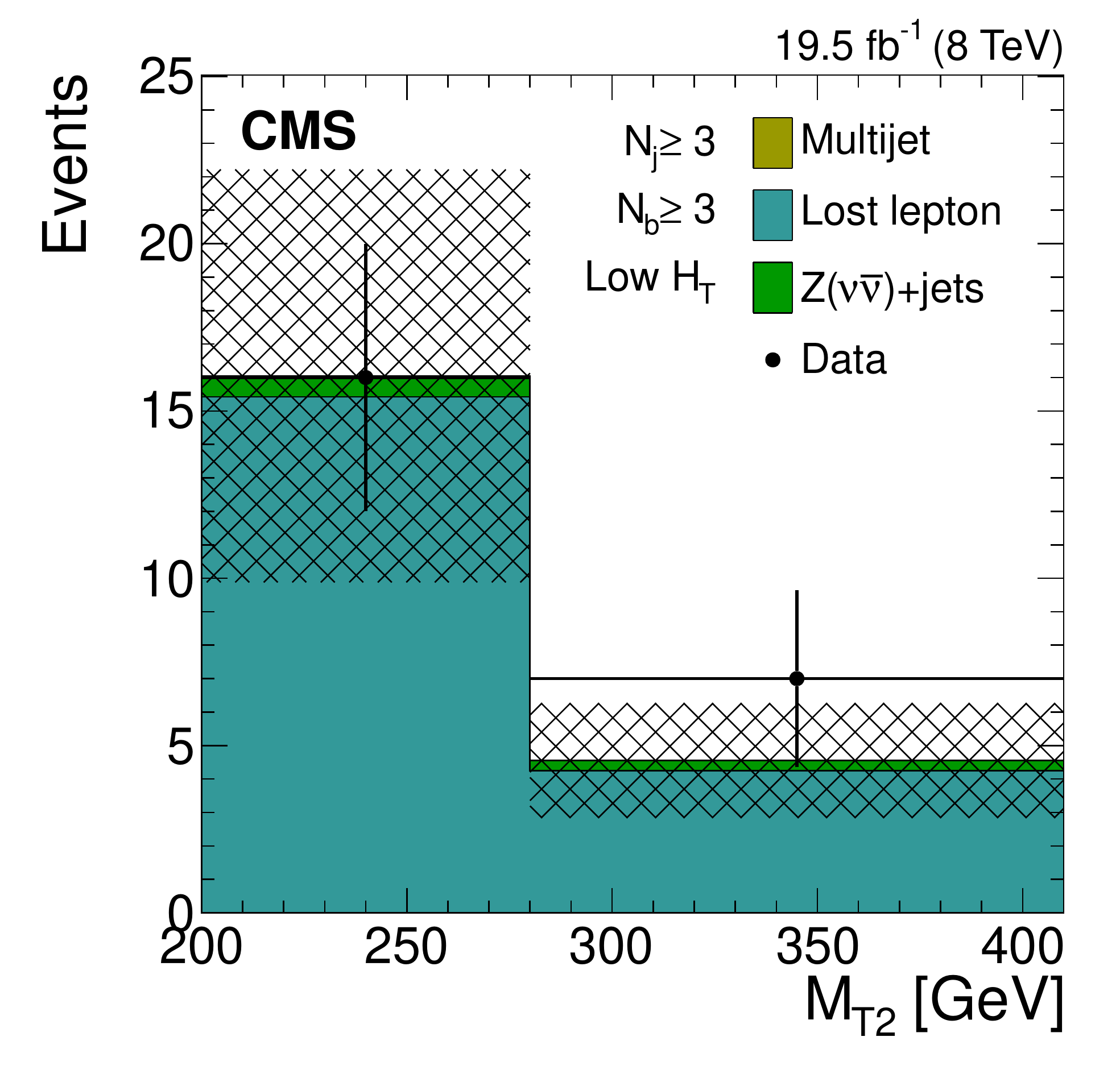}
\includegraphics[width=0.32\textwidth]{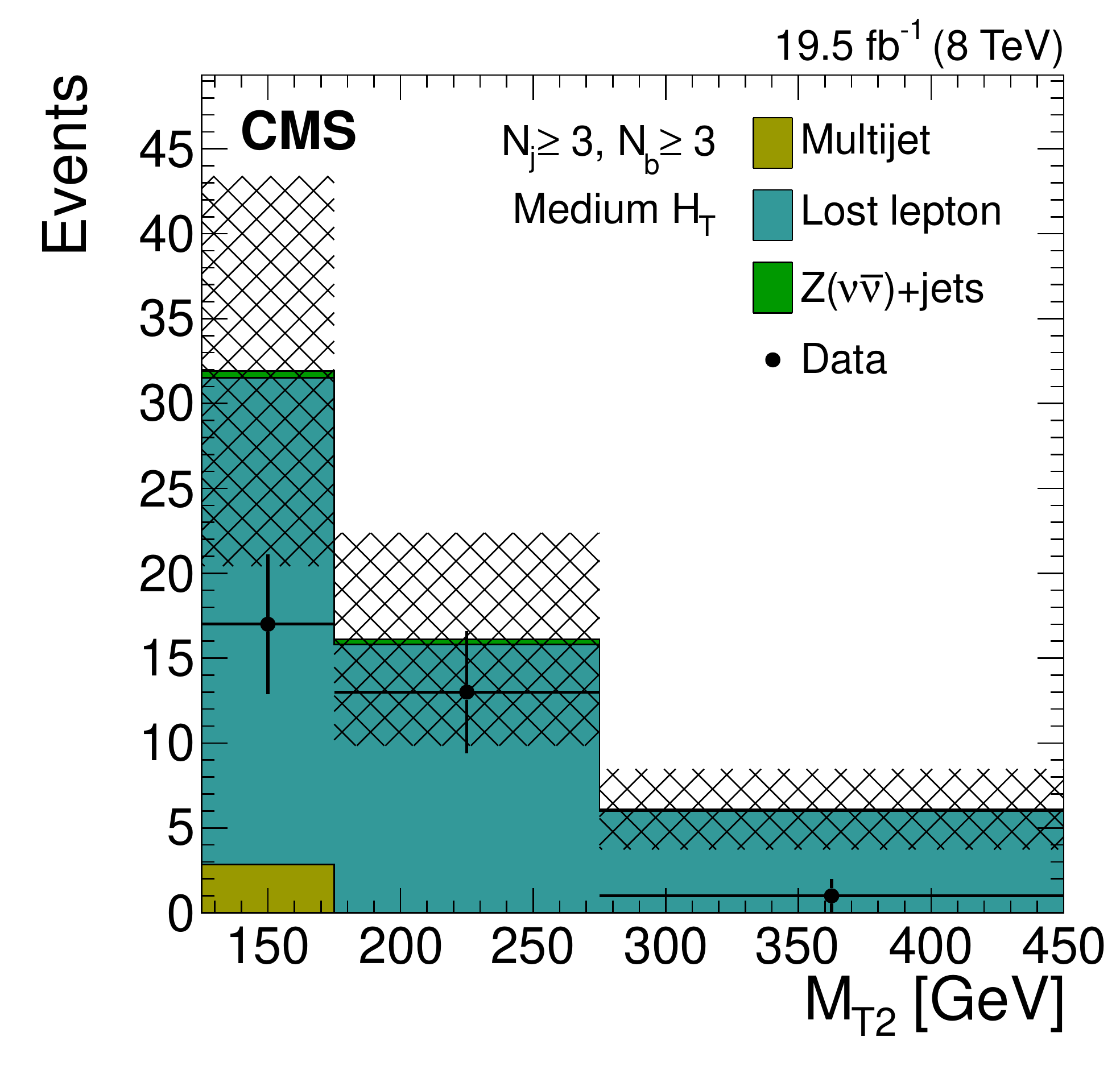}
\includegraphics[width=0.32\textwidth]{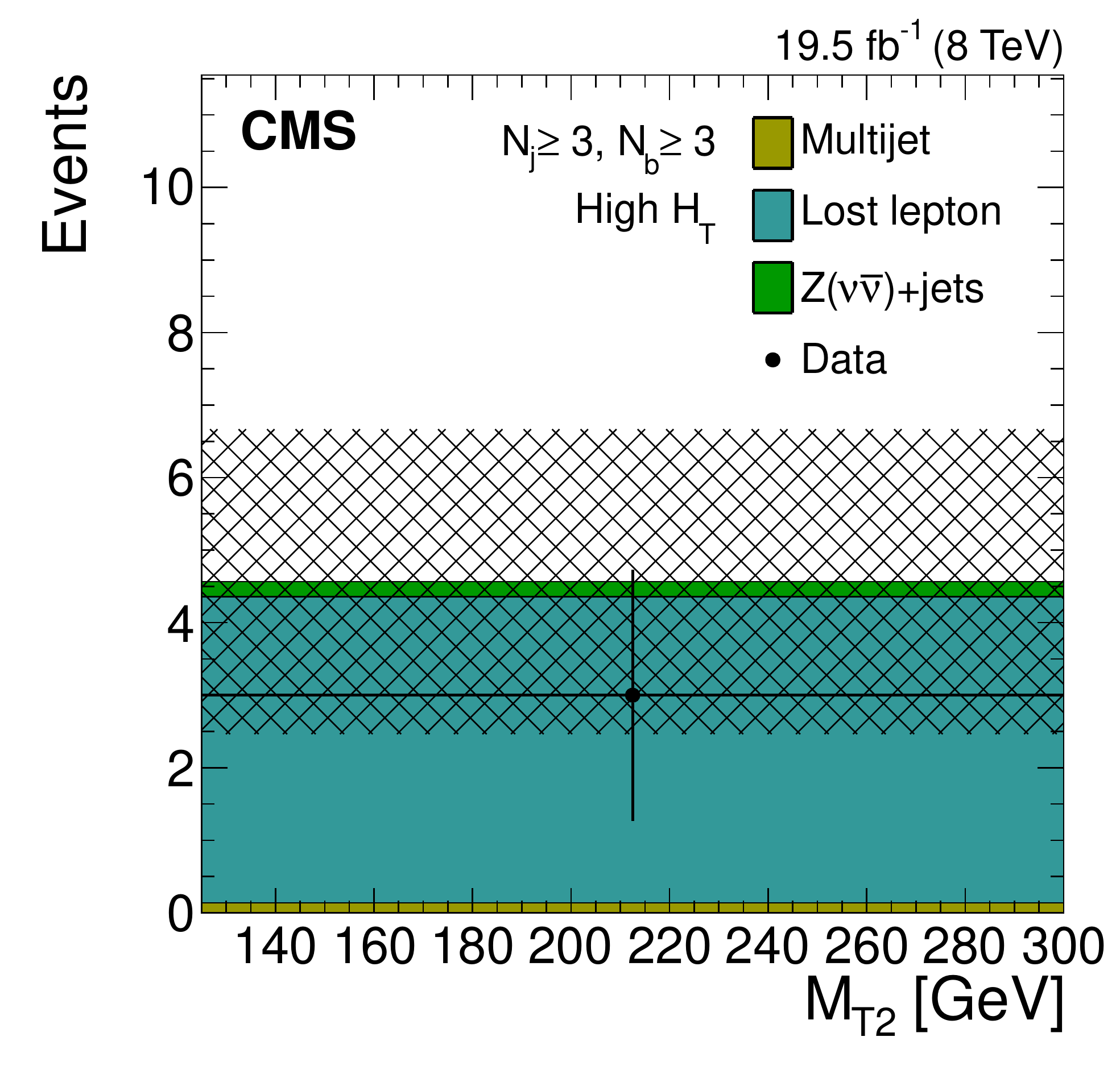}
\caption{
  Distributions of the \MTt variable for the estimated background processes and for data.
  Plots are shown for events satisfying the low-\HT (left), the medium-\HT (middle), and the
  high-\HT (right) selections, and for different topological signal regions (\NJ, \NB) of the
  inclusive-\MTt event selection.
  From top to bottom, these are $(\NJ\geq6,\NB=1)$, $(\NJ\geq6,\NB=2)$, $(\NJ\geq3,\NB\geq3)$.
  The uncertainties in each plot are drawn as the shaded band and do not include the uncertainty in the shape of the lost-lepton background.
}
  \label{fig:datadrivenMT2_3}
\end{figure}

For the inclusive-\MTt search,
the final event yields in all
signal regions
are shown in \mbox{Figs.~\ref{fig:datadrivenMT2_1}--\ref{fig:datadrivenMT2_3}}.
The comparison between observed and predicted yields is shown separately for different topological
regions and for the different \HT selections.
The total uncertainty of the background estimates is the quadratic sum of the statistical and systematic uncertainties
from the three categories of background.
The results are tabulated in Table~\ref{table:datadrivenMT2_short}.
The shape uncertainty in the estimation of the lost-lepton background is not included either for the figures or table.

\begin{table}
\centering
\topcaption{Event yields, for estimated background and data, in the signal regions of the inclusive-\MTt search.
  The uncertainties are the quadratic sum of statistical and systematic uncertainties.}
\label{table:datadrivenMT2_short}
\setlength{\tabcolsep}{3pt}
\small
\begin{tabular}{l|rr@{$\pm$}lr|rr@{$\pm$}lr|rr@{$\pm$}lr}
\hline
Signal
 & \multicolumn{4}{c}{Low-\HT region} & \multicolumn{4}{|c|}{Medium-\HT region} & \multicolumn{4}{c}{High-\HT region} \\
\rule[0pt]{0pt}{2.2ex}region & \MTt [\GeVns{}] & \multicolumn{2}{c}{Prediction} & Data & \MTt [\GeVns{}] & \multicolumn{2}{c}{Prediction} & Data & \MTt [\GeVns{}] & \multicolumn{2}{c}{Prediction} & Data \\
\hline
\multirow{9}{*}{\begin{minipage}{4.25em}$\NJ=2$,\\ $\NB=0$\end{minipage}}
 & 200--240 & 553&70 & 588  & 125--150 & 167&21 & 171  & 120--150 & $\:$21.9&4.9 & 18 \\
 & 240--290 & 395&53 & 451  & 150--180 & 128&17 & 104  & 150--200 & 19.4&4.3 & 18 \\
 & 290--350 & 288&40 & 318  & 180--220 & 85.8&11.3 & 91  & 200--260 & 14.5&3.4 & 10 \\
 & 350--420 & 236&52 & 232  & 220--270 & $\,$70.0&10.3 & 78  & 260--350 & 6.3&1.8 & 9 \\
 & 420--490 & 165&36 & 162  & 270--325 & 38.1&5.8 & 48  & 350--550 & 4.3&1.6 & 8 \\
 & 490--570 & $\,$68.9&15.5 & 61  & 325--425 & 43.4&10.1 & 45  & $>$550   & 3.0&1.4 & 6 \\
 & 570--650 & 17.3&4.3 & 19  & 425--580 & 21.3&4.7 & 29  &                    & \multicolumn{2}{c}{$ $} &         \\
 & $>$650   & 4.1&1.6 & 1  & 580--780 & 20.8&5.6 & 10  &                    & \multicolumn{2}{c}{$ $}  &         \\
 &                    & \multicolumn{2}{c}{$ $} &  & $>$780   & 3.5&1.4 & 2  &           & \multicolumn{2}{c}{$ $} &         \\
\hline
\multirow{6}{*}{\begin{minipage}{4.25em}$\NJ=2$,\\ $\NB\geq$1\end{minipage}}
 & 200--250 & 56.4&12.8 & 56  & 100--135 & 27.4&9.6 & 30  & 100--180 & 11.4&8.1 & 2 \\
 & 250--310 & 34.2&8.1 & 44  & 135--170 & 21.1&7.5 & 19  & $>$180   & 4.4&2.6 & 2 \\
 & 310--380 & 25.9&7.4 & 29  & 170--260 & 13.4&5.4 & 15  &                    & \multicolumn{2}{c}{$ $} &         \\
 & 380--450 & 19.9&5.8 & 13  & 260--450 & 7.3&3.5 & 7  &                    &  \multicolumn{2}{c}{$ $} &         \\
 & 450--550 & 12.6&3.8 & 15  & $>$450   & 3.4&1.7 & 9  &                    &  \multicolumn{2}{c}{$ $} &         \\
 & $>$550   & 2.6&0.8 & 3  &                    & \multicolumn{2}{c}{$ $} &   &    &  \multicolumn{2}{c}{$ $}  &         \\
\hline
\multirow{9}{*}{\begin{minipage}{4.25em}$\NJ=3$--5,\\ $\NB=0$\end{minipage}}
 & 200--240 & 979&108 & 1041  & 160--185 & 243&23 & 234  & 160--185 & 34.9&4.7 & 39 \\
 & 240--290 & 711&86 & 827  & 185--215 & 180&19 & 203  & 185--220 & 31.1&4.7 & 32 \\
 & 290--350 & 492&65 & 522  & 215--250 & 134&16 & 152  & 220--270 & 25.5&4.3 & 25 \\
 & 350--420 & 280&57 & 333  & 250--300 & 112&14 & 119  & 270--350 & 19.3&3.5 & 19 \\
 & 420--490 & 138&29 & 145  & 300--370 & 89.0&12.2 & 91  & 350--450 & 9.1&2.5 & 6 \\
 & 490--570 & 60.0&13.6 & 66  & 370--480 & 67.0&14.2 & 75  & 450--650 & 5.0&1.6 & 5 \\
 & 570--650 & 13.8&3.9 & 21  & 480--640 & 35.0&8.0 & 40  & $>$650   & 4.4&1.6 & 5 \\
 & $>$650   & 3.6&1.5 & 2  & 640--800 & 10.0&2.7 & 16  &                    & \multicolumn{2}{c}{$ $}  &         \\
 &                    & \multicolumn{2}{c}{$ $} &    & $>$800   & 3.4&1.5 & 4  &       &  \multicolumn{2}{c}{$ $}  &         \\
\hline
\multirow{6}{*}{\begin{minipage}{4.25em}$\NJ=3$--5,\\ $\NB=1$\end{minipage}}
 & 200--250 & 305&34 & 300  & 150--175 & 93.4&10.7 & 87  & 150--180 & 13.5&3.1 & 28 \\
 & 250--310 & 167&21 & 172  & 175--210 & 69.5&8.7 & 71  & 180--230 & 8.7&2.2 & 7 \\
 & 310--380 & 103&16 & 98  & 210--270 & 52.8&6.8 & 63  & 230--350 & 6.2&1.6 & 9 \\
 & 380--460 & 43.6&8.7 & 47  & 270--380 & 38.6&5.1 & 47  & $>$350   & 3.5&1.0 & 3 \\
 & 460--550 & 17.9&4.1 & 19  & 380--600 & 15.9&3.2 & 19  &               & \multicolumn{2}{c}{$ $}  &         \\
 & $>$550   & 4.0&1.1 & 4  & $>$600   & 3.6&0.9 & 4  &                    &  \multicolumn{2}{c}{$ $}  &         \\
\hline
\multirow{5}{*}{\begin{minipage}{4.25em}$\NJ=3$--5,\\ $\NB=2$\end{minipage}}
 & 200--250 & 91.1&22.0 & 97  & 130--160 & 42.4&7.5 & 53  & 130--200 & 6.8&2.3 & 9 \\
 & 250--325 & 52.7&13.7 & 39  & 160--200 & 26.5&5.5 & 29  & $>$200   & 2.9&1.1 & 6 \\
 & 325--425 & 18.6&5.8 & 16  & 200--270 & 15.4&3.7 & 19  &                  & \multicolumn{2}{c}{$ $} &         \\
 & $>$425   & 4.5&1.9 & 11  & 270--370 & 5.5&1.7 & 11  &                    & \multicolumn{2}{c}{$ $} &         \\
 &                    & \multicolumn{2}{c}{$ $} &  & $>$370   & 2.9&1.1 & 5  &         & \multicolumn{2}{c}{$ $}  &         \\
\hline
\multirow{5}{*}{\begin{minipage}{4.25em}$\NJ\geq 6$,\\ $\NB=0$\end{minipage}}
 & 200--280 & 50.8&8.9 & 56  & 160--200 & 38.5&6.2 & 44  & 160--200 & 12.1&2.9 & 12 \\
 & 280--380 & 14.7&3.1 & 16  & 200--250 & 19.3&3.6 & 34  & 200--300 & 10.1&3.2 & 7 \\
 & $>$380   & 7.3&2.3 & 8  & 250--325 & 14.1&2.8 & 23  & $>$300   & 4.5&1.7 & 2 \\
 &                    & \multicolumn{2}{c}{$ $} &  & 325--425 & 5.8&1.9 & 9  &           & \multicolumn{2}{c}{$ $} &         \\
 &                    & \multicolumn{2}{c}{$ $} &  & $>$425   & 2.3&0.8 & 4  &          & \multicolumn{2}{c}{$ $}  &         \\
\hline
\multirow{4}{*}{\begin{minipage}{4.25em}$\NJ\geq 6$,\\ $\NB=1$\end{minipage}}
 & 200--250 & 32.0&6.7 & 31  & 150--190 & 38.7&5.9 & 38  & 150--200 & 7.3&3.2 & 6 \\
 & 250--325 & 14.7&3.1 & 23  & 190--250 & 21.1&3.5 & 21  & 200--300 & 5.1&2.4 & 5 \\
 & $>$325   & 4.8&1.5 & 11  & 250--350 & 10.5&1.9 & 13  & $>$300   & 2.3&1.1 & 1 \\
 &                    & \multicolumn{2}{c}{$ $} & & $>$350   & 3.0&0.8 & 4  &           & \multicolumn{2}{c}{$ $}  &         \\
\hline
\multirow{4}{*}{\begin{minipage}{4.25em}$\NJ\geq 6$,\\ $\NB=2$\end{minipage}}
 & 200--250 & 12.0&4.3 & 15  & 130--170 & 41.0&7.0 & 54  & 130--200 & 10.6&6.0 & 10 \\
 & 250--300 & 4.6&1.6 & 13  & 170--220 & 19.4&3.8 & 28  & $>$200   & 4.7&2.9 & 2 \\
 & $>$300   & 2.8&1.0 & 6  & 220--300 & 10.4&2.1 & 8  &                    & \multicolumn{2}{c}{$ $} &         \\
 &              & \multicolumn{2}{c}{$ $} &  & $>$300   & 4.3&0.8 & 6  &             & \multicolumn{2}{c}{$ $}  &         \\
\hline
\multirow{3}{*}{\begin{minipage}{4.25em}$\NJ\geq 3$,\\ $\NB\geq 3$\end{minipage}}
 & 200--280 & 16.1&6.2 & 16  & 125--175 & 31.9&11.4 & 17  & $>$125   & 4.5&2.1 & 3 \\
 & $>$280   & 4.6&1.7 & 7  & 175--275 & 16.1&6.3 & 13  &                & \multicolumn{2}{c}{$ $}  &         \\
 &            & \multicolumn{2}{c}{$ $} &  & $>$275   & 6.1&2.4 & 1  &              & \multicolumn{2}{c}{$ $} &         \\
\hline
\end{tabular}
\end{table}

The level of compatibility between the data and the SM predictions is assessed by computing the pull value
for all signal regions, where the pull value is defined for each signal region bin as:
\begin{equation}
\label{eq:pull}
\mathrm{Pull} = \frac{N_\text{obs} - N_\text{bkg}}{\sqrt{\smash[b]{\sigma_\text{obs}^2 + \sigma_\text{bkg}^2}}},
\end{equation}
where $N_\text{obs}$ is the observed number of events, $\sigma_\text{obs}$ is its statistical uncertainty,
and $N_\text{bkg}$ is the background estimate with a total uncertainty of $\sigma_\text{bkg}$.
After the average pull over all the signal regions is calculated, pseudo-experiments are used to evaluate the probability to
observe an average at least as large as the average observed in data.
The probability is found to be 11\%.
Thus, the data are found to be in agreement with the SM predictions within the uncertainties.

\begin{figure}[!ht]
  \centering
\includegraphics[width=0.99\textwidth]{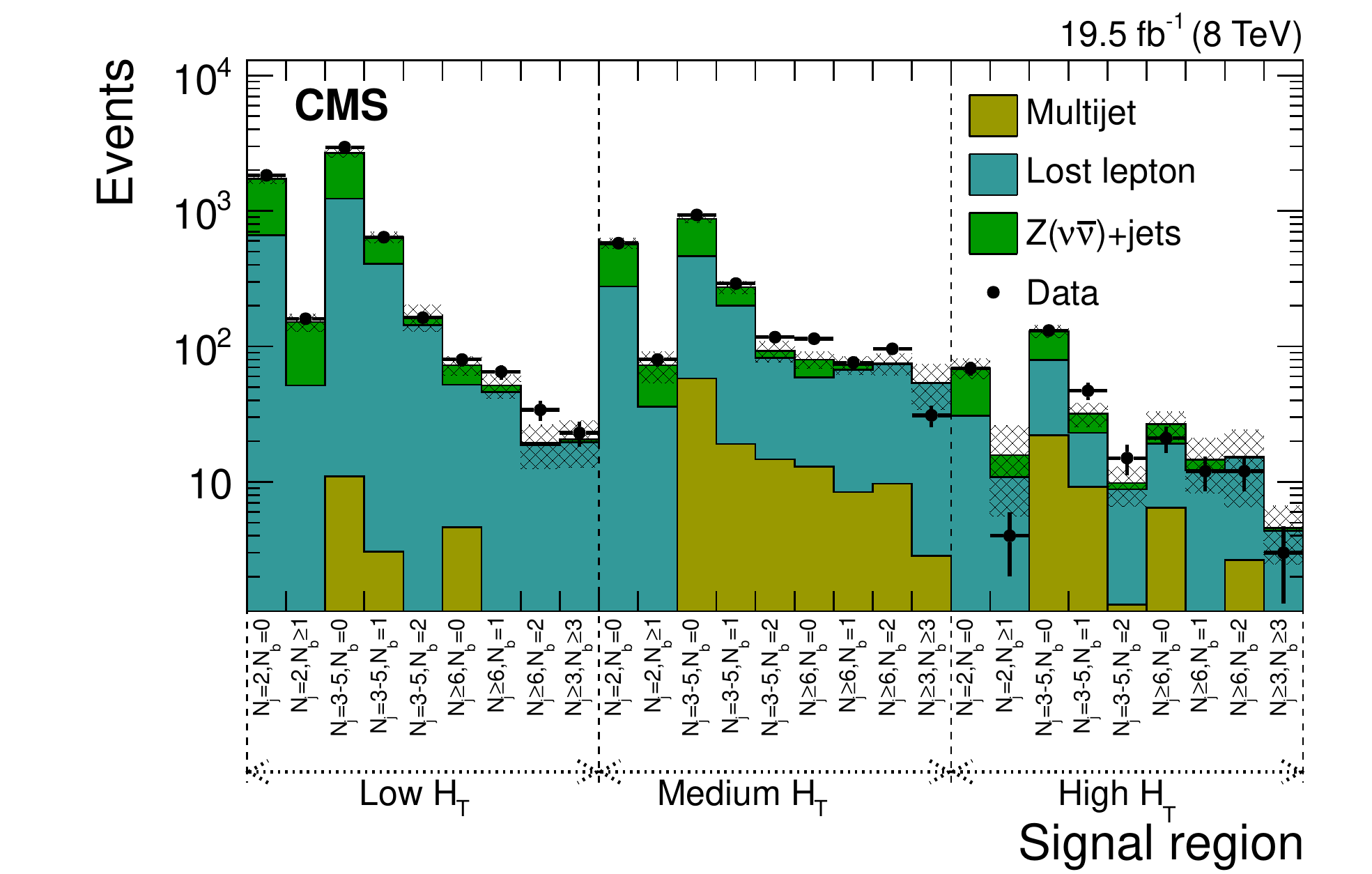}
  \caption{
    Event yields, for both estimated backgrounds and data, for the three \HT selections and all the topological
    signal regions of the inclusive-\MTt search.
  The uncertainties are drawn as the shaded band and do not include the uncertainty in the shape of the lost-lepton background.
}
  \label{fig:datadrivenMT2_summary}
\end{figure}

In order to present the results in a compact manner, the yields of all \MTt bins that belong to the same
topological region and that satisfy the same \HT selection are summed.
The resulting sums are presented in Fig.~\ref{fig:datadrivenMT2_summary}.

\subsection{Results for the \texorpdfstring{\MTt Higgs analysis}{MT2 Higgs analysis}}
\label{sec:data:higgs:results}

\begin{figure}[!htb]
\centering
\includegraphics[width=0.495\textwidth]{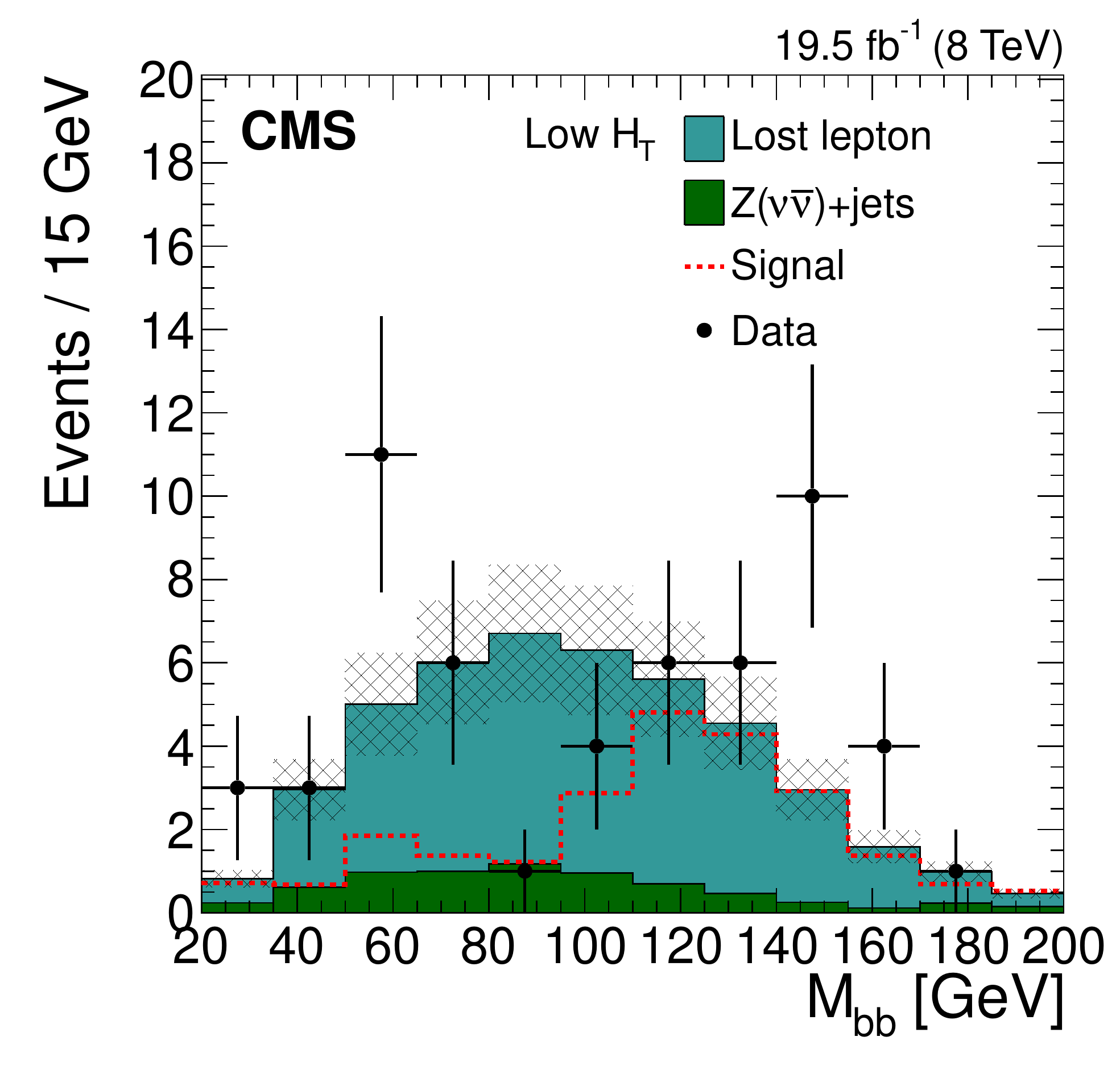}
\includegraphics[width=0.495\textwidth]{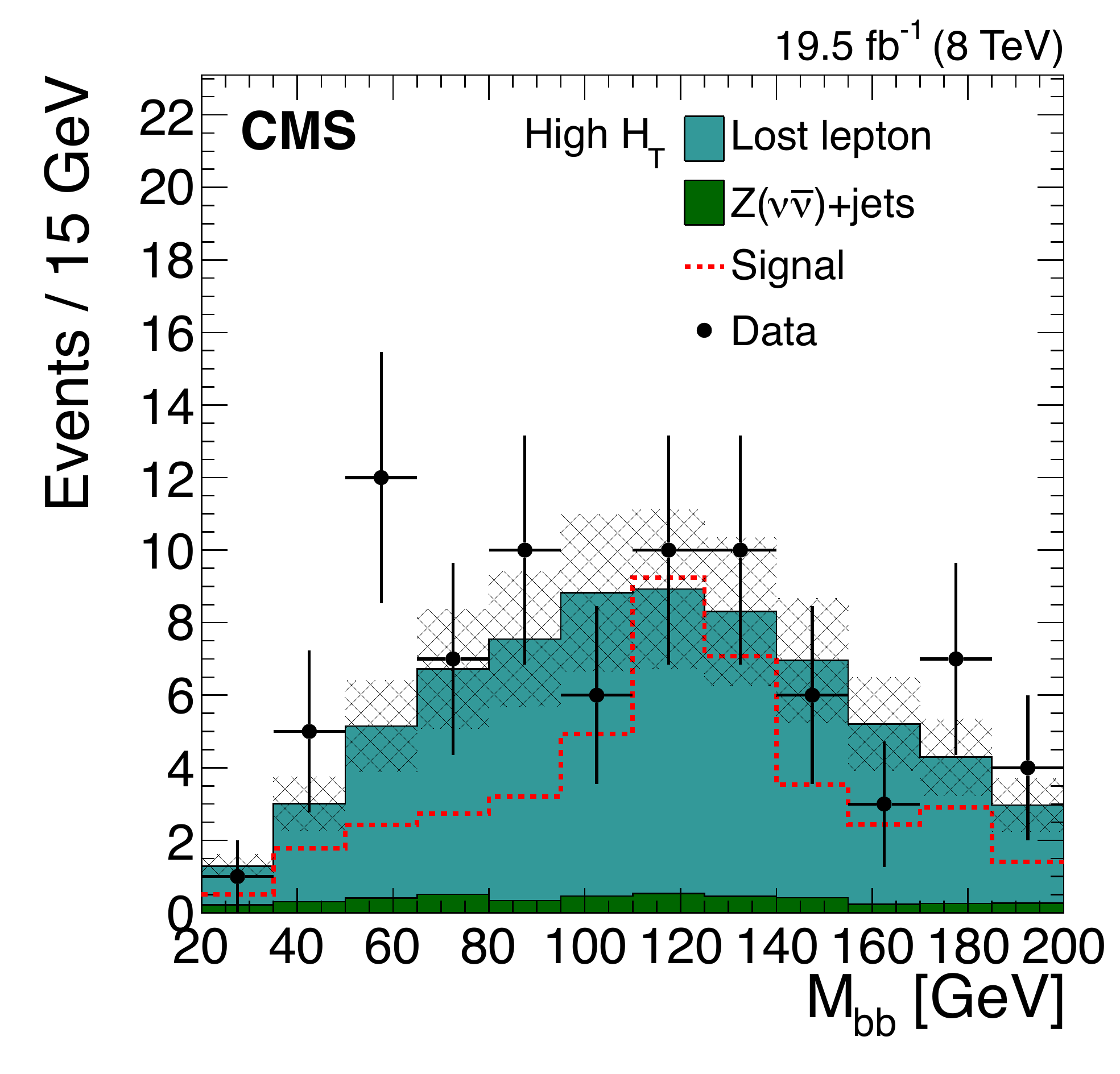}
\caption{
Distributions of the \mbb variable for the \wjets and \ttjets processes (\ie the lost-lepton background), the
\zinv background, data, and a possible SUSY signal. The distributions are shown for both the low- (left) and the high-\HT (right)
selections of the \MTt-Higgs search. The lost-lepton background is estimated from data control samples, while the
\zinv is evaluated using simulation.
  The uncertainties in each plot are drawn as the shaded band and do not include the uncertainty in the shape of the lost-lepton background.
The signal model consists of gluino pair production events with one of the two gluinos
containing an \Ph~boson in its decay chain. For this model it is assumed $\mgluino = 750\GeV$ and $\mlsp = 350\GeV$.
}
\label{fig:Higgs_finalFit}
\end{figure}

For the \MTt-Higgs analysis, the observed numbers of events in data and the predicted background yields are summarized
in Table~\ref{table:BkgandData_MT2Higgs_PAS} for the two different selections in \HT.
The background predictions and the data yields are shown for the different \mbb bins in
Fig.~\ref{fig:Higgs_finalFit} along with the distribution of events for a possible
SUSY scenario.
This scenario is based on gluino pair production in which one of the gluinos produces
one \Ph boson in its decay chain. More details about this signal scenario are
provided in Section~\ref{sec:exclusion.simple}.

\begin{table}[!htb]
\centering
\topcaption{Event yields for the \wjets and \ttjets processes (\ie the lost-lepton background), the
\zinv background, and data. Yields are shown for both the low- and the high-\HT selections of the
\MTt-Higgs search. The lost-lepton background is estimated from data control samples, while the
\zinv is evaluated using simulation.
}
\begin{tabular}{l|cc|cc}
\hline
Channel & Lost lepton & \zinv  & Total background  &  Data \\
\hline
Low-\HT & $37.1 \pm 9.0 $ & $6.9 \pm 6.9$ & $44.0 \pm 11.3$ & 55 \\
High-\HT & $64.8 \pm 16.4$ & 4.4 $\pm$ 4.4 & 69.2 $\pm$ 17.0 & 81 \\
\hline
\end{tabular}
\label{table:BkgandData_MT2Higgs_PAS}
\end{table}

\section{Systematic uncertainties}
\label{sec:systematics}

A summary of the range-of-effect for each source of uncertainty relevant for the background
prediction or signal efficiency is presented in Table~\ref{table:syst}.
While the systematic uncertainties in the background predictions have already
been discussed in Section~\ref{sec:bkgEstimation}, the dominant sources of
systematic uncertainties in the selection efficiencies of signal events are
described here.
\begin{table}[htb]
\centering
\topcaption{Summary of the different systematic uncertainties of the SM background predictions and of the signal efficiency. A given source of uncertainty can contribute
differently depending on the search region, and the typical ranges of effect are
shown. Sources of uncertainty that change the shape of the $\MTt$ distributions
in the inclusive-$\MTt$ analysis or the shape of the $\mbb$ distributions in
the $\MTt$-Higgs search
are marked with a cross in the last column.
}
\label{table:syst}
\begin{tabular}{cccc}
\hline
Process             & Source/Region             & Effect         & Shape \\\hline
\multirow{2}{*}{Multijet}   & $\MTt<200\GeV$ & 10--50\%  &\NA \\
                       & $\MTt \geq 200\GeV$ & 50--100\% &\NA \\\hline
\multirow{6}{*}{\wjets and Top} & Lost-lepton method (sys $\oplus$ stat) & 10--65\% &\NA\\
                       & b-tagging scale factor &\NA& x \\
                       & Jet energy scale         &\NA& x \\
                       & Matching scale            &\NA& x \\
                       & Renormalization and factorization scales  &\NA& x \\
                       & System recoil modelling  &\NA& x \\\hline
\multirow{4}{*}{\zinv} & Systematics on $Z(\nu\bar{\nu})/\gamma$ ratio ($\NB=0$--1) & 20--30\% &\NA\\
        & Systematics on $1\PQb/0\PQb$ ratio from $Z_{\ell\ell}$ ($\NB=1$) & 10--75\% &\NA\\
        & Statistics from $\gamma$+jets data ($\NB=0$--1) & 5--100\% &\NA\\
        & Simulation ($\NB\geq2$) & 100\% &\NA\\\hline
\multirow{6}{*}{Signal} & Integrated luminosity & 2.6\%  &\NA\\
        & Trigger efficiency & 1\%  &\NA\\
        & Parton distribution functions & 5--15\% &\NA \\
        & b-tagging scale factor & 5--40\% & x  \\
        & Jet energy scale         & 5--40\% & x  \\
        & System recoil modelling        & 10--20\%  & x \\
\hline
\end{tabular}
\end{table}

The corrections for the differences observed between the signal simulation and data due to the jet energy scale and b-tagging
efficiencies yield uncertainties in the signal yield of around 5\%, but these uncertainties can
become as large as 40\% in kinematically extreme regions. The uncertainty associated with
the corresponding correction to account for the \pt spectrum of the
recoil system reaches a maximum of 20\% for $\pt>250\GeV$.
The systematic uncertainty associated with the parton distribution functions
is evaluated following the prescription of the PDF4LHC
group~\cite{Alekhin:2011sk,pdf4lhc,Martin:2009iq,Ball:2010de,Lai:2010vv}, and
is found to have an effect of about 5\%, increasing to a maximum of 15\% for small
splittings between the parent particle mass and the LSP mass.
Additionally, uncertainties associated with the luminosity
determination~\cite{CMS-PAS-LUM-13-001} and the trigger efficiency are included.

\section{Statistical interpretation of the results}
\label{sec:stats}

This section describes the statistical procedure used to interpret the observed
event yields in order to set upper limits on the cross sections of potential signal processes.
A test of the background-only and signal+background hypotheses is performed using a modified frequentist
approach, often referred to as \CLs~\cite{Junk:1999kv}.

Signal regions
are
combined
through a joint likelihood function.
This function is constructed as the product of Poisson
probabilities for each bin of \NJ, \NB, \HT, and \MTt.
The Poisson probabilities are functions of
the number of observed events in each bin, $n_i$, and the predictions in
each bin, $\lambda_i$, where $i$ ranges from 1 to the number of bins, $N_{\text{bins}}$.  The
likelihood function is given by
\begin{equation}
\mathcal L = \prod_{i=1}^{N_{\text{bins}}} \frac{\lambda_i^{n_i}\re^{-\lambda_i}}{n_i!}.
\end{equation}
The prediction in each bin is a sum over the signal and background
contributions:
\begin{equation}
  \label{eq:mu}
  \lambda_i = \mu\, s_{i} + \sum_{j=1}^{N_\text{bkg}} b_{ij},
\end{equation}
where $b_{ij}$ is the background prediction in bin $i$ for background
source $j$, and $s_i$ is the signal prediction in bin $i$, scaled by the
signal-strength modifier $\mu$ to test other values of the signal
production cross section,
$\sigma=\mu \sigma_{\text{sig}}$, with
$\sigma_{\text{sig}}$ the nominal cross section for the signal model under
consideration.

The uncertainties are handled by introducing nuisance parameters $\theta$.
The signal and background
expectations, therefore, become dependent on $N_\text{sys}$ nuisance
parameters $\theta_m$, where $m=1\ldots N_\text{sys}$, \ie  $s=s(\theta_m)$ and $b=b(\theta_m)$.
All sources of uncertainties are taken to be either 100\%-correlated
(positively or negatively) or uncorrelated (independent), whichever is
found to be appropriate.
Incorporating the nuisance parameters, the likelihood function becomes:
\begin{equation}
\mathcal L (\text{data}|\mu,\theta) = \mathrm{Poisson}(\text{data}|\mu\,
s(\theta)+b(\theta)) p(\theta),
\end{equation}
where $p(\theta)$ is the probability density function
associated with the given systematic uncertainty.
In this equation, $\mathcal L (\text{data}|\mu,\theta)$ is the likelihood function for data for a given value of $\mu$ and $\theta$.

In order to test the compatibility of the data with the background-only
and signal+background hypotheses, a test statistic
$q_\mu$ \cite{Cowan:2010js} is constructed starting from the profile-likelihood ratio:
\begin{equation}
  \label{eq:testStat}
  q_\mu=-2\ln \frac{\mathcal L (\text{data}|\mu,\hat\theta_\mu)}
{\mathcal L (\text{data}|\hat\mu,\hat\theta)},\qquad
\mathrm{with}\; 0\le\hat\mu\le\mu,
\end{equation}
where ``data'' can be the actual data or the output of a pseudo-experiment. Both the
denominator and numerator are maximized.
In the numerator, the signal
parameter strength $\mu$ remains fixed and the likelihood is maximized for only the nuisance parameters,
whose values after the maximization are denoted $\hat\theta_{\mu}$.
In the denominator, the likelihood is maximized with respect to both $\mu$ and $\theta$,
and $\hat\mu$ and $\hat\theta$ are the values for which $\mathcal L$ is maximal.
The lower constraint
$0\le\hat\mu$ is imposed as the signal strength cannot be negative, while the
upper constraint guarantees a one-sided confidence interval (this means
that upward fluctuations of data are not considered as evidence
against the signal hypothesis).
The value of the test statistic for the actual observation
is denoted as $q_\mu^\text{obs}$.
This test statistic was chosen by the LHC Higgs Combination Group~\cite{ATLAS:2011tau}.

To set limits,
probabilities to observe an outcome at least as signal-like
as the one observed are calculated for the null (background-only) hypothesis
$H_0$ and for the test (signal+background) hypothesis $H_1$, for a given value of the
signal-strength modifier $\mu$, as:
\begin{equation}\begin{split}
\CLsb(\mu) =  & P(q_\mu\ge q_\mu^\text{obs}|H_1),  \\
\CLb(\mu)  =  & P(q_\mu\ge q_\mu^\text{obs}|H_0).
\end{split}\end{equation}
The \CLs quantity is then defined as the ratio of these probabilities:
\begin{equation}
  \CLs(\mu)=\frac{\CLsb(\mu)}{\CLb(\mu)}.
\end{equation}
In the modified frequentist approach, the value of $\CLs(\mu)$ is
required to be less than or equal to $\alpha$ in order to declare a $(1-\alpha)$
CL exclusion.
We set 95\% CL limits on the signal cross section
by finding the value of $\mu$ for which $\CLs(\mu)=0.05$.

In practice, the probability distributions of the background-only and the signal+background
hypotheses are determined from distributions of the test statistic constructed
from
pseudo-experiments.
Once the ensembles of
pseudo-experiments for the two hypotheses are generated, the observed \CLs
limit is calculated from these distributions and the actual observation
of the test statistic $q_\mu^\text{obs}$.
The expected \CLs limit is calculated
by replacing $q_\mu^\text{obs}$ by the expected median from the distribution of the
background-only hypothesis.
Further details on the procedure employed to compute the limits on the signal production
cross section are given in Ref.~\cite{ATLAS:2011tau}.

\section{Exclusion limits}
\label{sec:exclusion}

The 95\% CL upper limits on signal production cross sections
are computed following the \CLs formulation described in
Section~\ref{sec:stats}, using the results presented in
Section~\ref{sec:data:geq3jets.results} and the systematic uncertainties
summarized in Section~\ref{sec:systematics}.

\subsection{Exclusion limits on simplified models}
\label{sec:exclusion.simple}

In this section, we interpret the results of our search in terms of simplified
models \cite{Alves:2011wf},
which allow the exclusion potential of the data to be examined in the context of a large variety of models.

\begin{figure}[!htbp]
\centering
\includegraphics[width=0.48\textwidth]{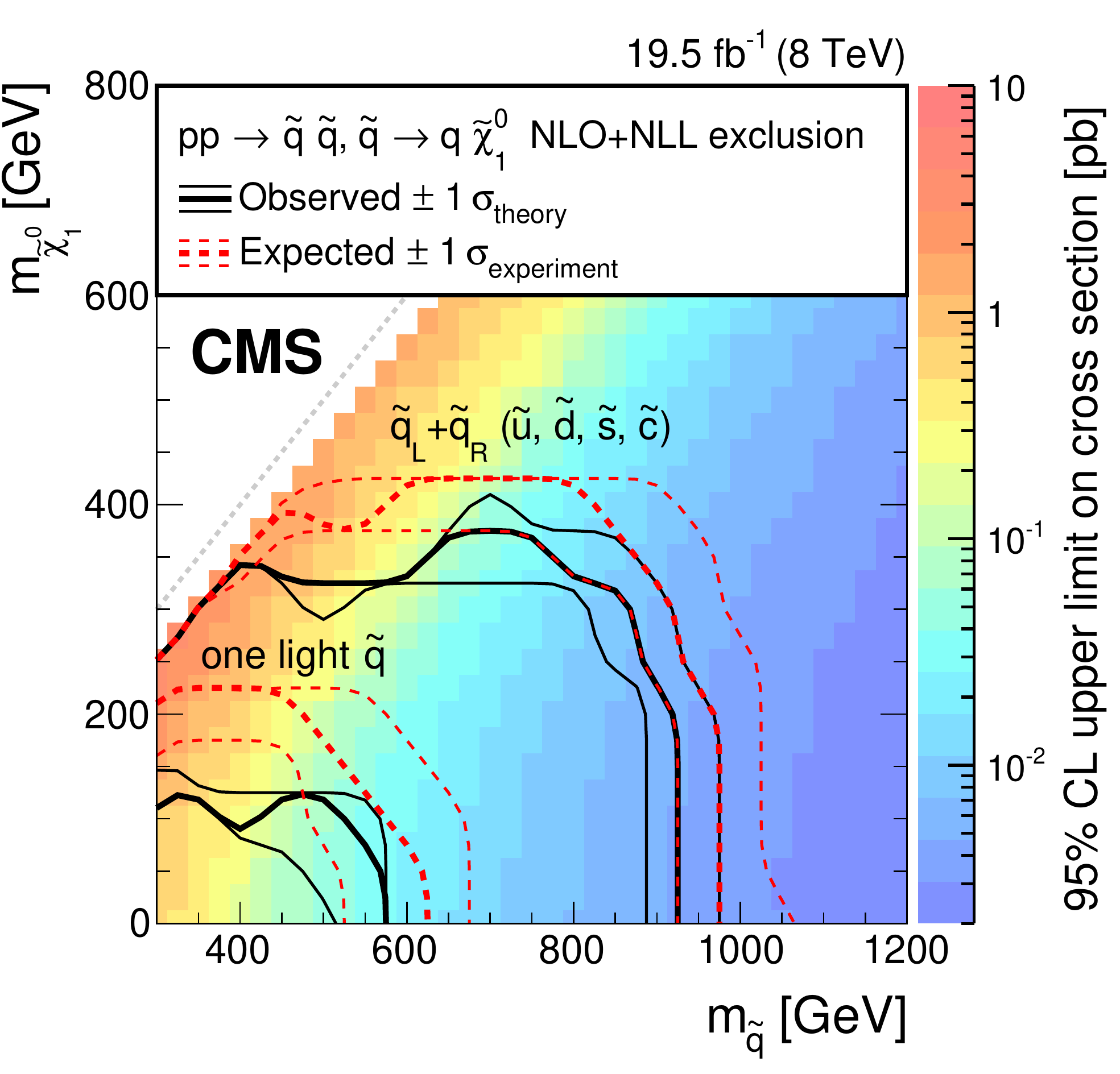}
\includegraphics[width=0.48\textwidth]{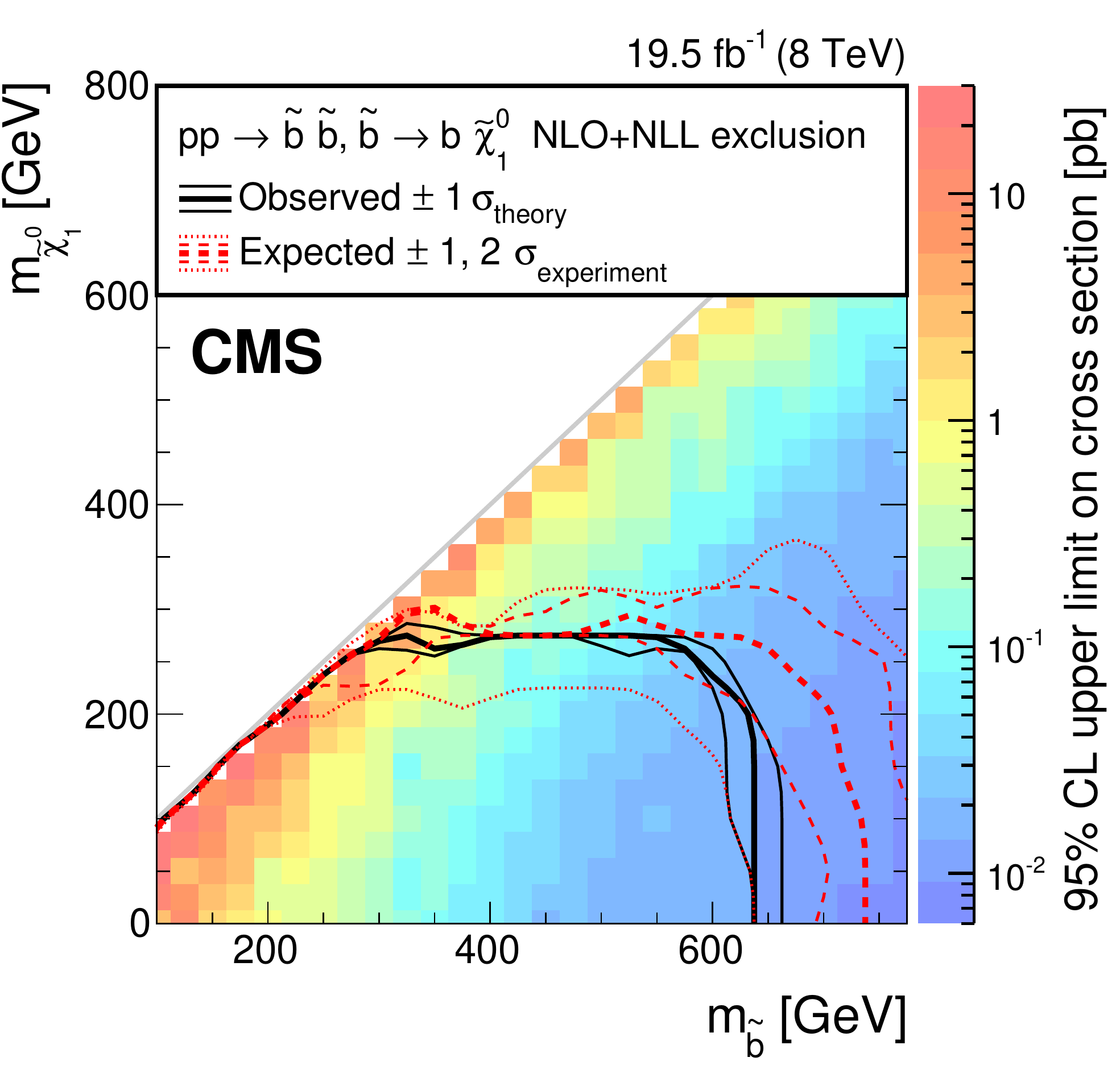}\\
\includegraphics[width=0.48\textwidth]{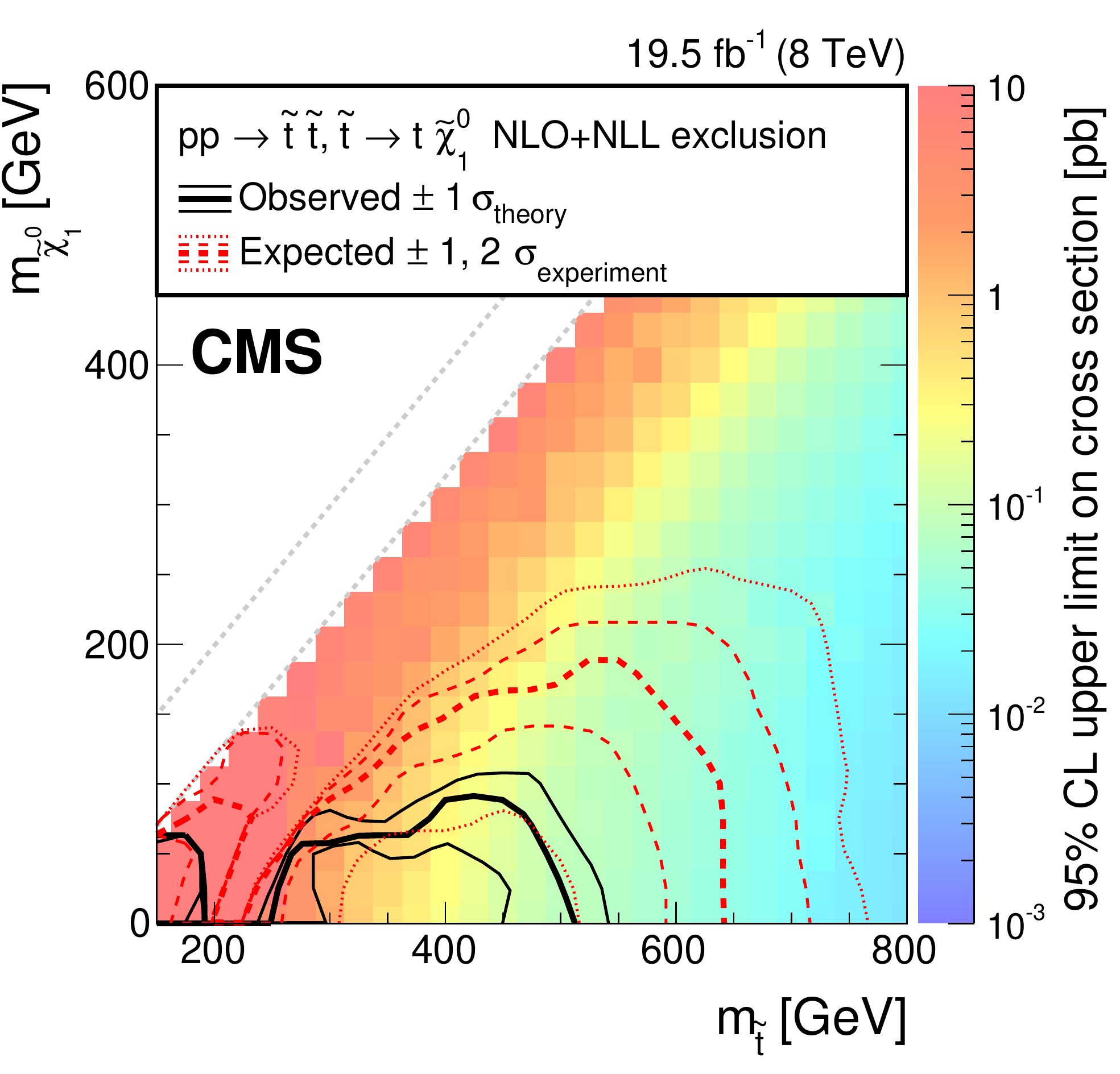}
\caption{Exclusion limits at 95\% CL for
  (upper left) direct squark production,
  (upper right) direct bottom-squark production,
  and
  (bottom) direct top-squark production.
  For the direct squark production, the upper set of curves corresponds to the
  scenario where the first two generations of squarks are degenerate and light,
  while the lower set corresponds to only one accessible light-flavour squark.
  For convenience, diagonal lines have been drawn corresponding to
  $\mlsp=m_{\PSQ,\PSQb,\PSQt}$ and
  $\mlsp=m_{\PSQ,\PSQb,\PSQt}-\left(m_{\PW} + \mb\right)$
  where applicable.}
\label{fig:exclSMS-direct}
\end{figure}

\begin{figure}[!htbp]
\centering
\includegraphics[width=0.48\textwidth]{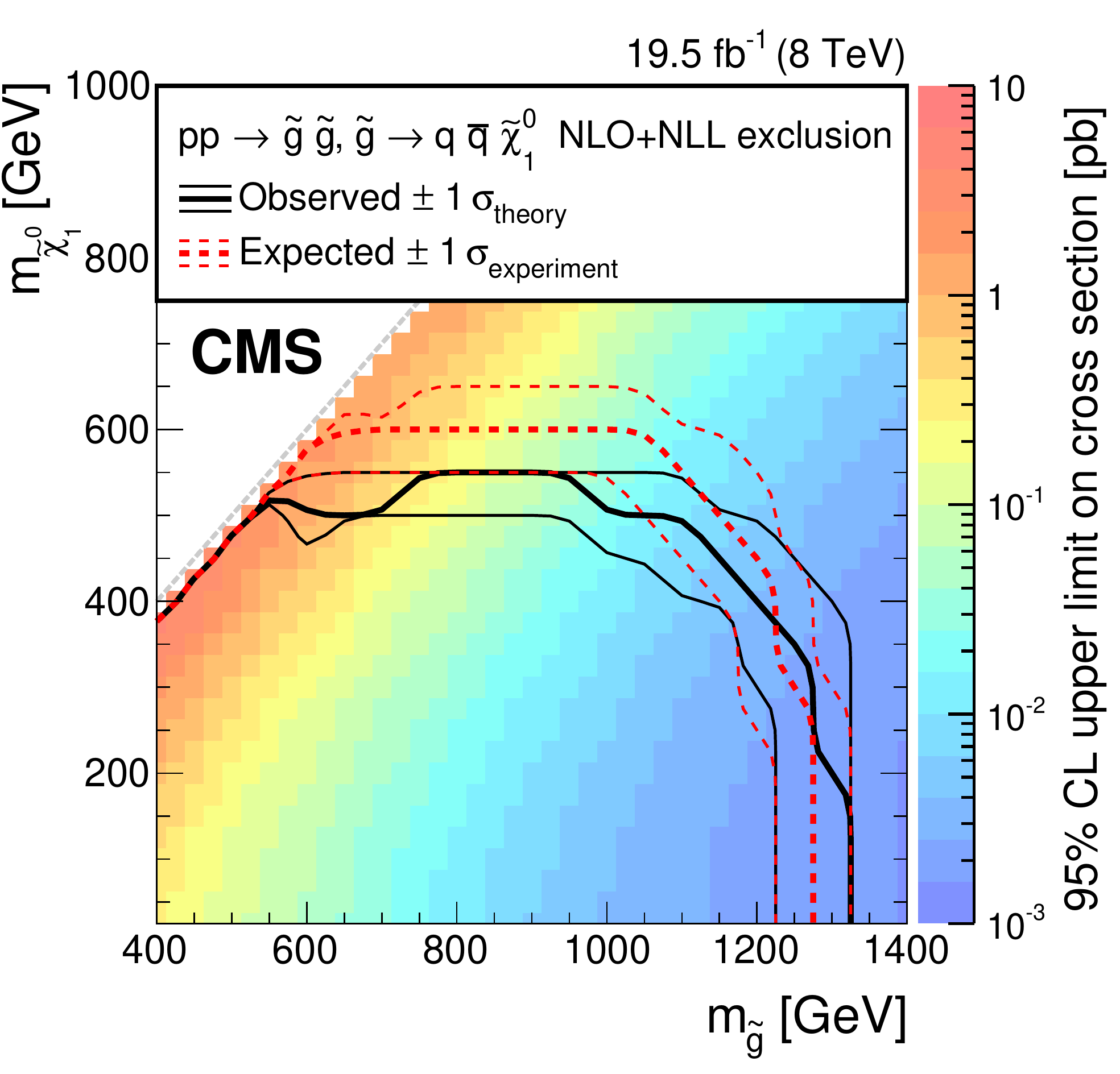}
\includegraphics[width=0.48\textwidth]{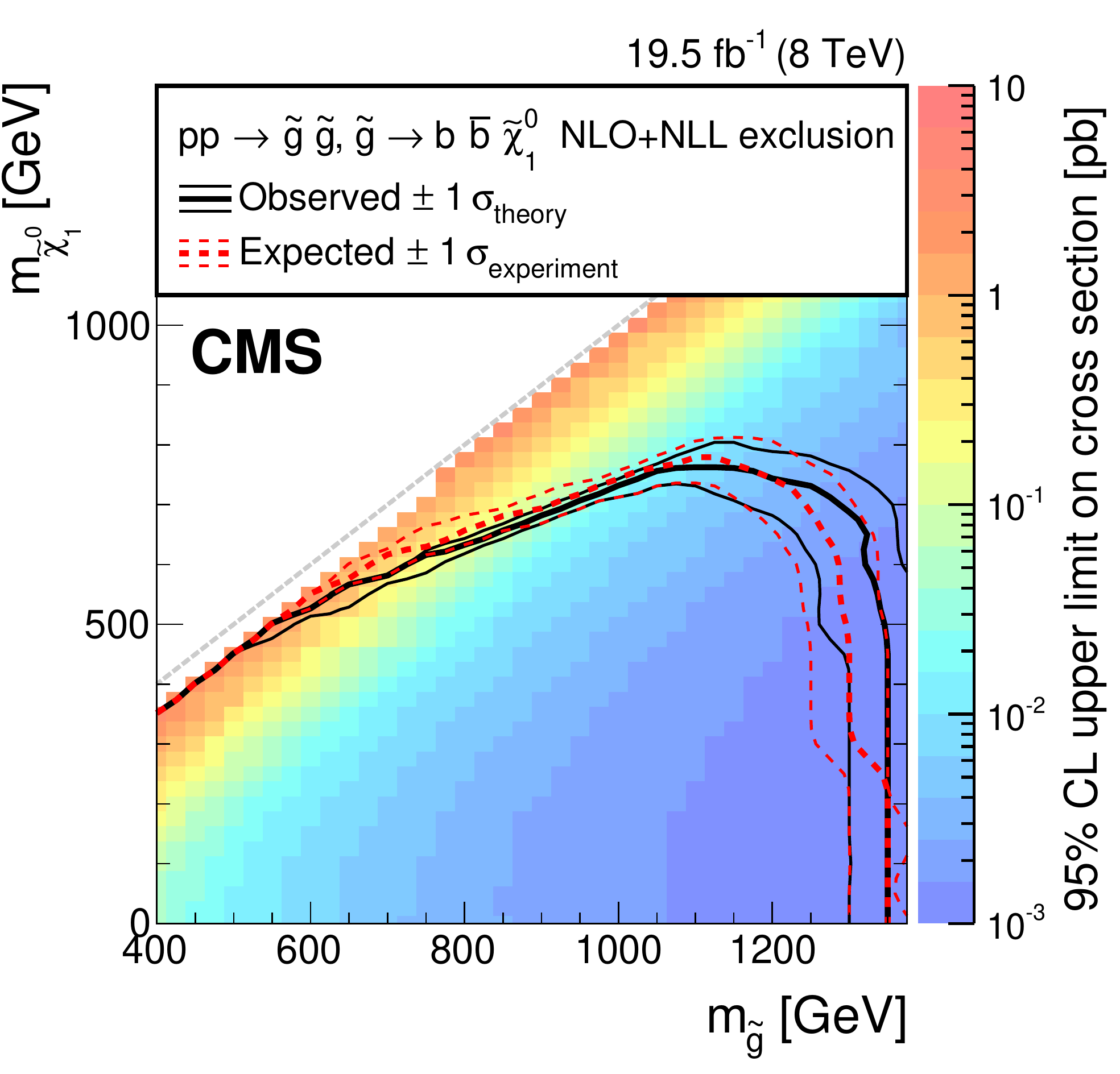}\\
\includegraphics[width=0.48\textwidth]{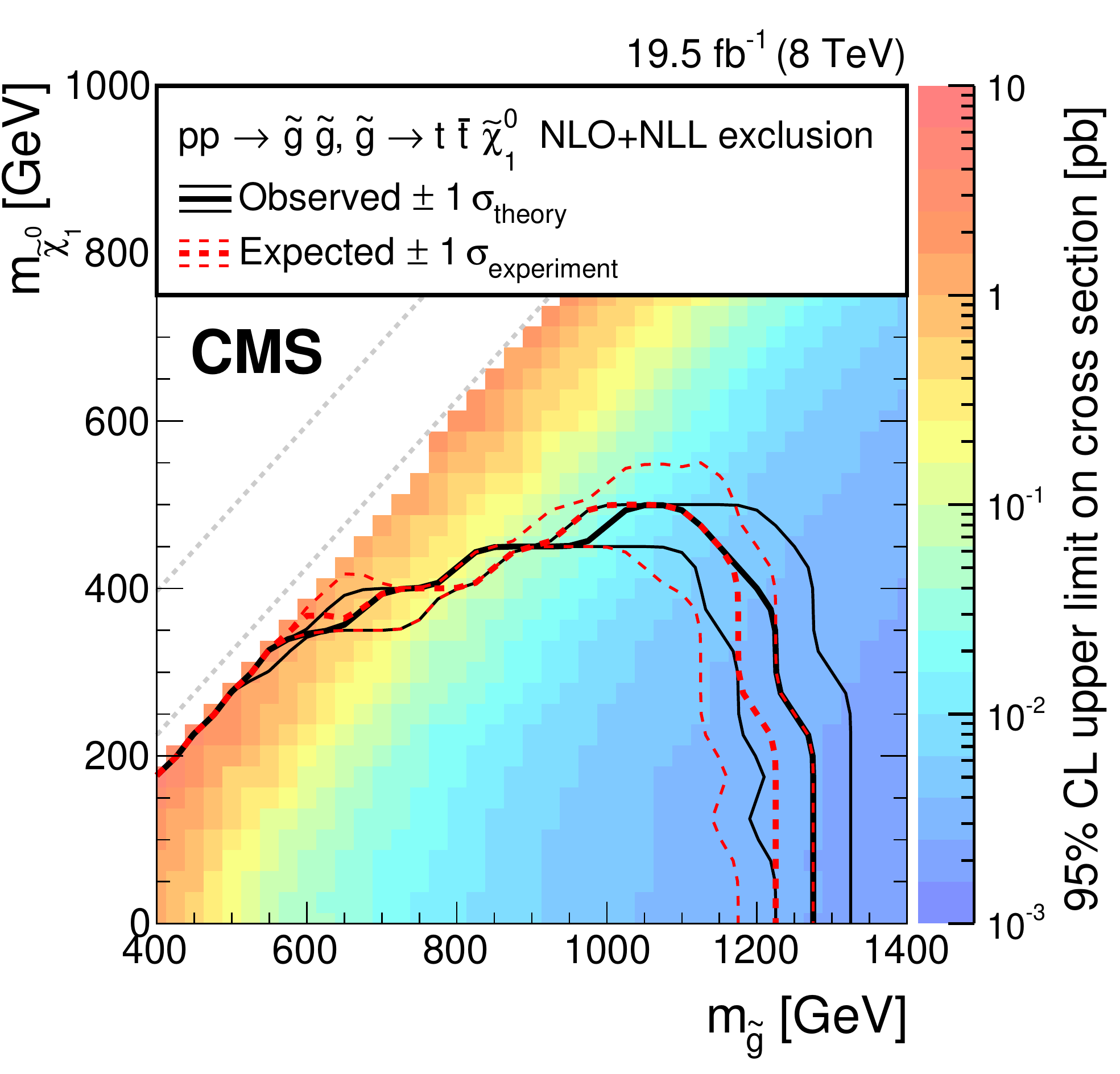}
\caption{Exclusion limits at 95\% CL for gluino mediated (upper left) squark production, (upper right)
  bottom-squark production, and (bottom) top-squark production.
  For convenience, diagonal lines have been drawn corresponding to
  $\mlsp=\mgluino$ and
  $\mlsp=\mgluino-\mtop$ where applicable.
}
\label{fig:exclSMS-gluino}
\end{figure}

The following list describes the simplified models
that are probed and the corresponding subsets of signal regions
from the inclusive-\MTt search that are used to set the limits:
\begin{itemize}
  \item direct pair production of squarks with $\PSQ \to \PQq \PSGczDo$. The
    topological regions that are used to probe this model are those defined by the selections
    $(\NJ=2,\NB=0)$, $(\NJ=2,\NB\geq1)$, $(3\leq \NJ\leq5,\NB=0)$, $(3\leq \NJ\leq5,\NB=1)$,
    and $(\NJ\geq6,\NB=0)$. Exclusions limits are shown in Fig.~\ref{fig:exclSMS-direct}~(upper left) for two scenarios:
    one assumes that the first two generations of squarks
    ($\PSQu_\mathrm{L}$, $\PSQu_\mathrm{R}$, $\PSQd_\mathrm{L}$, $\PSQd_\mathrm{R}$, $\PSQc_\mathrm{L}$, $\PSQc_\mathrm{R}$, $\PSQs_\mathrm{L}$, $\PSQs_\mathrm{R}$) are degenerate and light; the other requires that only one light-flavour squark be kinematically accessible.

  \item direct pair production of bottom squarks with $\PSQb \to \PQb
    \PSGczDo$. The signal regions that are used are those defined by $(\NJ=2,\NB\geq1)$,
    $(3\leq \NJ\leq5,\NB=1)$, and $(3\leq \NJ\leq5,\NB=2)$. The corresponding exclusion
    limits are shown in Fig.~\ref{fig:exclSMS-direct}~(upper right).

  \item direct pair production of top squarks with $\PSQt \to \PQt
    \PSGczDo$. The topological regions used to probe this model are those defined by
    $(3\leq \NJ\leq5,\NB=1)$, $(3\leq \NJ\leq5,\NB=2)$, $(\NJ\geq6,\NB=1)$, $(\NJ\geq6,\NB=2)$, and
    $(\NJ\geq3,\NB\geq3)$. The corresponding exclusion limits are shown in Fig.~\ref{fig:exclSMS-direct}~(bottom).

  \item gluino pair production with $\PSg\to \PQq\PAQq \PSGczDo$.
    The topological regions used to probe this model are those defined by
    $(3\leq \NJ\leq5,\NB=0)$, $(3\leq \NJ\leq5,\NB=1)$,
    $(\NJ\geq6,\NB=0)$, and $(\NJ\geq6,\NB=1)$. The corresponding exclusion limits are
    shown in Fig.~\ref{fig:exclSMS-gluino}~(upper left).

  \item gluino pair production, with $\PSg\to \bbbar \PSGczDo$.
    The topological regions used to probe this model are those defined by
    $(3\leq \NJ\leq5,\NB=1)$, $(3\leq \NJ\leq5,\NB=2)$,
    $(\NJ\geq6,\NB=1)$, $(\NJ\geq6,\NB=2)$, and $(\NJ\geq3,\NB\geq3)$.
    The corresponding exclusion limits are shown in Fig.~\ref{fig:exclSMS-gluino}~(upper right).

  \item gluino pair production, with $\PSg \to \ttbar \PSGczDo$.
    The topological regions used to probe this model are those defined by
    $(\NJ\geq6,\NB=1)$, $(\NJ\geq6,\NB=2)$, and $(\NJ\geq3,\NB\geq3)$.
    The corresponding exclusion limits are shown in Fig.~\ref{fig:exclSMS-gluino}~(bottom).
 \end{itemize}
All exclusion limits are obtained at NLO + next-to-the-leading-logarithm (NLL) order in $\alpha_s$.

For the direct pair production of top squarks,
the analysis is not sensitive to model points with $m_{\PSQt} - m_{\PSGczDo} = m_{\PQt}$, because the $\PSGczDo$ is produced at rest in the top-squark frame.

For all the considered models, the observed limits are compatible within one standard deviation with the expected limits,
with the exception of the limits on the direct pair production of
top squarks with $\PSQt \to \PQt \PSGczDo$, which are shown in Fig.~\ref{fig:exclSMS-direct}~(bottom).
A comparison between the background estimates obtained directly from simulation and those calculated
from the data control samples
suggests that the weaker-than-expected limits are not caused by an excess in the signal region,
but rather by a downward fluctuation in the lost-lepton control sample,
leading to a possible underestimate of the lost-lepton background.
Considering the large number of data control samples (there are 81 lost-lepton control regions),
the probability to observe a fluctuation as large as the one observed is $\sim$65\%.


\begin{figure}[!htb]
\centering
\includegraphics[width=0.48\textwidth]{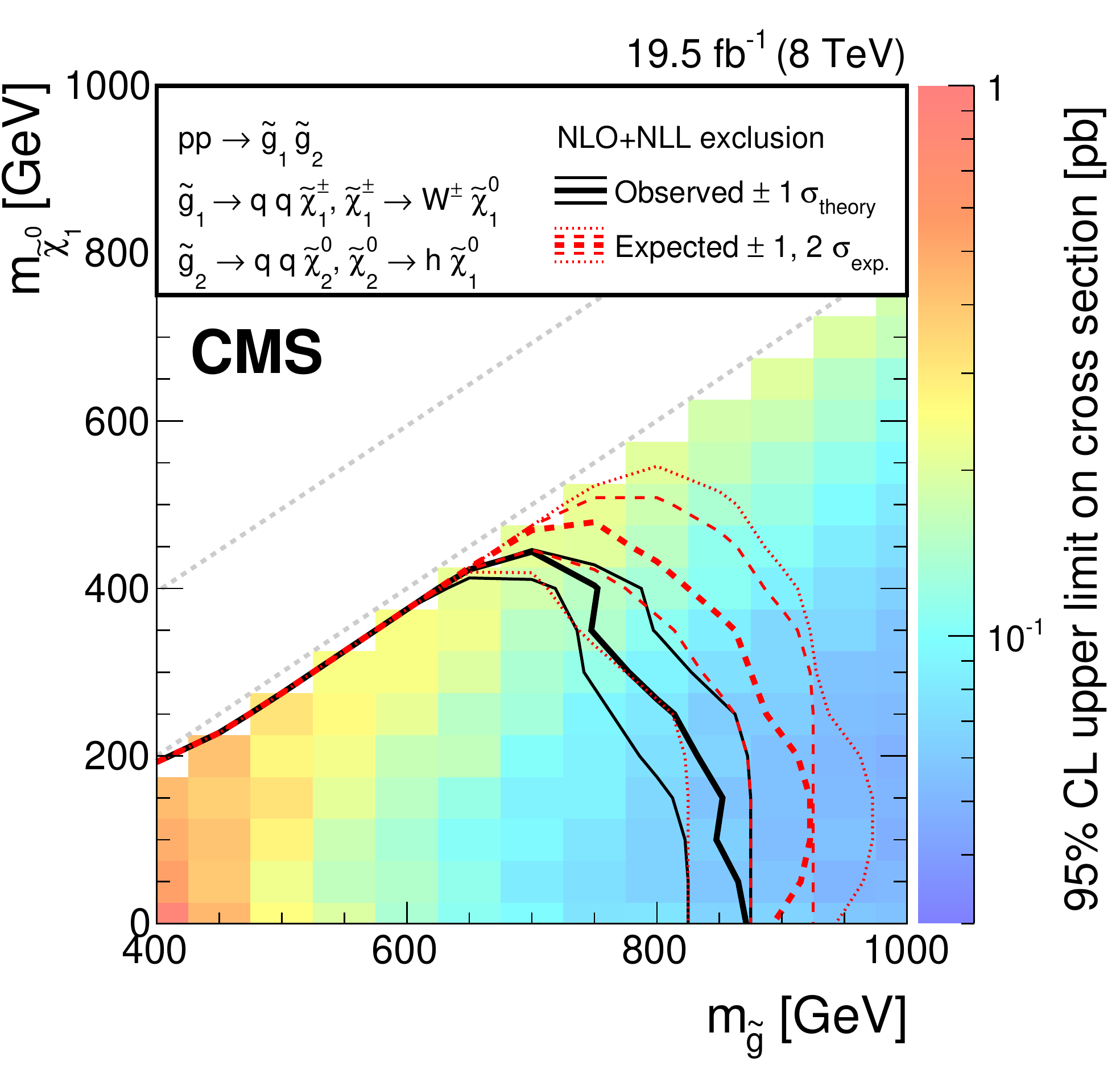}
\caption{Exclusion limits at 95\% CL for gluino pair production with one gluino decaying via $\PSg \to \PQq\PAQq\PSGczDt$, $\PSGczDt\to \Ph\PSGczDo$, while the other gluino decays via
 $\PSg\to \PQq\PQq' \PSGczDopm$,
 $\PSGczDopm\to \PW^\pm \PSGczDo$. For
 convenience, diagonal lines have been drawn corresponding to
 $\mlsp=\mgluino$ and
 $\mlsp=\mgluino-200\GeV$.}
\label{fig:exclSMS_t5wh}
\end{figure}

The results of the \MTt-Higgs search are used to probe the following model:
gluino pair production with one gluino decaying via $\PSg \to \PQq\PAQq\PSGczDt$, $\PSGczDt\to \Ph\PSGczDo$, and the other gluino decaying via
 $\PSg\to \PQq\PQq' \PSGczDopm$, $\PSGczDopm\to \PW^\pm \PSGczDo$.
In this scenario, the neutralino $\PSGczDt$ and chargino $\PSGczDopm$ are
    assumed to be degenerate, with mass $\mneutTwo=\mcharg=\mlsp + 200\GeV$. The corresponding exclusion limits are shown in Fig.~\ref{fig:exclSMS_t5wh}.

\subsection{Exclusion limit in the cMSSM{\slash}mSUGRA model}
\label{sec:exclusion.cMSSM}

\begin{figure}[!ht]
\centering
\includegraphics[width=0.8\textwidth]{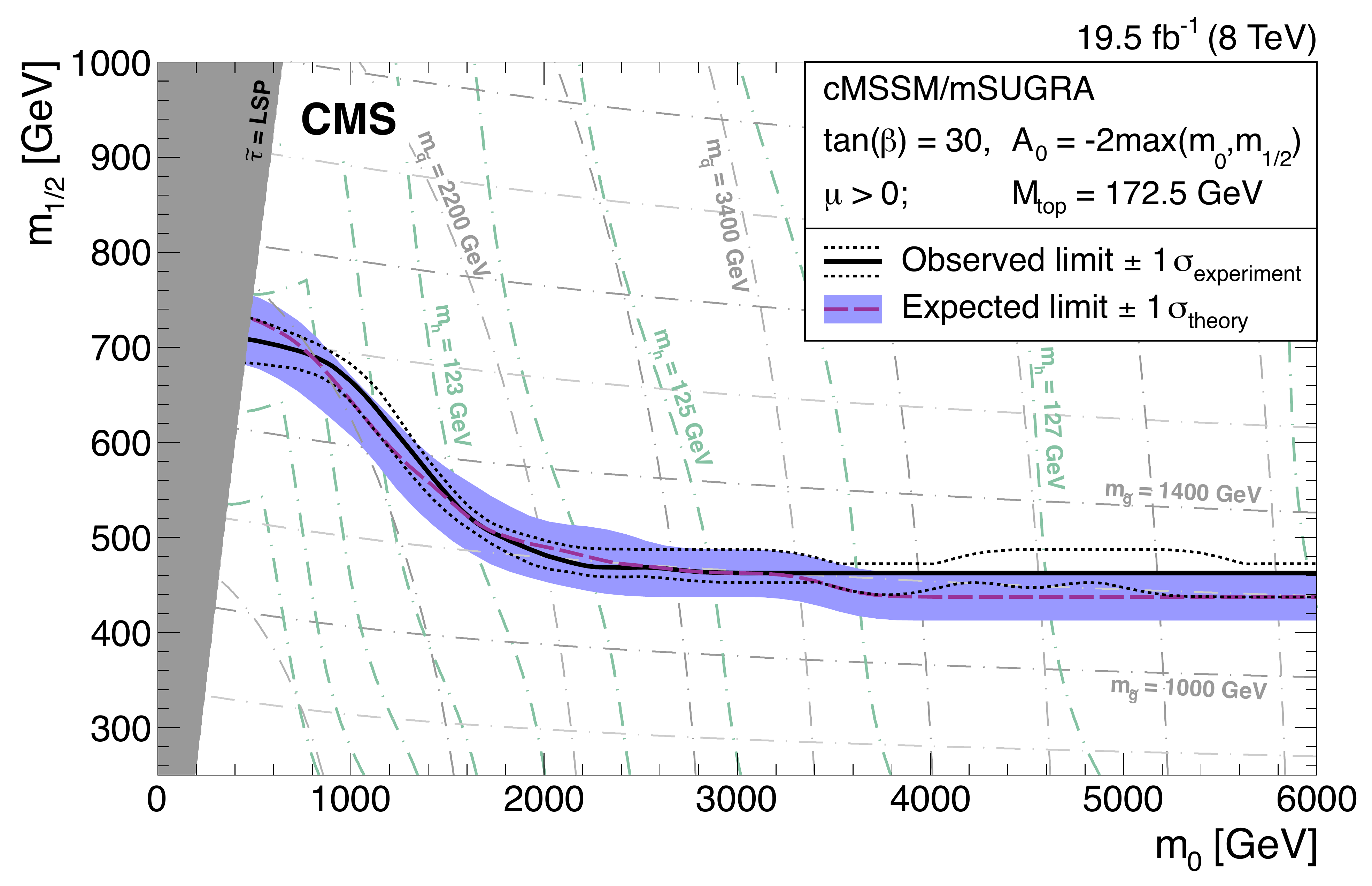}
\caption{Exclusion limits at 95\% CL as a function of $m_0$ and $m_{1/2}$ for the
cMSSM{\slash}mSUGRA model with $\tan\beta=30$, $A_0 = -2\max(m_0, m_{1/2})$, and
$\mu>0$. Here, \msq is the average mass of the first-generation squarks.
\label{fig:cMSSM:m0m12}}
\end{figure}
\begin{figure}[!htb]
\centering
\includegraphics[width=0.8\textwidth]{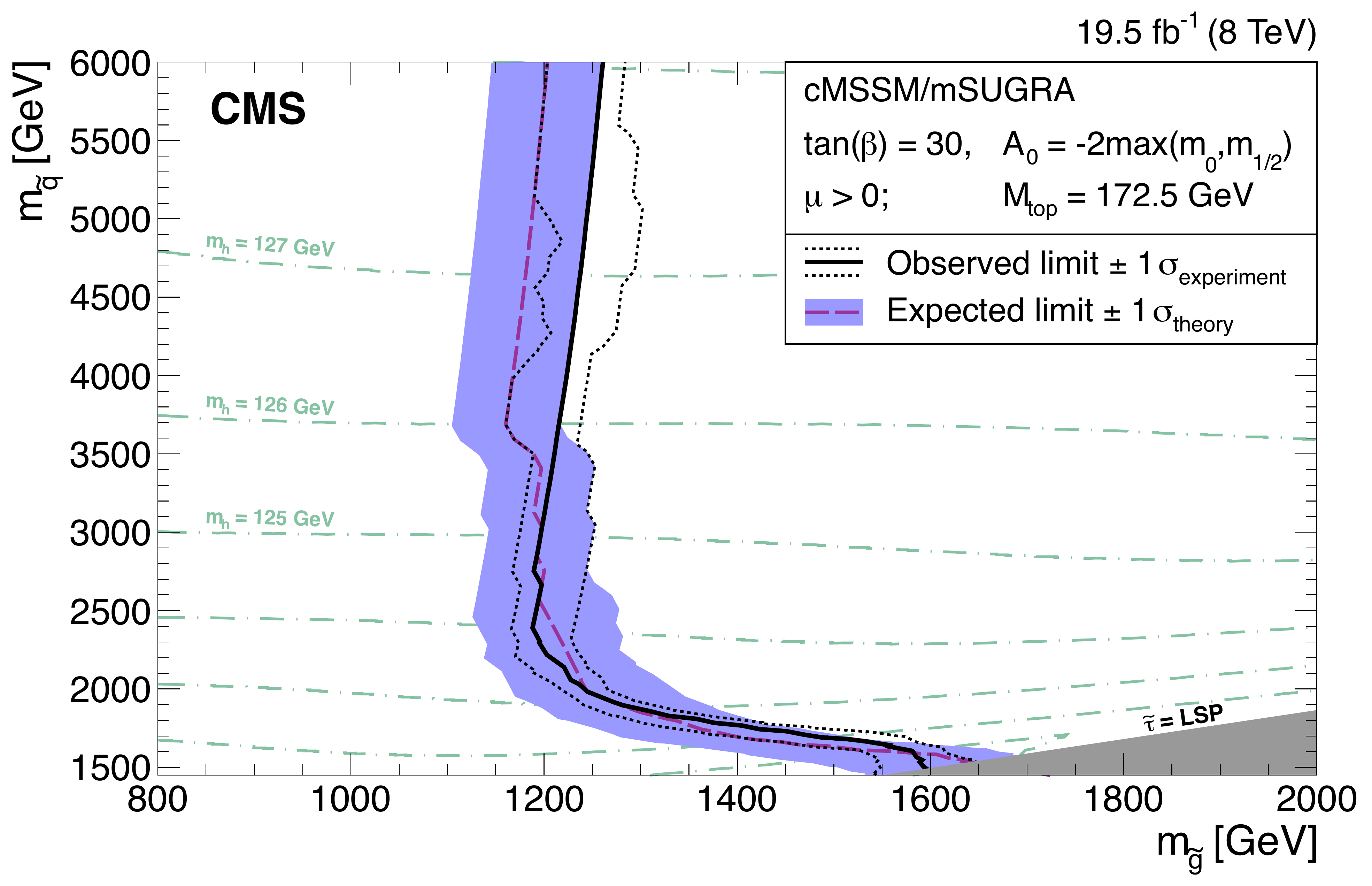}
\caption{Exclusion limits at 95\% CL as a function of \mgluino and \msq for the cMSSM{\slash}mSUGRA model with $\tan\beta=30$, $A_0 = -2\max(m_0, m_{1/2})$, and $\mu>0$. Here, \msq is the average mass of the first-generation squarks.
\label{fig:cMSSM:mgmq}
}
\end{figure}

We also provide an interpretation of our results in terms of the cMSSM{\slash}mSUGRA model.
The model has five free parameters: $m_0$, $m_{1/2}$, $A_0$, $\tan\beta$,
and $\sign{\mu}$.
In order to obtain an \Ph boson mass of about 125\GeV, the value
$A_0 = -2\max(m_0, m_{1/2})$ is chosen, as proposed in Ref.~\cite{Brummer:2012ns}.
Furthermore, we choose $\mu>0$ and $\tan\beta=30$.
Exclusion limits as a function of $m_0$ and $m_{1/2}$ are shown in Fig.~\ref{fig:cMSSM:m0m12}.
These limits
are presented in Fig.~\ref{fig:cMSSM:mgmq} as a function of \mgluino and \msq,
where \msq is the average mass of the first-generation squarks.

\begin{table}[!thb]
\centering
\topcaption{Summary of observed mass limits (at  95\% CL) for different SUSY simplified models and
for the cMSSM{\slash}mSUGRA model.
The limits quoted are the observed limits
using the signal cross section minus one standard deviation
($\sigma_\text{theory}$) of its uncertainty.
For the simplified models, the limit on the mass of the parent particle is quoted for $\mlsp = 0$,
while for the LSP the best
limit on its mass is quoted. The best limit on the mass splitting between the
parent particle mass and the LSP mass is also given.
Finally, the absolute limits on the squark and gluino masses
are quoted for the cMSSM{\slash}mSUGRA model.
}
\setlength{\tabcolsep}{4pt}
\begin{tabular}{lccc}
\hline
Simplified & Limit on parent particle & Best limit on & Limit on \\
model  & mass at $\mlsp = 0$ & LSP mass & mass splitting \\
\hline\\[-2.5ex]\hline
Direct squark production & & & \\
Single light squark & $\msq > 520\GeV$ & $\mlsp > 120\GeV$ & $\Delta m(\PSQ,\PSGczDo) < 200\GeV$ \\
8 degenerate light squarks & $\msq > 875\GeV$ & $\mlsp > 325\GeV$ & $\Delta m(\PSQ,\PSGczDo) < 50\GeV$ \\
\hline
Bottom squark & $m_{\PSQb} > 640\GeV$ & $\mlsp > 275\GeV$ & $\Delta m(\PSQb,\PSGczDo) < 10\GeV$ \\
\hline
Top squark & & & \\
$\mstop>\mtop+\mlsp$ & $m_{\PSQt} > 450\GeV$ & $\mlsp > 60\GeV$ & $\Delta m(\PSQt,\PSGczDo) < 230\GeV$ \\
$\mstop<\mtop+\mlsp$ & $m_{\PSQt} > 175\GeV$ & $\mlsp > 60\GeV$ & $\Delta m(\PSQt,\PSGczDo) < 90\GeV$ \\
\hline\\[-2.5ex]\hline
Direct gluino production & & & \\
\hline
$\PSg\to \PQq \PAQq \PSGczDo$ & $\mgluino>1225\GeV$ &  $\mlsp > 510\GeV$ & $\Delta m(\PSg,\PSGczDo) < 25\GeV$ \\
$\PSg\to \PQb \PAQb \PSGczDo$ & $\mgluino>1300\GeV$ &  $\mlsp > 740\GeV$ & $\Delta m(\PSg,\PSGczDo) < 50\GeV$ \\
$\PSg\to \ttbar \PSGczDo$ & $\mgluino>1225\GeV$ &  $\mlsp > 450\GeV$ & $\Delta m(\PSg,\PSGczDo) < 225\GeV$ \\
\hline
$\PSg_{1} \to \PQq \PAQq \PSGczDt ,\,\PSGczDt\to \Ph \PSGczDo$, & \multirow{2}{*}{$\mgluino>825\GeV$} & \multirow{2}{*}{$\mlsp > 410\GeV$} & \multirow{2}{*}{$\Delta m(\PSg,\PSGczDo) < 225\GeV$} \\
 $\PSg_{2}\to \PQq \PQq' \PSGczDopm ,\,\chi^{\pm}_{1}\to \PW^\pm \PSGczDo$ & & & \\
\hline\\[-2.5ex]\hline
cMSSM{\slash}mSUGRA model & Mass limit for $\msq = \mgluino$ & Gluino mass limit & Squark mass limit \\
\hline
 & $m_{\PSg,\,\PSQ} > 1550\GeV$ & $\mgluino > 1150\GeV$ & $\msq > 1450\GeV$ \\
\hline
\end{tabular}
\label{table:exclSMSsummary}
\end{table}

In Table~\ref{table:exclSMSsummary}, we summarize the exclusion limits
from Figs.~\ref{fig:exclSMS-direct}--\ref{fig:cMSSM:m0m12}.

\section{Summary}
\label{sec:conclusion}

A search for supersymmetry (SUSY) in hadronic
final states characterized by large values of unbalanced transverse momentum has been carried out using
a sample of $\sqrt{s} = 8\TeV$ pp collisions.
The data were collected by the CMS experiment at the CERN LHC
and correspond to an integrated luminosity of 19.5\fbinv.
An event selection based on the kinematic mass variable \MTt
has been employed to reduce the background from
standard model processes
and to enhance the sensitivity of the search
to a wide range of SUSY signatures.

Two related searches have been implemented.
The first is an inclusive search based on several signal regions defined by the
number of jets and b-tagged jets,
the hadronic energy in the event, and the value of the \MTt variable.
The second is a search for events that contain a Higgs boson in the decay chain of a heavy SUSY
particle.
Assuming that this boson decays to a bottom quark-antiquark pair in accordance with the branching fraction of the standard model Higgs boson,
this category of events has been investigated to seek an excess at 125\GeV in the invariant mass distribution of
the selected b-tagged jet pairs.

No significant excess over the expected number of background events has been observed, and 95\% confidence level exclusion limits
on several SUSY simplified models and on the cMSSM{\slash}mSUGRA model have been derived.
Mass limits have been conservatively derived using the theoretical signal cross
sections reduced by one times their uncertainty ($-1\sigma_\text{theory}$).
In the context of simplified models based on pair-produced gluinos,
each decaying into a quark-antiquark pair and a lightest SUSY particle (LSP) via an off-shell squark,
gluino masses have been probed up to 1225--1300\GeV depending on the squark
flavour.
For the direct pair production of the first- and second-generation squarks, each
assumed to decay to a quark of the same flavour and a light LSP, masses below
875\GeV have been probed under the assumption of eight degenerate light squarks.
If only a single squark is assumed to be light, this limit decreases to 520\GeV.
For the direct production of third-generation squark pairs, each
assumed to decay to a quark of the same flavour and a light LSP, masses up to
640\GeV  for bottom squarks and 450\GeV for top squarks have been probed.
In the cMSSM{\slash}mSUGRA scenario corresponding to $\tan\beta =
30$, $A_0=-2\max(m_0,m_{1/2})$, and $\mu>0$, absolute mass limits have been found to be:
$\msq > 1450\GeV$, $\mgluino > 1150\GeV$, and
$\msq=\mgluino > 1550\GeV$ when equal squark and gluino masses are assumed.

\begin{acknowledgments}
\label{sec:Acknowledgements}
We congratulate our colleagues in the CERN accelerator departments for the excellent performance of the LHC and thank the technical and administrative staffs at CERN and at other CMS institutes for their contributions to the success of the CMS effort. In addition, we gratefully acknowledge the computing centres and personnel of the Worldwide LHC Computing Grid for delivering so effectively the computing infrastructure essential to our analyses. Finally, we acknowledge the enduring support for the construction and operation of the LHC and the CMS detector provided by the following funding agencies: BMWFW and FWF (Austria); FNRS and FWO (Belgium); CNPq, CAPES, FAPERJ, and FAPESP (Brazil); MES (Bulgaria); CERN; CAS, MoST, and NSFC (China); COLCIENCIAS (Colombia); MSES and CSF (Croatia); RPF (Cyprus); MoER, ERC IUT and ERDF (Estonia); Academy of Finland, MEC, and HIP (Finland); CEA and CNRS/IN2P3 (France); BMBF, DFG, and HGF (Germany); GSRT (Greece); OTKA and NIH (Hungary); DAE and DST (India); IPM (Iran); SFI (Ireland); INFN (Italy); MSIP and NRF (Republic of Korea); LAS (Lithuania); MOE and UM (Malaysia); CINVESTAV, CONACYT, SEP, and UASLP-FAI (Mexico); MBIE (New Zealand); PAEC (Pakistan); MSHE and NSC (Poland); FCT (Portugal); JINR (Dubna); MON, RosAtom, RAS and RFBR (Russia); MESTD (Serbia); SEIDI and CPAN (Spain); Swiss Funding Agencies (Switzerland); MST (Taipei); ThEPCenter, IPST, STAR and NSTDA (Thailand); TUBITAK and TAEK (Turkey); NASU and SFFR (Ukraine); STFC (United Kingdom); DOE and NSF (USA).

Individuals have received support from the Marie-Curie programme and the European Research Council and EPLANET (European Union); the Leventis Foundation; the A. P. Sloan Foundation; the Alexander von Humboldt Foundation; the Belgian Federal Science Policy Office; the Fonds pour la Formation \`a la Recherche dans l'Industrie et dans l'Agriculture (FRIA-Belgium); the Agentschap voor Innovatie door Wetenschap en Technologie (IWT-Belgium); the Ministry of Education, Youth and Sports (MEYS) of the Czech Republic; the Council of Science and Industrial Research, India; the HOMING PLUS programme of Foundation for Polish Science, cofinanced from European Union, Regional Development Fund; the Compagnia di San Paolo (Torino); the Consorzio per la Fisica (Trieste); MIUR project 20108T4XTM (Italy); the Thalis and Aristeia programmes cofinanced by EU-ESF and the Greek NSRF; and the National Priorities Research Program by Qatar National Research Fund.
\end{acknowledgments}

\bibliography{auto_generated}

\cleardoublepage \appendix\section{The CMS Collaboration \label{app:collab}}\begin{sloppypar}\hyphenpenalty=5000\widowpenalty=500\clubpenalty=5000\input{SUS-13-019-authorlist.tex}\end{sloppypar}
\end{document}

%% file: SUS-13-019-authorlist.tex
\textbf{Yerevan Physics Institute,  Yerevan,  Armenia}\\*[0pt]
V.~Khachatryan, A.M.~Sirunyan, A.~Tumasyan
\vskip\cmsinstskip
\textbf{Institut f\"{u}r Hochenergiephysik der OeAW,  Wien,  Austria}\\*[0pt]
W.~Adam, T.~Bergauer, M.~Dragicevic, J.~Er\"{o}, M.~Friedl, R.~Fr\"{u}hwirth\cmsAuthorMark{1}, V.M.~Ghete, C.~Hartl, N.~H\"{o}rmann, J.~Hrubec, M.~Jeitler\cmsAuthorMark{1}, W.~Kiesenhofer, V.~Kn\"{u}nz, M.~Krammer\cmsAuthorMark{1}, I.~Kr\"{a}tschmer, D.~Liko, I.~Mikulec, D.~Rabady\cmsAuthorMark{2}, B.~Rahbaran, H.~Rohringer, R.~Sch\"{o}fbeck, J.~Strauss, W.~Treberer-Treberspurg, W.~Waltenberger, C.-E.~Wulz\cmsAuthorMark{1}
% ...
\vskip\cmsinstskip
% ...
1:~~Also at Vienna University of Technology, Vienna, Austria\\
2:~~Also at CERN, European Organization for Nuclear Research, Geneva, Switzerland\\
% ...
